\documentstyle[preprint,tighten,aps]{revtex}
\def\lapprox{\mathrel{\mathop
  {\hbox{\lower0.5ex\hbox{$\sim$}\kern-0.8em\lower-0.7ex\hbox{$<$}}}}}
\def\gapprox{\mathrel{\mathop
  {\hbox{\lower0.5ex\hbox{$\sim$}\kern-0.8em\lower-0.7ex\hbox{$>$}}}}}
\def\mathrm{\mbox}
\def\eg{{\em e.g.}\/}
\def\etal{{\em et al.}\/}

\def\ie{{\em i.e.}\/}
\def\text{\textstyle}
\def\chiq{$\chi ^2$}
\def\csi{\xi}
\def\tsole{t$_\odot$}

\def\lum{L$_\odot$}
\def\raggio{R$_\odot$}
\def\ksole{K_\odot}

\def\fiI{\Phi_{\mbox{\footnotesize{k}}}}
\def\fin{\Phi_{\mbox{\footnotesize{N}}}}
\def\fio{\Phi_{\mbox{\footnotesize{O}}}}
\def\fip{\Phi_{\mbox{\footnotesize{p}}}}
\def\fipp{\Phi_{\mbox{\footnotesize{pp}}}}
\def\fipppep{\Phi_{\mbox{\footnotesize{pp+pep}}}}
\def\fibe{\Phi_{\mbox{\footnotesize{Be}}}}
\def\ficno{\Phi_{\mbox{\footnotesize{CNO}}}}
\def\fibecno{\Phi_{\mbox{\footnotesize{Be+CNO}}}}
\def\fib{\Phi_{\mbox{\footnotesize{B}}}}
\def\fipep{\Phi_{\mbox{\footnotesize{pep}}}}
\def\fitot{\Phi_{\mbox{\footnotesize{tot}}}}
\def\fii{\Phi_i}
\def\fiint{\Phi_{\mbox{\footnotesize{int}}}}
\def\fimin{\Phi_{\mbox{\footnotesize{min}}}}

\def\fibessm{\Phi_{\mbox{\footnotesize{Be}}}^{SSM}}
\def\fibssm{\Phi_{\mbox{\footnotesize{B}}}^{SSM}}

\def\beqa{\begin{eqnarray}}
\def\eeqa{\end{eqnarray}}
\def\beq{\begin{equation}}
\def\eeq{\end{equation}}
\def\d{\mbox{d}}
\newcommand{\tHe}{$^3$He~}
\newcommand{\treHe}{$^3$He~}
\newcommand{\qHe}{$^4$He~}
\newcommand{\sBe}{$^7$Be~}
\newcommand{\oB}{$^8$B~}
\newcommand{\sLi}{$^7$Li~}
\newcommand{\dC}{$^{12}$C~}
\newcommand{\qN}{$^{15}$N~}

\newcommand{\sO}{$^{16}$O~}
\newcommand{\quaN}{$^{14}$N~}
\newcommand{\qO}{$^{15}$O~}
\newcommand{\tN}{$^{13}$N~}
\newcommand{\svtt}{$\langle \sigma v \rangle _{33}$}
\newcommand{\svtq}{$\langle \sigma v \rangle _{34}$}

\def\ga{$\gamma$}

\def\to{\rightarrow}
\newcommand{\unitan}{$\cdot 10^9 $ cm$^{-2}$ s$^{-1}$}

\newcommand{\unitasb}{$10^6 $ cm$^{-2}$ s$^{-1}$}
\newcommand{\unitanb}{$10^9 $ cm$^{-2}$ s$^{-1}$}
\newcommand{\unitaerg}{$10^{33}$ erg/cm$^{2}$/s}
\def\Sgamin{{\mbox{S}}_{Ga}^{min}}

\def\Sclint{{\mbox{S}}_{Cl,int}}
\def\Sclpep{{\mbox{S}}_{Cl,pep}}
\def\Sclbecno{{\mbox{S}}_{Cl,Be+CNO}}

\def\sclb{\sigma_{Cl,B}}

\def\sclpep{\sigma_{Cl,pep}}

\def\sclI{\sigma_{Cl,k}}

\def\sgab{\sigma_{Ga,B}}
\def\sgabe{\sigma_{Ga,Be}}
\def\sgapp{\sigma_{Ga,pp}}

\def\sgacno{\sigma_{Ga,CNO}}

\def\sgap{\sigma_{Ga,p}}
\def\sgaI{\sigma_{Ga,k}}
\def\sgai{\sigma_{Ga,i}}
\def\qpp{Q_{pp}}
\def\qp{Q_{p}}

\def\qb{Q_{B}}

\def\qcno{Q_{CNO}}
\def\qbe{Q_{Be}}
\def\qi{Q_{i}}
\def\qI{Q_{k}}

\def\enecno{\langle E_\nu \rangle _{\mbox{\footnotesize{CNO}}}}

\def\enebe{\langle E_\nu \rangle _{\mbox{\footnotesize{Be}}}}

\def\enei{\langle E_\nu \rangle _i}
\def\O{\Omega}
\def\ii{i}
\def\Rapporate{R}
\def\RO{\omega}
\def\alphaO{\alpha}
\def\Phii{\Phi}
\def\alphai{\alpha}
\newcommand{\ssm}{ \frac{\Delta X}{X}  }
\newcommand{\tre}{\left. \frac{\delta X}{X} \right |_{-3\%} } 
\newcommand{\sette}{\left. \frac{\delta X}{X} \right |_{-7\%} } 
\newcommand{\boh}{\left. \frac{\delta X}{X} \right |_{-13\%} } 
\def\Sppzero{S_{pp}}
\def\fpp{f_{\mbox{\footnotesize{p+p}}}}
\def\ftt{f_{\mbox{\footnotesize{$^3$He+$^3$He}}}}
\def\ftq{f_{\mbox{\footnotesize{$^3$He+$^4$He}}}}
\def\fpbe{f_{\mbox{\footnotesize{p+$^7$Be}}}}
\def\fpn{f_{\mbox{\footnotesize{p+$^{14}$N}}}}
\def\fhh{f_{\mbox{\footnotesize{He+He}}}}

 \newcommand{\laij}{$\lambda_{ij}$~}

 \newcommand{\spp}{$S_{pp}$}
  
\begin{document}
\preprint{\vbox{\noindent
To appear in Physics Reports
 \hfill astro-ph/9606180\\
          \null\hfill  INFNFE-10-96\\
          \null\hfill  INFNCA-TH9604}}
\title{Solar neutrinos: beyond standard solar models}
\author{
         V.~Castellani$^{1,2,3}$\cite{email1},
         S.~Degl'Innocenti$^{4,5}$\cite{email2},
         G.~Fiorentini$^{6,5}$\cite{email3},
         M.~Lissia$^{7,8,}$\cite{email4}
         and B.~Ricci$^{5,}$\cite{email5}
       }
\address{
$^{1}$Dipartimento di Fisica dell'Universit\`a di Pisa, I-56100 Pisa, Italy\\
$^{2}$Osservatorio Astronomico di Collurania, I-64100 Teramo, Italy\\
%$^{3}$Universit\`a dell'Aquila, I-67100 L'Aquila, Italy \\
$^{3}$Istituto Nazionale di Fisica Nucleare, LNGS, I-67017 Assergi (l'Aquila),
     Italy\\
$^{4}$Max-Planck Institut for Astrophysics, D-85740 Garching bei M\"{u}nchen, 
     Germany\\
$^{5}$Istituto Nazionale di Fisica Nucleare, Sezione di Ferrara, 
      via Paradiso 12, I-44100 Ferrara, Italy\\
$^{6}$Dipartimento di Fisica dell'Universit\`a di Ferrara, 
       via Paradiso 12, I-44100 Ferrara, Italy\\
$^{7}$Istituto Nazionale di Fisica Nucleare, Sezione di Cagliari, 
      via Negri 18, I-09127 Cagliari, Italy\\
$^{8}$Dipartimento di Fisica dell'Universit\`a di Cagliari, 
      via Ospedale 72, I-09124 Cagliari, Italy\\
        }
\date{June 1996}
\maketitle                 % Produces the title.
\begin{abstract}
After a  short survey  of the physics of solar neutrinos, giving an overview
of  hydrogen
burning reactions, predictions of standard solar models and results of solar
neutrino experiments,  we discuss the solar-model-independent  indications in
favour of non-standard neutrino properties. The experimental results look 
to be in contradiction with each other,  even disregarding some experiment:
unless electron neutrinos disappear in their trip from the sun to the earth, 
the fluxes of
intermediate energy neutrinos (those from \sBe electron capture and  from  the
CNO cycle) result to be unphysically negative, or anyway extremely reduced 
with respect to standard solar model predictions. Next  we  review extensively
non-standard solar models built as attempts to solve the solar neutrino puzzle.
The dependence of the central solar temperature on  chemical
composition, opacity, age and on the values of the astrophysical S-factors for
hydrogen-burning reactions is carefully investigated.  Also, possible
modifications of the branching among the various pp-chains in view of
nuclear physics uncertainties are examined. Assuming {\bf standard}
neutrinos, all  solar models examined fail in reconciling theory with
experiments, even when the physical and chemical inputs are radically changed 
with respect to present knowledge and even if some of the experimental results
are discarded.
\end{abstract}
\newpage
\tableofcontents 
\newpage
%%%%%%%%%%%%%INTRODUCTION%%%%%%%%%%%%
%
\addcontentsline{toc}{section}{Introduction}  
\section*{Introduction}
In a prophetical paper of 1946, Bruno Pontecorvo
\cite{Pontecorvo1946}
reviewed the arguments for the existence of neutrinos (at that time 
 based mainly on  conservation laws) and stated:\\
 
``{\em Direct proof of the existence of the neutrino \ldots must be based 
on experiments, the interpretation of which does not require the law of 
conservation of energy, i.e., on experiments in which some characteristic 
process produced by \underline{free neutrinos} \dots is observed}.'' \\

Then he pointed out the relevance  of inverse beta decay for neutrino 
detection and outlined the  physics potential of neutrinos from nuclear 
reactors. He also mentioned solar neutrinos:
``{\em The neutrino flux from the sun is of the order of 10$^{10}$ 
neutrinos/cm$^2$/sec. The neutrinos emitted by the sun, however, are 
not very energetic \ldots}''.
In 1946  Pontecorvo was not  optimistic about solar neutrino detection. 
Since then, solar neutrino physics has made enormous progress, 
with a  significant acceleration during the last decade.

From the seventies to the first half of the eighties the only result was 
that of the Chlorine experiment \cite{Davis}, essentially implying
an observed $^8$B neutrino flux lower than that predicted by
Standard Solar Models (SSMs).  
Many physicists were not convinced of the existence of a solar neutrino 
problem, mainly because  of the uncertainties in 
estimating  the flux of the rare, high energy  $^8$B neutrinos.
In 1989 \cite{Kamioka} results from the KAMIOKANDE experiment 
were presented, confirming the low $^8$B neutrino flux.
However, as  observed by Bahcall and Bethe \cite{BahcallBethe1990}, the 
comparison between 
KAMIOKANDE and Chlorine data implied an additional puzzle.
 As the Chlorine experiment is also sensitive to   $^7$Be neutrinos,
if one subtracts from the Cl signal the $^8$B contribution as implied by
KAMIOKANDE data (for standard neutrinos), no room is left 
for $^7$Be neutrinos. Thus the  problem   started involving also these 
less energetic neutrinos, for which the theoretical predictions are much 
more robust. 
At the beginning of the nineties, the results of Gallium 
experiments (GALLEX \cite{Gallex} and SAGE \cite{Sage}) became available;
for these the signal
 is expected to depend primarily  on pp neutrinos, but 
to a lesser extent
also on $^7$Be neutrinos and on $^8$B neutrinos.
 In both Gallium experiments the signals were again significantly smaller than 
 the SSM predictions.
The exposure of the Gallium 
detectors to a $^{51}$Cr neutrino source \cite{calibration,calibrationSage}
was important for confirming quantitatively the sensitivity of the apparatus
for neutrino detection.

By combining the 
 results of the various experiments, it became  possible
 to obtain relevant information on neutrinos independently of solar models.
 As we shall see,  the solar neutrino experiments tell us that the following
 assumptions are 
most likely contradictory:

\begin{itemize}

	\item Neutrino properties are correctly described by 
 the minimal standard model of electroweak interactions 
(\ie~``standard'' neutrinos).

	\item Solar energy arriving onto earth  is produced in the sun
 by nuclear reactions of the form:
 
 \centerline{$ 4p+2e^- \rightarrow $\qHe$+2\nu_e $  .}

\end{itemize}

Since there is  little doubt that the sun, as many other stars, is powered
by the conversion of hydrogen into helium, then 
 experiments point towards some non-standard neutrino 
properties and --- as  was the case in 1946 ---
energy conservation
becomes  again a key tool, albeit an indirect one, for 
the study of neutrinos.

This paper attempts to answer the following questions:

i) 
which information on neutrino properties can be obtained directly 
from solar neutrino experiments, independently --- as much as 
possible --- of solar models?

ii) 
how much room --- if any --- is still open for an astrophysical or 
nuclear physics solution of the solar neutrino puzzle?

As the SSM has to be abandoned in face of the clear 
disagreement with all experiments,  we shall present  
a systematic analysis of non-standard solar models (\ie~models
where some input parameter is varied beyond its estimated
uncertainty), in order to explore the possibility
 of reconciling theory and experiments.

As we are writing this paper, a generation of solar neutrino experiments has 
been essentially  completed.
 In preparation for the exciting future  experiments 
(SUPERKAMIOKANDE~\cite{superkam}, BOREXINO~\cite{borex}, 
SNO~\cite{sno}, HELLAZ~\cite{hellaz}, \ldots),
 it is  time to summarize what we have 
learnt so far.

In the minimal standard model of electroweak interactions, neutrinos are 
massless and consequently there is no mixing among weak flavour states. 
Also their magnetic moments vanish and they are stable. 
The experimental hint of non-standard neutrino properties 
(\ie~masses, mixing, magnetic moments, possible decay schemes, \ldots) thus 
seems to indicate some physics beyond the standard model.
The present  situation looks to us very  similar to the one described by 
Pontecorvo in 1946: just change the word {\em existence} to 
{\em non-standard properties}, in the sense that solar neutrino 
experiments together with energy 
conservation strongly point towards non-standard neutrino properties.
Much as in 1946, the future of neutrino physics now demands 
experiments  yielding decisive  {\bf direct}  evidence of 
non-standard neutrino properties. 

\subsection*{The plan of this paper}

In the first section we present a short and  simple introduction
to the field for non experts, reviewing: i) hydrogen 
burning in the sun and neutrino production, ii) the  predictions of 
Standard Solar Models (SSM) and  iii) the results of solar neutrino 
experiments.

In the next section we discuss the information on  neutrino fluxes 
which can be derived directly from experiments, almost independently of 
solar models, assuming standard neutrinos. This is the main 
part of our review, as it is meant to demonstrate that, 
{\bf for standard neutrinos }:

\begin{itemize}

\item the available experimental results appear to be
mutually inconsistent,
 even dismissing one out of the four experiments.

\item Even neglecting these inconsistencies,
the flux of intermediate energy neutrinos (Be+CNO)
 as derived from experiments is significantly smaller than
 the prediction of SSM's.
 
\item The different reductions of the  $^7$Be and 
$^8$B neutrino fluxes with respect to the SSM predictions
are essentially in contradiction with the 
fact that both $^7$Be and $^8$B neutrinos originate from the same parent 
$^7$Be nuclei. 
\end{itemize}
 
In Sec.~\ref{cap3} we describe a strategy for building non-standard solar 
models leading to reduced $^7$Be and $^8$B neutrino fluxes. 
  There are essentially two ways:
 i) producing models with smaller central solar 
 temperatures;
 ii) playing with  the nuclear  cross sections which determine the branches 
     of the fusion chain.
 In the same section we introduce some algorithms for discussing the 
 non-standard models, to be presented in the subsequent sections:
  we essentially seek parameterizations of the results (neutrino fluxes 
 and physical quantities characterizing the solar interior, \ldots) in 
 terms  of the input quantity which is being varied.  
  
In section~\ref{cap4} we examine non-standard solar models with a central
temperature $T_c$ different from the SSM prediction.
Clearly the central temperature is not a free parameter, its value
being fixed by the equations of stellar structure and by the physical 
inputs one is relying on. We present a systematic analysis for the 
variations of several physical inputs (S$_{pp}$, 
opacity tables, chemical composition, age, \ldots) affecting $T_c$.
An interesting feature common to the large majority of the models that
we examined, is that the radial temperature profiles 
 appear largely unchanged, aside from a scaling factor which can be 
evaluated, \eg, as the ratio of central temperatures. Thus, the 
profile is the same for a given value of $T_c$ independently of
which parameter is varied in order to obtain that central temperature.
This allows one
to compare experimental data with the theoretical predictions as a 
function of just one parameter, namely $T_c$.

Section~\ref{cap5} is devoted to the r\^{o}le of the nuclear cross sections 
which can influence the branching ratios of the  p-p chain. We discuss in 
detail the r\^{o}le  of the He+He reactions and we consider possible effects 
of to screening of nuclear charges by the 
stellar plasma. We also discuss the r\^{o}le of the p+$^7$Be reaction.

In section~\ref{cap6} we compare the non-standard 
solar models just built with experimental data.
We also consider hybrid models, where 
both  $T_c$ and nuclear cross sections are varied simultaneously.
All our attempts to reconcile theory and experiments having failed,
we conclude that
\begin{itemize}
\item for standard neutrinos, the 
present experimental data look in disagreement not 
only with the Standard Solar Model, but with any solar model that we are 
able 
to build.
\end{itemize}

In the Appendix A, we specify 
our Standard Solar Model in some detail. We concentrate here on our 
own calculations  as  we obviously have for these a more detailed knowledge 
of the 
physical and chemical inputs and of outputs of the code. In fact,
several groups have produced in the last few years updated and 
accurate standard solar model calculations 
\cite{BU89,BP92,BP95,P94,TCL,CL,SDF,CESAM,CD,RCVD96,DS96}.
 In this paper we shall not give a systematic presentation of these.
We also do not describe solar neutrino experiments nor 
helioseismology. All these matters are excellently covered in other 
review works, \eg,

\noindent
- Bahcall's book \cite{Bahcall1989} and papers in Review of Modern 
  Physics~\cite{BU89,BP92,BP95}, all extremely useful, centered 
  around standard solar models.

\noindent
- The report by Koshiba~\cite{Koshiba}, providing a fascinating journey
  through the experimental techniques of neutrino astrophysics.

\noindent
- The review by Turck-Chi\'eze \etal~\cite{TCreport}, again centered around  
  standard solar models, including a clear discussion of helioseismology in 
  relationship to the solar neutrino problem.

\noindent
- A recent reprint collection~\cite{collection}, presenting most of the 
 significant papers on solar neutrinos, and an excellent bibliography.
\newpage
%%%%%%%%%%%%%%%%%%  SECTION ONE  %%%%%%%%%%%%%%%%%%
%
\section{Overview}
\label{cap1}
The discovery of the enormous energy stored in nuclei early led 
astrophysicists to speculate that reactions among nuclear species were 
the source of the energy in stars:\\

``{\em Certain physical investigations in the past year make it probable to my 
mind that some portion of sub-atomic energy is actually being set free in 
a star. \ldots  Aston has \ldots shown \ldots 
that the mass of the helium atom is 
less than the sum of the masses of the four hydrogen atoms which enter into 
it. \ldots Now mass cannot be annihilated, and the deficit can only represent 
the energy set free in the transmutation. \ldots If  five per cent of a 
star's mass consists initially of hydrogen atoms, which are gradually 
being combined to form more complex elements, the total heat liberated 
will more than suffice  for our demands, and we need look no further for 
the source of a star's energy \ldots} 
(Eddington 1920)~\cite{Eddington1920}\footnote{In the same paper,
a few line ahead, one reads: 
``{\em If indeed the sub-atomic energy in the stars is being freely used to 
maintain their great furnaces, it seems to bring a little nearer to 
fulfilment our dream of controlling this latent power for the 
well-being of the human race --- or for its suicide.}''}.\\

After the discovery of the tunnel effect in 
1928~\cite{Gamow1928,CondonGurney1929}, Atkinson and Houtermans presented 
the first qualitative theoretical approach to the 
problem~\cite{AtkinsonHoutermans1929}. From the nuclear data available to 
them and the already known fact that hydrogen is the most abundant 
element in the sun and the universe, they concluded that the energy-producing 
mechanism involved primarily hydrogen, as suggested by Eddington. 
As a matter of fact, the fusion of hydrogen nuclei to form helium is still 
the only known process that can supply the required power for the long 
 solar life. 

Hydrogen burning can be represented 
symbolically  by the ``fusion'' reaction:
\begin{equation}
 \label{eq1}
		4p + 2e^- \rightarrow {\mbox{\qHe}} + 2\nu_e  \, .
\end{equation}
The total energy 
released in 
Eq. (\ref{eq1})
 is   Q= 26.73 MeV and  
only a small part of it (about 0.6 MeV) is carried away by the two neutrinos.

From the solar radiative flux at the earth, called the solar constant
$K_\odot=8.533(1\pm 0.004) \cdot 10^{11}$MeV cm$^{-2}$ s$^{-1}$  \cite{BP95},
 one immediately derives the total flux $\fitot$ 
  of electron neutrinos  arriving on the earth,
 if they are not lost {\em en route} (by neglecting the small fraction
 of energy carried by neutrinos):
\begin{equation} 
\label{eq2}
\Phi_{tot}\approx  2K_\odot/Q \approx 7\cdot 10^{10} 
{\mbox {cm$^{-2}$ s$^{-1}$}} \quad .
\end{equation}     

 The definite proof of nuclear energy 
production in the solar core lies in the detection of solar neutrinos.
 Since the cross sections for neutrino detection depend strongly
  on neutrino energy,
 the energy spectrum of solar neutrinos
has to be known and that requires a detailed knowledge of the reactions
summarized by Eq.~(\ref{eq1}).

\subsection{Hydrogen burning reactions}

In stellar interiors, nuclear interactions  between charged particles
(nuclei) occur at collision energies $E$ well below the height of the 
Coulomb barrier and are only possible due to the tunnel effect,
with a characteristic probability \cite{Gamow1928}:
\begin{equation}
\label{Ptunnel}
P\approx exp \left [ -Z_i Z_j e^2 2\pi /\hbar 
\sqrt{ \frac{\mu_{ij}}{2E}} \right ] 
\end{equation}
where  $Z_{i,j}$ are the atomic numbers of the colliding nuclei and 
 $\mu_{ij}$
 is their reduced mass.
The lightest elements thus react most easily and the conversion of 
H into He
 is the first 
chain of 
nuclear reactions which can halt
  the contraction of
a new-born stellar structure 
and settle stars in the so-called Main Sequence 
phase.

The mechanisms for hydrogen burning were first elucidated in the late 
1930s independently by von Weizs\"{a}cker and Bethe and Critchfield  
\cite{Weizsacker1937,BetheCritchfield1938,Bethe1939};
from their works it became clear 
that two different sets of reactions could convert sufficient hydrogen 
into helium to provide the energy needed for a star's luminosity, namely 
the proton-proton (p-p) chain and the CNO bi-cycle,
  which 
we are going to review briefly (see \cite{Rolfs} for a more extended 
presentation).

At the temperature and density characteristic of the solar interior, 
hydrogen burns with the largest probability  (98\%) through the pp chain. 
This develops through the nuclear reactions presented in 
Fig.~\ref{catenapp}, where we show the different possible branches.

The chain starts with a weak interaction process, either the pp or the pep
reaction.  The mechanisms for these two reactions are very similar 
and hence the ratio  of pep to pp reaction probabilities is  fixed 
almost independently of  the details of 
the  solar model;  at the 
densities of the solar interior  ($\rho  \lapprox$150 g/cm$^3$)
the three body process has a 
smaller probability.
The  deuterons produced quickly burn into \treHe as a consequence of the much
larger cross section of the electromagnetic process 
d+p$\rightarrow$\treHe+$\gamma$ .

After  $^3$He production, several branches  occur.
The hep reaction turns out to have a negligible rate and will not be 
discussed further. Even if the $^3$He abundance remains  a  factor 10$^4$ 
lower than that of  $^4$He  in the energy production region,
the $^3$He+$^3$He $\rightarrow$ \qHe +2p 
reaction is favoured with respect to  $^3$He+$^4$He$\rightarrow ^7$Be+$\gamma$, 
because the  probability of the reaction mediated by  strong interaction is
four orders of magnitude larger than that of the electromagnetic process
and because the tunnelling probability through the Coulomb barrier for 
the case of the lighter \treHe nuclei is a factor ten larger.

With the $^3$He+$^3$He reaction the chain reaches one of the possible
terminations (pp-I chain).  
On the other hand, after the $^3$He+$^4$He reaction, the chain branches 
again at the $^7$Be level, due to the competition between electron and 
proton capture. Note that, at the temperatures of the solar 
interior, the electron capture (although it is a weak process)
dominates over the 
electromagnetic reaction since it has no Coulomb barrier. 
The chain passing through the $^7$Be+e$^-$ reaction is called pp-II chain, 
and the one involving the $^7$Be+p reaction is called pp-III chain.
  
To conclude this overview of the pp chain we remark that the 
basic process in any case  is the aggregation of four protons to form 
a \qHe nucleus, the isotopes $^2$H, $^3$He, $^7$Be, $^7$Li and  $^8$B 
being intermediate products which are created and destroyed along the chain, 
their number densities staying constant (and small) when the chain reaches
equilibrium.  

Regarding the neutrino production we note that:

\noindent
i) in the pp-I chain only pp or pep neutrinos  are produced, the latter 
with a small probability (0.2\% in the standard solar model).

\noindent
ii) In the pp-II chain one pp (or pep) and one $^7$Be neutrino are emitted.

\noindent
iii) In the pp-III chain one pp (or pep) and one $^8$B neutrino are emitted.

Figure~\ref{catenaCNO} shows the main reactions in  the CN and NO cycles 
which become efficient at  rather high temperatures. The overall 
conversion of four protons to form a \qHe nucleus is achieved with 
the aid of C, N and O nuclei.
 The total energy release is clearly the same as in the pp chain.
The ratio of the astrophysical S-factors\footnote{We denote as 
$\sigma_{ij}$ the cross section for the reaction between nuclei with 
atomic mass numbers $i$ and $j$, $v_{ij}$ is the relative velocity,
$E$ is the collision energy and 
the astrophysical S-factor is:
$S_{i,j}=\sigma_{ij} exp [2\pi Z_i Z_j e^2 /(\hbar v_{ij})] E$.}
for the two reactions \qN(p,\qHe)\dC and \qN(p,$\gamma$)\sO 
being about a factor 
1000, the r\^{o}le of NO cycle is generally marginal.
In the sun essentially only the CN cycle (left side of Fig.~\ref{catenaCNO}) 
is contributing to the energy production.

For each fusion 
of four protons into \qHe  via the CN cycle one neutrino 
from the decay of \tN and another from that of \qO are produced.
Note that the efficiency of the process is determined by the reaction
with the smallest cross section, which clearly is  p+$^{14}$N 
$\rightarrow ^{15}$O +$\gamma$, an electromagnetic process with
the largest Coulomb repulsion.

\subsection{Stellar structures and standard solar models}
\label{structure}

According to the theory of stellar structure and evolution, the condition 
for hydrostatic equilibrium jointly with the conservation of energy and the 
mechanism for energy transport determine  the physical structure of a star 
and its evolution from four main inputs, namely:

\noindent i) 
The initial chemical composition.

\noindent ii) 
The equation of state for stellar matter.

\noindent iii) 
The radiative opacity, $\kappa$, as a function of
density $\rho$, temperature $T$ and chemical composition.

\noindent iv) 
The energy production per unit mass and time $\epsilon$, again as a 
function of $\rho$, $T$ and chemical composition.

For the equation of state (EOS) one has to evaluate the ionization degree 
and the population of excited states for all nuclear species. In addition, 
one has to take into account several physical effects of the stellar plasma
(like Coulomb interaction and/or electron degeneracy) which
introduce deviations from the ``perfect gas law'' prediction.
The study of EOS has improved along many years and accurate tabulations
are available (see \cite{refEOS} for a recent discussion).

The radiative opacity $\kappa$
is directly connected with the photon mean free path, $\lambda= 1/
\kappa \rho$. All throughout the internal radiative region 
$\kappa$  governs the temperature gradient through
 the well known relation 
\cite{CoxGiuli1968}:
\begin{equation}
\frac{dT}{dr} = - \frac{3}{4ac} \frac{ \rho \kappa}{ T^3} F
\end{equation}
where $F$ is  the electromagnetic energy flux. The evaluation of 
 $\kappa$ as a function of the composition of the gas and of its 
physical conditions requires a detailed knowledge of all the processes 
important for radiative flow (elastic and inelastic scattering, 
 absorption and emission, inverse bremsstrahlung, \ldots)
and in turn a detailed evaluation of  the atomic levels 
 in the solar 
interior. As a consequence, the evaluation of
$\kappa$ is a  rather complex task, 
but the results have been continuously
improving over the years. At present one relies mainly  on the 
calculations of the OPAL group at the Lawrence Livermore National Laboratories 
(see, \eg~Refs.~\cite{IGL90,IGL92,RG92}).

Expressions for (the nuclear contribution) to $\epsilon$ are derived 
essentially from the tables of nuclear reaction rates, originally 
compiled and continuously updated by Fowler, until his death 
(see Ref.~\cite{CF88} and references therein).

According to Stix \cite{Stix}, ``{\em The standard model of the sun 
could be defined as the model which is based on the most plausible 
assumptions}.'' This means that physical and chemical
inputs are chosen at the central values of 
experimental/observational/theoretical results.
We prefer to call such a model the Reference Solar Model (RSM);
throughout this review we use as such the ``best model with diffusion'' 
of Bahcall and Pinsonneault~\cite{BP95} denoted as BP95.
More generally,  let us {\bf define a Standard Solar Model (SSM) as one which 
reproduces, within uncertainties, the observed
properties of the sun, by adopting a set of physical and chemical
inputs chosen within the range of their  uncertainties}.

The actual sun has a mass 
M$_\odot=(1.98892\pm0.00025)\cdot10^{33}$g~\cite{AstronomicalAlmanac1990}
and a radius  R$_\odot=(6.9598\pm 0.0007)\cdot10^{10}$cm~\cite{Allen}.
Denoting by X, Y and Z the relative mass abundances of H, He 
and heavier elements respectively (X+Y+Z = 1), the present ratio of 
heavy elements (metallicity) to hydrogen in the photosphere is   
(Z/X)$_{photo}=0.0245(1\pm0.061)$\cite{GN93,BP95}. At the age 
t$_\odot = (4.57\pm0.01)$ Gyr \cite{BP95}, the sun is producing a luminosity 
L$_\odot=3.844(1\pm0.004)\cdot10^{33}$ erg/sec \cite{BP95}.
 
If a complete information about the (initial) photospheric composition 
was available and if the theory of stellar models was
capable to predict  stellar radii firmly, then there would
be no free parameter. The theoretical model should account
for the solar luminosity and radius at the solar age, without tuning 
any parameters.

The photospheric helium abundance is however  not strongly constrained 
by direct observations, since no helium line exists at the photospheric 
depth, see Ref.~\cite{AG89} for a critical discussion
(helioseismology can however provide some indirect information, see later).
Furthermore, the present observed photospheric composition is different 
from the initial one due to diffusion and gravitational settling.
Moreover, the radii (but only the radii) of stars with convective 
envelopes (\eg~the sun) depend on the assumption made about the 
convective transport, as we shall discuss later on. As a consequence, one 
has the freedom of adjusting some parameters in the course of the calculation.

In order to produce a standard solar model, one studies the evolution
of an initially homogeneous solar mass up to the sun age.
 To obtain L$_\odot$, R$_\odot$ and (Z/X)$_{photo}$ at 
\tsole~, one can tune three parameters:
the initial helium abundance Y, the metal abundance Z
and a third quantity related to the efficiency of convection
(see below).

The effects of these parameters can be  understood simply. The luminosity
of the sun (more generally of any star in the main sequence) 
depends in a rather sensitive way on the initial helium content Y; 
increasing  it, the initial sun is brighter and a given luminosity
is reached in a shorter time.
Since  the ratio Z/X is constrained by observational data,
Y and Z cannot be 
chosen independently: if Y increases, Z must 
decrease. 

To get the proper radius R$_\odot$, one  adjusts the efficiency of the external 
convection, which dominates the energy transport in the outer layers of 
the sun (about the outer 30\% of the solar radius). The precise 
 description of the convection in the external part of the sun 
is an essentially unsolved problem, and the process is commonly 
described in terms of a phenomenological model, the so-called ``mixing 
length theory'', see \eg~Ref.~\cite{CoxGiuli1968}.
Following this approach, we define the mixing length $l$ 
as the distance over which a moving unit of gas can be identified 
before it mixes appreciably. This length is related to the pressure 
scale height $H_p$=1/ ($d$lnP/$d$r) through
\begin{equation}
l= \alpha \, H_p \, ,
\end{equation}		
where $\alpha$ is assumed to be independent of the radial coordinate
and it is used as free parameter.
By varying  $\alpha$, one varies the mixing length and thus the 
convection efficiency, which determines the solar radius. 
Thus if $\alpha$ is increased, convection becomes more efficient, 
the temperature gradient  smaller
and the surface temperature  higher. Since the solar luminosity is
fixed, the radius has to decrease.

In the last few years, one of the principal improvements of SSMs was the
treatment of element diffusion
\cite{NOER,COX,Loeb,PM,BP92,CD,P94,Thoule,BP95,DS,DS96,RCVD96}. 
Noerdlinger (1977) first included the effects of He diffusion on the 
evolution of the sun, while Cox, Guzic and Kidman (1989) were the first 
to take into account the settling of heavy elements. The stronger pull 
of gravity on helium and heavier elements causes them to diffuse slowly
downward, relative to hydrogen.
As a result, if diffusion is taken into account, helium and 
heavier elements in the solar photosphere are depleted with respect to 
their abundance in the original mixture.

We remark that, as a strong simplification, the sun is taken as 
spherically symmetric, \ie~all physical quantities vary only
radially.
This relies on the assumption that the internal rotation is sufficiently 
slow~\cite{helionature1995} and the internal magnetic fields are 
sufficiently weak so that the corresponding forces are negligible.

All in all, it looks that
 a solar model has three (essentially) free parameters, $\alpha$,
Y, and (Z/X)$_{in}$, and produces three numbers that can be directly measured:
the present radius, luminosity and heavy element content of the photosphere.
This may not look as a big accomplishment.
At this stage, one's confidence in the Standard Solar Models actually rests
on the success of stellar evolution theory to describe many, and
more complex, evolutionary phases in good agreement with observational data. 

In recent years however, helioseismology has added important data
 on the solar structure, which provide severe constraints
and tests of standard solar model calculations.
Helioseismology can accurately determine the depth of the convective
zone and the speed of sound $c_b$ at the transition radius
R$_b$ between convective and
radiative transport~\cite{sismo}:
\begin{eqnarray}
\label{ranges}
\mbox{R}_b/\mbox{R}_\odot &=& 0.710-0.716 \\ \nonumber
 c_b&=& (0.221-0.225) \, \mbox{Mm/s}
\end{eqnarray}
The indicated range for R$_b$ has been  confirmed recently~\cite{RCVD96}.

Actually, within the present uncertainties, the information on R$_b$
and on $c_b$ are not independent, since to the per cent level, $c_b$
and R$_b$ are related through \cite{sismo}:
 \begin{equation}
\label{rbcb}
c_b^2\simeq \frac{2}{3} \frac{G {\mbox{M}}_\odot}{{\mbox{R}}_\odot}
\left ( \frac{{\mbox{R}}_\odot}{{\mbox{R}}_b} -1 \right )
\end{equation}

 Several determinations of the helium photospheric
abundances have been derived from inversion
(deconvolution) of helioseismological data,
yielding (see Table~\ref{elios}):
\begin{equation}
\label{yrange}
 {\mbox{Y}}_{photo}= 0.233-0.268 \quad .
\end{equation} 
One should note that the small errors often quoted generally
reflect the observational frequency errors only. 
The results actually depend on the method of inversion and on the starting 
physical inputs, \eg~the EOS. Such a dependence has been recently 
studied in Ref.~\cite{RCVD96}.

{\bf With these  additional constraints, Eqs. (\ref{ranges}) and (\ref{yrange}),
there are essentially
 no free parameters  for SSM builders.}

\subsection{Results of  standard solar models}
\label{sec13}

Table~\ref{Modelli} shows the main results of  recent  
solar models produced by various authors; 
{\bf all  of them include microscopic 
diffusion of helium and heavy elements}, using (slightly) different 
physical and chemical inputs, which are summarized in
Table~\ref{modelliinput}.

 By varying Y and $\alpha$, and adjusting (Z/X)$_{in}$ within 
 the observed range,
  all models are able to 
 reproduce  \lum, \raggio and (Z/X)$_{photo}$ at the solar age.

 The comparison with helioseismological measurements, which provide 
 additional constraints, is interesting, see Fig. \ref{figbarbara} from
\cite{Barbara}.
Only some of the recent solar models with microscopic diffusion are in 
agreement with the helioseismological constraints. With the notation of 
Table \ref{Modelli}, they are P94 \cite{P94}, RVCD96 \cite{RCVD96}, 
BP95 \cite{BP95}  and FRANEC96 \cite{Ciacio}.
On the other hand, all models without diffusion fail.
The importance of diffusion for achieving agreement with
helioseismological data is discussed in appendix A.

{\bf  We shall consider as Standard Solar Models
(SSMs) only those models which pass the helioseismological tests}.
We recall that, among these,
{\bf we shall refer to BP95 ``best model with diffusion'' \cite{BP95} as to 
the Reference Solar Model (RSM)}. 

From Table~\ref{Modelli} one sees that calculated central temperatures
are in agreement at the per cent level.

As a matter of fact, this shows that building up a stellar
model is by now a well-established and reliable procedure, the small
differences being essentially the result of small variations in the
assumed input physics. At the same time, this agreement indicates the 
consensus about the treatment of the rather sophisticated physics needed 
to account for the behaviour of stellar structures.

 Concerning neutrinos, the  solar model determines
the internal distribution of temperature, density and H abundance
and, thus, the efficiency of the various chains.
In this way one gets the fluxes at earth of the different
neutrino components ($\fipp, \fipep, \fibe, \ldots$). From 
Table~\ref{Modellibis} one notes that:

%\noindent
i) 
the fluxes of pp neutrinos  are stable to a few percent.
As they are the most abundant ($\fipp\approx \fitot$) by far, 
their flux is  directly related to the solar 
luminosity.

%\noindent
ii)
The ratio
\beq
\label{csi}
\csi= \fipep/(\fipep + \fipp)
\eeq
 is almost independent of solar models, as it is weakly sensitive to 
the solar temperature and density.
 
iii) 
After the pp, the \sBe neutrinos are the most abundant, their flux 
accounting for about 8\% of the total, and again the prediction is
rather stable  among
the different calculations with the exception of DS96.

iv) 
The spread of the calculated \oB neutrino fluxes is much larger
 showing the larger sensitivity to the different
physical inputs used in different solar models. 

v) 
Concerning neutrinos from the CN cycle, again their prediction is
somehow model-dependent\footnote{In \cite{DS} Shaviv
 presented values of $\fin$ and $\fio$ an
order of magnitude  smaller than those of all other calculations. This
feature is no more present in the more recent calculation by the
same author \cite{DS96}.}.

 For a full  CN cycle  one should clearly have the same 
 number of \tN and \qO neutrinos. Actually,
 the reaction \quaN(p,\ga)\qO is too slow 
in the solar region below
 10$^7$ K and the chain has 
 not reached equilibrium, still favoring the transformation of \dC into \quaN.
 This is  why 
 \tN neutrinos are slightly more abundant than the  \qO neutrinos,
 \ie, if one defines
  \begin{equation}
  \label{eta}
  \eta=\fin /  (\fin+\fio) \, ,
  \end{equation}
  one finds $\eta$ slightly larger than 0.5. Note that
$\eta =0.53$ in most calculations.
Table~\ref{Modellibis} does not include results on $^{17}$F neutrinos, 
which are two orders of magnitude less abundant than \tN or 
\qO neutrinos\footnote{
The reader will note that the model  DS96 \cite{DS96} is the one
yielding the smallest prediction on \sBe, CNO and  \oB neutrino fluxes
(Table \ref{Modellibis}) and the smallest central temperature. We remark
that this model is not a standard solar model
according to our definition, 
since it does not satisfy the helioseismological constraints.
In addition the astrophysical S-factors 
used in \cite{DS96} seem to us not acceptable, see Sects. \ref{subsec4spp}
 and  \ref{cap5}.}.

We find useful to group some neutrinos according to their origin, putting
pp and pep neutrinos in the same group  
   \begin{equation}
  \fip=\fipp+ \fipep \, ,
  \end{equation}
and into another  the neutrinos from the CN cycle:
   \begin{equation}
  \ficno=\fin+\fio \, .
  \end{equation}
  Following the common terminology, we call these latter the ``CNO 
  neutrinos'', although they are neutrinos just from the CN cycle.
  
 Figure~\ref{profilo} shows the production region in the sun for the main
neutrino components. Because of the strong temperature dependence,
\oB neutrino production is peaked at a very small distance from the solar 
center (R$\approx$0.04 R$_\odot$).
The same holds for \tN and \qO neutrinos, which are produced 
with equal rate  up to R$\approx$0.14 R$_\odot$. The secondary peak at
 R$\approx$0.16 R$_\odot$
is due to \tN neutrinos only, in a region where the CN cycle
is marginally active, C and N have not reached  their
equilibrium values and nuclear reactions are transforming
the more abundant $^{12}$C into $^{14}$N. 
The \sBe neutrino production peaks at R$\approx$0.06 R$_\odot$, 
whereas for pp+pep
neutrinos it  peaks at R$\approx$0.1 R$_\odot$
 and 
extends over a large portion of the solar
core.

\subsection{Solar neutrino spectrum and the predicted neutrino signals}
  
{\bf The energy spectrum of each component
(pp, pep,\sBe, \oB, \ldots) is determined by kinematics and/or 
nuclear physics and it is (essentially) independent of solar physics}. 

The  pp neutrinos have a continuous energy spectrum extending
up to $E_\nu = 0.42$ MeV \cite{BU89}. For a  pep neutrino, its  momentum has to 
match that of the produced deuteron and consequently its energy 
is uniquely determined by kinematics, giving $E_\nu$= 1.442 MeV. 
The energy of  $^7$Be neutrino  depends on the state of \sLi
produced in the electron capture reaction. One has $E_\nu$ = 0.861 MeV
with 90\% probability (\sLi ground state), and 
$E_\nu= 0.383$ MeV with  10\% probability (first excited state).
The energy spectrum of $^8$B neutrinos  extends up to  $E_\nu =15$ MeV
 \cite{BU89,spettroboro}. 
Apart for the very rare hep neutrinos, $^8$B neutrinos are thus the 
 most energetic ones. 
In the CN cycle, the neutrinos from $\beta ^+$ decay of 
\tN and \qO  have  end points
at 1.20 and 1.73 MeV, respectively.
 
 In Table~\ref{SIGMA} we give the average energy for each neutrino 
 component. The average energies for the p and CNO neutrinos
have been calculated by taking the values of $\eta$ and $\csi$ as in the 
RSM.

The solar neutrino spectrum predicted by standard solar models is 
shown in Fig.~\ref{spettro} (from Ref.~\cite{Bahcall1989}). 
Note that the pp neutrinos, which are the most abundant and the most
reliably predicted ones,
are also those with the lowest energy and thus are most difficult to
detect, since neutrino cross sections  increase
with neutrino energy.  On the other hand, $^8$B neutrinos, which are
rare and delicate to predict, are those with the highest energy so
that their detection is  comparatively ``easy''.

The  solar neutrino signal in radiochemical experiments is expressed as the 
reaction  rate  per target atom. A suitable unit is 
the Solar  Neutrino Unit (SNU), defined as one reaction  per second per 
 10$^{36}$ 
atoms. A 1 SNU signal means that in a target containing 10$^{31}$ atoms 
(of the order of several hundreds tons) there is about one reaction per day;
this already gives an idea of the low counting rate
 of these radiochemical experiments.
  The signal $S$ 
 is expressed in terms of the neutrino interaction cross section 
 $\tilde{\sigma}(E_\nu)$ and the  differential neutrino flux $d\Phi/dE_\nu$:
 \begin{equation}
 S = \int _{E_{th}} dE_\nu \tilde{\sigma}(E_\nu) d\Phi/dE_\nu \, ,
\end{equation}
where $E_{th}$ is the energy threshold for the detection.
In the case of standard neutrinos,  the spectrum can be written
for each component 
 as 
\begin{equation}
d\fii/dE_\nu = \fii df/dE_\nu \, ,
\end{equation}
where $df/dE_\nu$ is normalized to unity  and it is fully determined from 
nuclear physics and kinematics.
The signal can thus be expressed as the sum of contributions arising 
from each neutrino component
\begin{equation}
\label{totalsign}
S=\sum_i \sigma_i \fii   \, ,
\end{equation}
where
\beq
\label{sigmamedio}
\sigma_i =\int dE_\nu \tilde{\sigma}(E_\nu) df/dE_\nu
\eeq
is the detection cross section averaged over the spectrum of 
the $i$-th neutrino component, and  is again  determined  purely
from nuclear physics.

{\bf In summary, nuclear physics is responsible for $\sigma_i$, whereas
solar physics determines $\fii$}.

 Of course $\sigma_i$ depends on the reaction which 
one is considering and we present in Table~\ref{SIGMA} the quantities 
relevant for the two reactions so far used in the experiments
\begin{equation}
\label{sedici}
	\nu_e+^{37}Cl \rightarrow e^- +  ^{37}Ar \quad ({\mbox{ E$_{th}$=0.814 
	MeV}})
\end{equation}
and 
\begin{equation}
\label{diciassette}
	\nu_e+^{71}Ga \rightarrow e^- + ^{71}Ge \quad ({\mbox {E$_{th}$=0.233 
	MeV}}) \, ,
	\end{equation} 
where the threshold energies are indicated in 
parenthesis\footnote{Note that $\nu_\mu$ and $\nu_\tau$ cannot 
induce reactions (\ref{sedici}) and (\ref{diciassette}) as the 
 corresponding charged lepton 
($\mu$ or $\tau$) would require neutrino energies orders of magnitude 
larger than those appropriate for the sun.}.

The cross
sections increase with the  neutrino energy. This is a general 
feature of weak interaction cross sections at energies well below
the Fermi energy scale. In addition, this trend is more evident at low 
energies  
due to the presence of reaction thresholds, so that only neutrinos with 
high enough energy can induce the process. 
Note that  the \oB neutrinos, which are the most energetic ones, have 
cross sections three order of magnitude larger than the others, and 
this compensates for their significantly smaller flux.

The contributions ($\sigma_i  \fii$) to the signal in the Chlorine and 
Gallium experiments  from each component of the neutrino flux are 
shown in Table~\ref{contribution}, as predicted by the RSM.
For  Chlorine  most of the signal comes from
\oB neutrinos, whereas for Gallium the largest contribution 
is from pp neutrinos. This is a consequence of the quite different energy 
thresholds for the two reactions.

Table~\ref{Modellibis} shows the total signals in the Chlorine 
and Gallium detectors,
\begin{equation}
\label{segnali}
S_{Ga}= \sum_i \sigma_{Ga,i} \fii \quad \quad
S_{Cl}= \sum_i \sigma_{Cl,i} \fii \, ,
\end{equation}
as predicted by the different SSMs.

For the Chlorine signal the predicted values
vary  among the SSM within 20\%,
mainly due to the variations of the \oB flux.
The  spread
of  predictions for the Gallium experiments
 is  smaller,  reflecting the rather stable estimates
on pp and \sBe neutrino fluxes.

Radiochemical experiments are not the only way to detect
neutrinos.  As an alternative approach one can use the
neutrino-electron elastic scattering
\beq
\label{scattering}
\nu+ e^- \rightarrow \nu+e^- \quad ,
\eeq
detecting the scattered electron.

Whereas for neutrino interactions with nuclei the
cross section depends on nuclear physics and it is thus subject to
some uncertainties, the cross section for scattering on electrons
is well known from elementary particle physics.
Both charged and neutral current interactions occur for electron 
neutrinos. For $\nu_\mu$ and $\nu_\tau$ {\bf only} the neutral current 
interaction contributes and this results in a reduced cross section,
\eg~for KAMIOKANDE $\sigma_{\nu_\mu} \approx (1/7) \sigma_{\nu_e}$~.
If one detects electrons with energy $E_e$ larger
than some minimal value $E_o$, the signal is:
\begin{equation}
S(E_e>E_o)= \int_{E_o} dE_e dE_\nu \frac{d\sigma (E_\nu, E_e)}{dE_e} \frac 
{d\Phi (E_\nu)}{dE_\nu} \, ,
\end{equation}
where $d\sigma/dE_e$ is the differential neutrino-electron elastic 
cross section, which depends on the energy of the scattered electron $E_e$.
Because of background limitations, $E_o$ is generally 
larger than a few MeV, so that only \oB neutrinos are detected.

{\bf Assuming standard neutrinos, the signal can thus be translated into an
effective (energy integrated) \oB neutrino flux}, and data are generally
presented in that form. Note, however, that this ``experimental flux''
can be taken as the true \oB neutrino flux only for standard neutrinos: 
should $\nu_e$ transform into $\nu_\mu$, the detection cross section 
would be different.

\subsection{A look at  experimental results}
\label{seclook}

So far we have results from four solar neutrino experiments; see 
Table~\ref{EXPE} for a summary and Refs.~\cite{Bahcall1989,Koshiba}
for detailed reviews.

The KAMIOKANDE  experiment \cite{Kamioka}, located in the Japanese Alps,
detects  the Cerenkov light emitted by electrons that are scattered in 
the 
{\bf forward} direction by solar neutrinos, through the reaction
(\ref{scattering}).

 This experiment, being sensitive to 
the neutrino direction, is the prototype of neutrino telescopes and 
is the only real-time experiment so far. 

The experiment is  only sensitive to the high energy neutrinos 
($E_\nu \geq$ 7 Mev) from \oB decay.
The solar neutrino spectrum deduced from KAMIOKANDE  is in 
agreement (within  large uncertainties) with that of neutrinos 
from \oB decay in the laboratory \cite{Kamioka}.
Assuming  that the spectra are 
the same (\ie~standard $\nu_e$), one gets for the 
\oB neutrino flux the result shown in 
Table~\ref{EXPE}. This corresponds to about 500  neutrino events 
collected in a 5.4 year data-taking period.
Although there is no really sound way of combining statistical and 
systematical errors, we use the usual recipe and combine them quadratically
(see next section);
here and in the following , we take
\begin{equation}
\label{ska1}
\Phi_{KA} = (2.73\pm0.38) \cdot 10^6 {\mbox{cm$^{-2}$s$^{-1}$}} \, .
\end{equation}
This value is less than one half that predicted by the RSM.

 All other experiments use 
radiochemical techniques. The $^{37}$Cl experiment of Davis and 
coll.~\cite{Davis} has been the first operating solar neutrino detector. The 
reaction used for neutrino detection is the one 
proposed by Pontecorvo in 1946 \cite{Pontecorvo1946}:
\begin{equation}
	\nu_e+ {}^{37}\text{Cl} \rightarrow e^- +  {}^{37}\text{Ar} \, .
\end{equation}    

The energy threshold being 0.814 MeV, the experiment is 
sensitive mainly to \oB neutrinos, but also to \sBe neutrinos. The 
target, containing $10^5$ gallons of perchloroethylene, is located in the 
Homestake gold mine in South Dakota.
Every few months a small sample of $^{37}$Ar (typically some fifteen atoms!)
is extracted from the tank and these radioactive atoms are counted in 
low background proportional counters. The result, averaged over more than 20 
years of operation, is~\cite{Davis}
\begin{equation}
\label{Scl}
	S_{Cl} = 2.55\pm0.25  \quad {\mbox {SNU}} \, .
\end{equation} 
The theoretical expectation is higher by a factor three.  For almost 20
years this discrepancy has been known as the ``Solar Neutrino Problem". 
{\bf About 75\% of the total theoretical rate
is due to  \oB neutrinos}, see 
Table~\ref{contribution}, and 
hence it was for a long time believed that the
discrepancy was due to the difficulty in predicting this rare source.

Two radiochemical solar neutrino experiments using $^{71}$Ga are operating:
GALLEX, located at the Gran Sasso laboratory in Italy and using 30 tons 
of Gallium in an aqueous solution, and SAGE, in the Baksan valley 
in Russia, which uses 60 tons of gallium metal. The neutrino 
absorption reaction is
\begin{equation}
	\nu_e+^{71}Ga \rightarrow e^- + ^{71}Ge \, .
\end{equation}  

The energy threshold is $E=0.233$~MeV, 
and most of the signal arises from pp neutrinos with a significant
contribution  from \sBe neutrinos as well (see  Table \ref{contribution}).
In each experiment, the rate of neutrino interactions in the Gallium tank 
is about 0.5 event per day.
The Germanium atoms are removed chemically from the Gallium and the 
radioactive decays of  $^{71}$Ge  (half-life=11.4 days) are detected in small 
proportional counters.  The results of the two experiments, see 
Table~\ref{EXPE}, can be combined (see next section) to give
\begin{equation}
\label{Sga}
	S_{Ga} =  74\pm8 \quad {\mbox {SNU}},
\end{equation} 
which we shall use as the representative value of the Gallium signal.
Again the value is almost a factor two below the theoretical prediction.

An overall efficiency test of the GALLEX 
detector has been  performed by using an intense $^{51}$Cr neutrino 
source, ($61.9\pm 1.2$) PBq~\cite{calibration}\footnote{The preliminary 
results on $^{51}$Cr neutrino source measurements by SAGE collaboration
have been presented  recently in \cite{calibrationSage}.}.
The source, produced via neutron irradiation of about 36 Kg of 
cromium enriched in $^{50}$Cr, primarily emits 756 keV neutrinos. 
It was placed for a period of a few months inside the GALLEX tank, 
to expose the gallium chloride target to a known neutrino flux.
The number of observed neutrino events agrees with expectation to the
10\% level.
 This result
``{\em  (a) provides an overall check of GALLEX, indicating that there are no 
significant experimental artifacts or unknown errors at the 10\% level that 
are comparable to the 40\% deficit of observed solar neutrino signal
(b) directly demonstrates for the first time, using a man-made neutrino 
source the validity of the basic principles of radiochemical methods used 
to detect rare events \ldots}''~\cite{calibration}. And last but not least:
``{\em because of the close similarity in neutrino energy spectra 
from $^{51}$Cr and from the solar \sBe branch, this source experiment 
also shows that the Gallium detector is sensitive to \sBe neutrinos with 
full efficiency.}''~\cite{calibration}.

\subsection{A remark on errors}
\label{secerror}

Because of the subtleties of experiments,  of
 the difficulty in estimating errors of theoretical 
evaluations and in face of the persistent discrepancy between experimental 
results and standard theory (standard solar models and standard neutrinos),
the discussion about the meaning of errors is lively at almost any 
conference on solar neutrinos. In this section we intend to define the 
errors we adopt, describe what we want to do with these errors, provide
some justification to our choice and outline its limits. 

Concerning experimental errors, we  add in quadrature statistical and 
systematical errors. We tentatively use this global error as an 
indicator of the distance between the true and the measured values of a 
physical quantity, according to the rules  of Gaussian  statistics.

Essentially we are assuming that the systematical uncertainty is 
a variable which can fluctuate according to statistical rules, so that we 
consider statistical and systematical errors as independent of each 
other and on the same footing.
This choice seems to be crazy (and actually it would be in many cases), 
however it has some justification in the present context.

As an example, consider measuring  the period of a pendulum 
using a watch. This watch is guaranteed to measure time better than
one second per day: one does not know if it is fast or slow, but it 
does not gain or loose  more than 1s/day. 
This systematical error cannot by traced to a single origin (otherwise 
it would be eliminated), but arises from many different sources, 
essentially the mechanical tolerances on each of the many components of 
the watch. Each of these sources gives a small contribution, positive
as well as negative, to the final error. Also each contribution
is statistically independent from the others.
Under these conditions one can reasonably suppose that, if one produces
a large number of watches of the same kind, the systematical errors
are distributed according to a gaussian with null average,
\ie~the distribution is the same as for statistical fluctuations.

Clearly one does not gain any accuracy by repeating the measurements 
over several oscillations, in exactly the same conditions with the same watch.
 On the other hand if one is using N different watches of the same kind  
(each one with its  systematical error $\epsilon _o= \pm1$s/day) and 
combine the results, accuracy is improved, as the chance of having  
a fast and a slow watch are equal.
In other words, in this example systematical errors can be treated 
similarly to statistical fluctuations.

The situation, at least in radiochemical experiments, is somehow similar.
For instance the 33  runs of GALLEX are actually 33 independent 
experiments: after each run, the $^{71}$Ge atoms are extracted in 
specific conditions, which are independent of the run number, and placed 
in a new counter. The radon content of the atmosphere where extraction 
occurs fluctuates around some mean value, within some systematical 
uncertainty. All counters have approximately the same efficiency, again 
within some systematical uncertainty, but one is better and another is worse
than the estimated mean efficiency. Thus the systematical error of the 
full experiment is significantly reduced with respect to that 
pertaining to each individual run and,  it can approximately be determined 
by statistical considerations. 

The same procedure used to combine different runs can be used in principle
to combine two different experiments like SAGE and GALLEX. 
The resulting systematical error is
 \begin{equation}
 1/\epsilon^2(syst) = 1/\epsilon^2_{GALLEX}(syst) 
                    + 1/\epsilon^2_{SAGE}(syst) \, ,
\end{equation}
in analogy to the statistical one
  \begin{equation}
 1/\epsilon^2(stat) = 1/\epsilon^2_{GALLEX}(stat) 
                    + 1/\epsilon^2_{SAGE}(stat) \, ;
\end{equation}
and the global error on Gallium result becomes
\begin{equation}
\label{somme}
\epsilon^2= \epsilon^2(syst) +\epsilon^2(stat) \, .
 \end{equation}
This is how Eq.~(\ref{Sga}) has been obtained.

The situation is  more complex  when considering errors of 
theoretical results. When these errors correspond to uncertainties 
on some measured quantities which are used as input in the theoretical 
calculations, we are back to the previous case.
Often, however, additional 
uncertainty is added due to extrapolations of experimental data, 
\eg~when estimating  nuclear cross sections at energies of  astrophysical 
interest. In other cases (\eg~opacity tables) the only information
comes from theoretical calculations. In these cases one can only resort to 
a comparison among different calculations. A total theoretical range 
$\epsilon_{TTR}$ is defined by Bahcall as ``{\em the range in values of 
published state-of-the-art calculations}''~\cite{Bahcall1989,BU89}, and 
this might be considered as the equivalent of an experimental 
three sigma error, \ie~the probability of being outside this range
is about 0.3\%. This looks plausible, how many sigmas being clearly a 
matter of taste. 
We generally adhere to Bahcall's choice. For consistency of notations, as we 
always show experimental 1$\sigma$ errors, we shall add to the theoretical
quantities an error
\begin{equation}
  \label{teorico}
   \epsilon_{theo}=\frac{1}{3} \epsilon_{TTR} \, .
\end{equation}
  
All this clearly means that {\bf treating errors given by equations of the 
form (\ref{somme}) or (\ref{teorico}) as if they were indicators of 
statistical (Gaussian) fluctuations is just a rough approximation, or 
even an undue simplification}. 

The confidence levels obtained in this way are thus to be taken 
{\em cum grano salis}, or much  more.
\newpage
%%%%%%%%%%%%%%%%% SECTION TWO %%%%%%%%%%%
%
\section{Neutrino fluxes almost independently of solar models}
\label{cap2}

All experiments give signals significantly smaller than those
predicted by the RSM and by other SSMs, for standard neutrinos.
Is this a problem of SSMs or is there anything deeper? 
Insight on this question can be gained by looking at properties of
neutrino fluxes which are largely independent of solar models.
In essence, one is asking if experimental data are in agreement with the 
following assumptions:

 A) 
the solar luminosity is supported by H burning reactions; 

 B) 
electron neutrinos do not disappear in their travel from sun 
 to earth (\ie~standard neutrinos).

The first assumption looks rather innocent, the second one being really 
the hypothesis to be tested. We shall follow this approach by discussing:
a lower bound on the Gallium signal for standard neutrinos 
(Sec.~\ref{2min});
the relationship between KAMIOKANDE and Chlorine signals 
(Sec.~\ref{clorokam});
the  information on intermediate energy neutrinos derived by 
using the full set (or any subset) of the data (Sec.~\ref{bounds}).

We shall show that the experimental results look mutually inconsistent
if the above two assumptions hold true.
The point is that the flux of intermediate energy neutrinos
($E_\nu \approx 0.5$--2 MeV):
\begin{equation}
\fiint=\fibe+\ficno  
 +\fipep
\end{equation}
would have to be negative!

We shall see that the same occurs if we disregard any of the 
experiments. Actually {\bf we do not know of any sound argument to doubt the 
experimental results, but we shall entertain this possibility, 
just as a working hypothesis}, to test the consistency of the 
assertion: neutrinos are standard and (some) experiments are correct.
We would like to make this point as clear as possible: when saying
``disregard one experiment'' or ``assume it is  wrong'' we are
advocating some hypothetical, large, unknown systematic error, just for the
sake of saving/testing the hypothesis of standard neutrinos. This
is clearly a desperate hypothesis, at least for $^{51}$Cr-calibrated
Gallium detectors.

The inconsistencies will be derived by treating the 
different components of the 
neutrino flux as essentially free parameters. Additional 
information can be gained  from the physical  relationships 
among the neutrino components,  as  imposed  by the development of 
the fusion chains.
For example, pep neutrinos accompany pp neutrinos, with a proportion 
which has to be essentially the same in any solar model. Also, if the CN cycle 
is efficient, there will be about as many \tN as \qO neutrinos. By using 
these additional, however weak, assumptions one gets  more severe 
hints against the hypothesis of standard neutrinos and/or tighter constraints
on \sBe and CNO neutrino fluxes (see Secs.~\ref{solveequation} and 
\ref{where}). Quantitative
statements about the chances of standard neutrinos and upper
bounds on intermediate energy neutrino fluxes are presented
in Sec.~\ref{probabilities}. These bounds appear in conflict with
the results of SSM, particularly for \sBe neutrino flux
(Sec.~\ref{standard}).

Futhermore, since \sBe and  \oB neutrinos both originate from
\sBe nuclei, there is a  relationship between them which 
looks again in  contradiction with the  experimental data, for
standard neutrinos (Sec. \ref{bebo}).

All these (essentially) solar model independent arguments point towards 
some non-standard neutrino properties, the main ingredient being the
luminosity constraint A). In Sec.~\ref{luminosity} we relax
even this hypothesis, still finding indications towards non-standard
neutrinos. On the same grounds, we discuss in Sec.~\ref{universal}
the case of  universal neutrino oscillations in the framework of a solar model 
independent approach. 

Before presenting the discussion let us advance a warning.
We regard this section as the most
relevant one 
of the present
review 
and  we therefore tried to be as detailed as possible.
For a first reading  we recommend Secs. \ref{lumcons} and ~\ref{2min},
where the basic ideas are explained, and Secs.~\ref{solveequation} and
\ref{where}, which contain the main results, and the concluding remarks 
at the end of this section~\ref{cap2}.

\subsection{The solar luminosity constraint}
\label{lumcons}

The solar luminosity constraint, essentially given
in Eq. (\ref{eq2}), will be a major ingredient of this section, so that
it is important to express it more precisely and to discuss it accuracy.

Assuming that all solar energy 
originates from nuclear reactions, and that all these
reach completion, \ie~:
\begin{equation}
\label{eq2bis}
	4p + 2e^- \rightarrow {\mbox{\qHe}} + 2\nu_e  \quad ,
\end{equation}
one immediately gets the following constraint for the neutrino fluxex $\fii$:
\begin{equation}
\label{lum1}
   K_{\odot}=\sum_i \left( \frac{Q}{2} - \langle E_\nu \rangle _i \right ) 
\Phi_i \, ,
\end{equation}
where $Q=26.73$ MeV and $\enei$ is the average energy carried by 
$i$-th neutrinos. The coefficients
\begin{equation}
\label{qf}
\qi= \frac{Q}{2} -\enei
\end{equation}
represent the average electromagnetic energy released per emitted 
neutrinos, and can be calculated  from Table \ref{SIGMA}.
Except for the rare \oB neutrinos, $\enei$ is of the order of half MeV, 
so that Eq. (\ref{eq2})
--- corresponding to Eq. (\ref{lum1}) when $\enei$ is neglected ---
 is accurate to the five per cent level.

Concerning the accuracy of Eq. (\ref{lum1}), the following comments 
are in order, see
 \cite{BK}\footnote{In the same 
 paper one derives direct upper bounds on neutrino fluxes
by using Eq. (\ref{lum1}) }:

i) gravitational energy generation is neglected. According to standard 
solar model
calculations, this causes an error of about $3\cdot 10^{-4}$ \cite{BU89}, 
negligible
in comparison with the uncertainty on $K_\odot$.

ii) The abundance of \tHe nuclei is assumed to be in equilibrium, which is not
strictly correct. In the outer region of the solar core \tHe is continually
produced, but the temperature is too low to burn \tHe at the equilibrium rate. 
Thus the pp chain is not always completed. Similar considerations hold for the 
CN cycle. All this gives, according again to standard solar model calculations, 
corrections to Eq. (\ref{lum1}) of the order of  4$\cdot 10^{-4}$ \cite{BK}.

All in all, Eq. (\ref{lum1}) should be accurate to better than  1\%.

Finally, we remark that one assumes the sun to be at thermal equilibrium when
using Eq. (\ref{lum1}) to relate the {\bf present} solar constant 
to the {\bf  present} nuclear energy production rate and
{\bf present} neutrino fluxes. The assumption of thermal 
equilibrium (stationary sun) looks quite reasonable to us. For the reader who is
willing to abandon it, we have anyhow prepared Sect. \ref{luminosity}.

\subsection{The minimal Gallium signal for standard neutrinos}
\label{2min}

For standard neutrinos, one can find a lower bound to the Gallium signal by
the following considerations \cite{Bahcall85}:

i) 
Since the pp neutrinos are the least energetic ones, they have
the smallest cross section in  Gallium detectors, 
\ie~$\sigma_{Ga,i} \geq \sigma_{Ga,pp}$. Thus 
from Eq.~(\ref{segnali}) one derives
\begin{equation}
\label{Sgamaior1}
S_{Ga}\geq \sigma_{Ga,pp} \sum_i \fii= \sigma_{Ga,pp} \fitot \, ,
\end{equation}
 \ie~the minimum signal is obtained by assuming that all neutrinos are
from the pp reaction.

ii) 
The  solar luminosity, namely the electromagnetic energy released 
per unit time, is obtained with a minimum number of fusions when neutrinos 
carry away the least energy. 
The  minimum total flux $\fimin$ is thus obtained when all neutrinos are 
from pp, and it is thus related to the solar constant $K_\odot\,$:
\begin{equation}
\label{fitot}
\fitot \geq \fimin=\frac{K_\odot} {\qpp} \, ,
\end{equation}
where the term in the denominator is the average 
electromagnetic energy released per emitted pp neutrino. This gives thus:
\begin{equation}
\label{Sgamin1}
   \Sgamin= \sigma_{Ga,pp} \frac{K_\odot} {\qpp}
      \, .
   \end{equation}    
In other words, the minimum Gallium signal is obtained when all neutrinos 
are from the pp-reaction. {\bf  Any solar model will give a prediction not 
smaller than $\Sgamin$, provided only that the {\em present}
electromagnetic energy production rate in the sun equals 
the {\em presently} observed solar luminosity}.

By using the values previously given  \cite{BP95}, 
$K_\odot= (0.853\pm0.003)\cdot 10^{12}$ MeV 
cm$^{-2}$ s$^{-1}$
and $\sigma_{Ga,pp}=(1.18\pm0.02)\cdot 10^{-9}$ SNU cm$^2$s , one derives:
\begin{equation}
\label{sgamin2}
     \Sgamin =   77\pm 2 \quad {\mbox{SNU}} \, ,
 \end{equation}    
 where the error essentially comes from the uncertainty on $\sigma_{Ga,pp}$.
 The central value of this prediction is  already 3 SNU above 
that of the experimental result (74$\pm$8 SNU),
however  
the distance between the two values is well within the errors.
Gallium results do not violate the lower limit for solar neutrinos.

However, this is far from being a satisfactory result. If pp burning
is at work, it appears difficult to switch off the pep reaction.
The pep neutrinos being more energetic, the minimal signal clearly increases. 
In Ref.~\cite{Bahcall85} the value $\Sgamin=79$ SNU is obtained under the
physically very plausible assumption that together with pp neutrinos also
pep neutrinos are created, with the ratio 
$\csi=\fipep/(\fipp+\fipep) \approx 2.5\cdot 10^{-3}$ as given by the 
 solar model of the same Ref.~\cite{Bahcall85}.
We did not consider pep contribution as we wanted to use minimal 
assumptions; should we follow the approach of Ref.~\cite{Bahcall85}, with 
the cross sections given in Table \ref{SIGMA}
and $\csi=2.36\cdot10^{-3}$
we  would find, in place of Eq.~(\ref{sgamin2}), 
$\Sgamin=80\pm2$  SNU\footnote{The small difference with respect to the 
result of Ref.~\cite{Bahcall85} originates from the slightly different, 
more recent values for  $\sigma_{Ga,pp}$,
$\sigma_{Ga,pep}$  and $\csi$ which we are using.},
which is still consistent with experimental result at 1$\sigma$ level.

However the situation gets somehow puzzling if one takes into 
account the other measurements.
For example, KAMIOKANDE does observe the energetic \oB 
neutrinos, implying a \oB contribution to the Gallium signal of about 
7$\pm$ 2 SNU, for
standard neutrinos.
All this means that for standard neutrinos the signal in Gallium 
experiments has to exceed 87$\pm$3 SNU.
The difference between  this minimal 
expectation  and the measured Gallium signal gets now larger than the error.

By considering the above information, one finds that the contribution of \sBe 
and  CNO neutrinos is:
\begin{equation}
\label{SBECNO}
\mbox{S}_{Be+CNO}= (-13 \pm 8.5) \, \mbox{SNU} \, .
\end{equation}
Therefore, it should not exceed  13 SNU, if the observed signal 
has to agree with with the minimal estimate to the three sigma 
level. This is significantly smaller than the  predictions  of all SSMs 
(about 50 SNU, including 38 SNU arising from \sBe neutrinos, see
Table \ref{contribution}).

In conclusion, Gallium results are still  consistent with standard 
neutrinos, essentially implying that the vast majority of them are 
from the  pp reaction and that  intermediate energy neutrinos are 
definitely fewer than in the SSMs.

The recently proposed Gallium neutrino observatory GNO \cite{GNO} which 
foresees a hundred ton target and improved detection techniques, aims 
at a measurement of the Gallium signal with an accuracy of 5 \%. This would
provide a significant progress in the``minimal Gallium signal'' test.

\subsection{Chlorine and KAMIOKANDE}
\label{clorokam}

As soon as the KAMIOKANDE data were available, 
assuming standard neutrinos, it became clear that they 
presented a new puzzle, when compared with the Chlorine results.
Let us divide the Chlorine signal in two parts, corresponding 
respectively to intermediate energy  neutrinos and  to \oB neutrinos:
\begin{equation}
{\mbox{S}}_{Cl}= \Sclint + {\mbox{S}}_{Cl,B} \, .
\end{equation}

 KAMIOKANDE only detects high energy events ($E_\nu\geq 7$ MeV);
assuming  
 standard neutrinos this is enough to tell the energy integrated flux of 
 $^8$B neutrinos, and thus their expected contribution to the Chlorine signal, 
 ${\mbox{S}}_{Cl,B}$, can be 
 calculated, by using the estimated value of $\sigma_{Cl,B}$ (see Table 
 \ref{SIGMA}):
 \begin{equation}
 {\mbox{S}}_{Cl,B}= \sclb \fib^{Kam} =2.98\pm0.41\quad {\mbox{SNU}} \, ,
 \end{equation}
 where most of the error comes from uncertainties on the KAMIOKANDE signal.
By comparison  with the measured Chlorine signal S$_{Cl}=2.55\pm0.25$ SNU, 
one sees that the \oB neutrinos, as measured by KAMIOKANDE, should 
produce a signal in Chlorine larger than observed. So little space is left 
for the  intermediate energy neutrinos:
 \begin{equation}
\label{eq28}
 \Sclint = -0.48\pm 0.49 \quad {\mbox{SNU}} \, .
 \end{equation}

Note that the central value is negative, however Eq. (\ref{eq28})
 is clearly consistent
with zero.  
At ``three sigma'' level 
one has  $\Sclint<1$ SNU, this limit being a factor two
smaller than the RSM prediction, see Table \ref{contribution}.

Let us remark that the above estimate on $\Sclint$ is independent
of the stationary sun hypothesis A). 

 The present uncertainty  on $\Sclint$ depends
essentially on the systematical  error of the  KAMIOKANDE result.
A better determination of $\Sclint$ would require a strong reduction 
of the systematical error in SUPERKAMIOKANDE with respect to KAMIOKANDE. 
In this case the ultimate uncertainty would be that due to the Chlorine 
experiment, implying  
$\Delta \Sclint \approx 0.25$ SNU, which would provide  significant 
information on intermediate energy neutrinos.

\subsection{The space for intermediate energy  neutrinos}
\label{bounds}

 An extension of the minimal Gallium signal argument 
of Sec.~\ref{2min}  can elucidate the r\^{o}le of various 
experiments in providing 
solar model independent information on neutrino fluxes.
We already known that experimental results
are smaller than the expectation from SSM and consequently the
fluxes 
of intermediate  and high energy neutrinos
should be smaller than theoretically predicted.
The relevant question is: how large  neutrino fluxes are allowed?

We have three equations relating the solar constant, 
the Gallium and Chlorine signals with the fluxes $\fii$:
\begin{equation}
\label{lum2}
   K_{\odot}=\sum_i \qi
   \Phi_i \, ,
\end{equation}
\begin{equation}
\label{sga2}
{\mbox{S}}_{Ga}= \sum_i \sigma_{Ga,i} \Phi_i \, ,
\end{equation}
\begin{equation}
\label{scl2}
{\mbox{S}}_{Cl}= \sum_i \sigma_{Cl,i} \Phi_i \, ,
\end{equation}
and the \oB neutrino flux as measured by KAMIOKANDE
\begin{equation}
\label{ska}
     \fib^{Kam}= (2.73\pm0.38)\cdot 10^6  {\mbox{cm$^{-2}$s$^{-1}$}}.
\end{equation}

In order to determine the maximal allowed fluxes of intermediate
energy neutrinos let us assume that {\bf only one} flux 
$\fiI$ (k=Be, N, O or pep) is non vanishing and attribute to $\fiI$
all the signal in the Chlorine and GALLEX experiments pertaining to the 
intermediate energy neutrinos.

For the Chlorine experiment, one has just to subtract from the
experimental value the \oB contribution, \ie~the maximal
flux $\fiI$ satisfies
\begin{equation}
\label{sclmaior}
{\mbox{S}}_{Cl} 
 =\sclI \fiI + \sclb \fib \, .
\end{equation}

For handling the information provided by Gallium experiments, we derive 
$\fipp$ from the luminosity equation. In terms of the average electromagnetic
energy  $\qi$ released per emitted neutrino, see Eq. \ref{qf},
one has from Eq.~(\ref{lum2}):
\begin{equation}
\label{pplum}
\fipp= \frac{\ksole}{\qpp} - \frac{\qb}{\qpp} \fib - \sum_{int} \fii 
\frac{\qi}{\qpp}
\end{equation}
 so that
\begin{equation}
\label{sgatutto}
{\mbox{S}}_{Ga}= 
\frac{\ksole}{\qpp} \sgapp +
     ( \sgab - \frac{\qb}{\qpp}  \sgapp ) \fib +
   \sum_{int}  ( \sgai -\frac{\qi}{\qpp} \sgapp  )\fii \, .
\end{equation}

All terms in the brackets are of course positive 
as the minimum signal is obtained  when all neutrinos are from pp.

Again isolating the k-th component, one obtains the maximal allowed flux from:
\begin{equation}
\label{sgamaior}
{\mbox{S}}_{Ga}
= 
\sgapp \frac{\ksole}{\qpp} + 
       ( \sgab -\frac{\qb}{\qpp} \sgapp   ) \fib+
      ( \sgaI -\frac{\qI}{\qpp} \sgapp   ) \fiI \quad .
  \end{equation}
The reader recognizes, for the limiting case $\fiI=\fib=0$, the 
inequality found in the previous section.

Clearly this exercise is most interesting for  \sBe neutrinos, as they 
are predicted to be the second most abundant ones.
By using the cross sections in Table \ref{SIGMA} and 
taking k=Be, from Eqs.~(\ref{sclmaior}, 
\ref{sgamaior}) one gets:
\begin{equation}
\label{sclmin}
{\mbox{S}}_{Cl}=0.24 \fibe +1.11 \fib
\end{equation}
and
\begin{equation}
\label{sgamin}
{\mbox{S}}_{Ga}= 77 + 6.19 \fibe +2.43 \fib  \, ,
\end{equation}
where the signals are in SNU, $\fibe$ in  $10^9$ cm$^{-2}$ s$^{-1}$ and 
$\fib$ in $10^6$ cm$^{-2}$ s$^{-1}$\footnote{We note 
that errors on the cross sections are negligible with
respect to the uncertainties on experimental signals.}.

In this way the experimental information can be presented
in the ($\fib , \fibe$) plane, see Fig.~\ref{intb}. 
Dashed lines are obtained by adopting in  Eqs.~(\ref{sclmin})
and (\ref{sgamin})  the central values of experimental results. 
 Full lines correspond to the experimental results
$\pm 1 \sigma$. Even by allowing for such a spread, still the curves
intersect in the unphysical region ($\fibe \leq 0$).

In Fig. \ref{intb} we also introduce the information from 
KAMIOKANDE experiment, Eq.~(\ref{ska}), and the following conclusions 
can be drawn:
 
i) 
the  $1\sigma$ allowed areas intersect in the unphysical ($\fibe<0$) region.
The same holds if one disregards either the Chlorine or the 
KAMIOKANDE experiment. Should one disregard Gallium result, still at 
$1\sigma$ one has  a very small value $\fibe \leq 1.0$\unitan.

ii) in order to stay within 2$\sigma$ from each experimental result
the beryllium flux has to be $\fibe<1.5$\unitan, \ie~a factor 
three at least smaller than the  predictions from SSMs.

Similar considerations hold for the other intermediate energy neutrinos, 
see Table \ref{limiti}. In any case, the best fit point is in the 
unphysical region and no point
can be found in the physical region which is
within 1$\sigma$ from each experiment.

Roughly speaking, the hypothesis of standard neutrinos (which
requires positive fluxes) disagrees with each of the three independent 
experimental results by at least 1$\sigma$. The probability for such 
a situation is at most $(0.32)^3$, \ie~a few percent.
Little space seems to be left for intermediate energy neutrinos.

\subsection{Four equations and four unknowns}
\label{solveequation}

So far the only assumption about the sun was Eq.~(\ref{lum2}),
that connecting neutrino fluxes to the solar constant. 
Stricter constraints on the intermediate energy neutrinos can be 
obtained by using some additional, albeit very weak, 
hypothesis. Theoretically, we know that pep neutrinos must accompany pp 
neutrinos and also that the CN cycle is (almost) at equilibrium when 
it is efficient in the sun.
The precise values of $\csi=\fipep/(\fipp+\fipep)$ and
$\eta=\fin/(\fin+\fio)$ are not important for the following
discussion and we  shall use 
$\csi=2.36 \cdot 10^{-3}$ and
$\eta=0.53$,
according to the RSM \cite{BP95}.

In this way we reduce our unknowns to just four variables:
\begin{equation}
\label{antesistemone}
\fip=\fipp+\fipep\, , \quad \ficno=\fin+\fio \, ,
\quad \fibe \quad {\mbox{and}}  \quad
\fib \, ,
\end{equation}
\ie~as many as the available pieces of information.
With the numerical values in Table \ref{SIGMA}, from Eq.~(\ref{lum2}) 
(after dividing both sides by $Q/2$) and Eqs.~(\ref{sga2}) and 
(\ref{scl2}) one gets:
\begin{mathletters}
\begin{eqnarray}
\label{sistemone1}
63.85&=&0.980\fip +0.939\fibe+0.937\ficno+0.498\cdot10^{-3}\fib
\\
&&\nonumber \\
\label{sistemone2}
S_{Ga}&=&1.23\fip+7.32\fibe+8.72\ficno+2.43\fib \\
&&\nonumber \\
\label{sistemone3}
S_{Cl}&=&0.38\cdot10^{-2}\fip+0.24\fibe+0.41\ficno+1.11\fib  \\
&&\nonumber \\ 
\label{sistemone4}
\fib^{Kam} &=& \fib \,  , 
\end{eqnarray}
\end{mathletters}
where all fluxes are in units of \unitanb\ but for the \oB 
flux which is in units of \unitasb\ and signals of radiochemical 
experiments are in SNU.
Uncertainties on the numerical coefficients (arising from errors
on the estimated neutrino cross sections) have been omitted, since 
 errors on the experimental signals are dominant.

We thus  determine the four fluxes:
\begin{eqnarray}
\label{inversione}
\fip &=& 1.90 S_{Cl} - 0.229 S_{Ga} - 1.56 \fib^{Kam} +83.1 \nonumber\\
\fibe  &=&-10.6 S_{Cl} + 0.57 S_{Ga} +10.4 \fib^{Kam} -43.1  \nonumber\\
\ficno &=& 8.6 S_{Cl} - 0.33 S_{Ga} - 8.8 \fib^{Kam} + 24.4 \nonumber \\
\fib   &=& \fib^{Kam} \, .
\end{eqnarray}
inserting the signals given in Table \ref{EXPE}, one 
finds\footnote{When errors on the neutrino cross section are taken into
account, the errors quoted in Eq. (\ref{soluzione}) are slightly
enlarged, becoming $\pm 2.2,\, \pm 7.2 $ and $\pm 5.1$, respectively.}:
\begin{eqnarray}
\label{soluzione}
\fip&=&(66.7\pm2.0){\mbox{\unitan}}\nonumber \\
\fibe&=&(0.4\pm6.6){\mbox{\unitan}} \nonumber \\
\ficno&=&(-2.\pm4.8){\mbox{\unitan}} \, .
\end{eqnarray}

It is remarkable that, for {\bf standard neutrinos} and with minimal
and quite reasonable
assumptions about solar models (\ie~the values of $\csi$ and $\eta$),
the main components of the solar neutrino flux are fully determined
from available experiments. {\bf  In other words, a solar neutrino 
spectroscopy is already at hand and it could be used to study the 
solar interior, if we know enough about neutrinos}.

We see, for example, from Eq.~(\ref{soluzione}) that the pp+pep flux is
determined for standard neutrinos with an accuracy of about 3\%.
This result might be surprising when considering that all experiments
have no more than 10\% accuracy. The point is that the total flux is
fixed by the luminosity, whereas the cross section  depends crucially
on neutrino energy, so that approximately ($\sgabe >> \sgapp$):
\begin{equation}
 \left ( \frac{\Delta \fip}{\fip} \right ) \approx 
  \left ( \frac{\Delta S_{Ga}}{S_{Ga}}  \right ) \frac{\sgapp}{\sgabe}  \,\, .
\end{equation}

The result on $\ficno$ clearly shows --- again for standard neutrinos ---
that the CN cycle cannot be the main energy source in the sun,
otherwise one should have:
\begin{equation}
  \ficno \approx \frac{2K_\odot}{Q} \approx 60 \cdot 10^9 
{\mbox{ cm$^{-2}$ s$^{-1}$}}  \, ,
\end{equation}
in violent contradiction with Eq.~(\ref{soluzione}). 
If the CN cycle were to dominate solar energy production, then
the Gallium signal would be
\begin{equation}
 S_{Ga} \approx \frac{2K_\odot}{Q} \sgacno \approx 550 \mbox{ SNU} \, ,
\end{equation}
an order of magnitude larger than the actual result.

\subsection{Where are Be and CNO neutrinos?}
\label{where}

One notes that the central value of $\ficno$ in Eq.~(\ref{soluzione})
is  negative \ie~unphysical. 
In view of the estimated errore,
this does not seem to be a problem. However, there is a strong correlation 
between 
$\ficno$ and $\fibe$, so that {\bf if $\ficno$ is forced to be positive, then
$\fibe$ becomes negative, and {\em vice versa} }.

In order to understand what is going on, and to clarify the r\^{o}le of 
each experimental result, let us once more reduce the number of 
unknowns. We start again from the basic equations 
(\ref{lum2})--(\ref{ska}) and use the following tricks, similar to 
those used previously:

(a) we group the neutrino fluxes as in Eq. (\ref{antesistemone}), so that 
we are left with the four variables $\fip,\fibe,\ficno$ and $\fib$.

(b) Since $\enecno \geq \enebe$, the corresponding cross section is
 larger than that of \sBe neutrinos. Thus the minimal CNO signal is 
obtained with the replacements
\begin{equation}
\enecno \rightarrow \enebe \quad {\mbox{and}} \quad
\sigma_{CNO} \rightarrow \sigma_{Be} \quad .
\end{equation}
This corresponds to dropping terms containing 
$\ficno$ in equations~(\ref{sistemone1}--\ref{sistemone4})
and replacing $\fibe$ with $\fibe+\ficno$,  
so that only the combination $\fibe+\ficno$ enters.
%*
We remark that such a substitution  represents also a safe approach, 
since the theoretical value of $\sgabe$ has essentially been verified (to 
the 10\% level) by the GALLEX neutrino source experiment~\cite{calibration}, 
whereas only theoretical predictions exist for $\sigma_{Ga,CNO}$.

One can eliminate $\fip$ by using the luminosity equation~(\ref{sistemone1}),
and each experiment provides a constraint on 
$\fibecno$ and/or $\fib$: 
\begin{eqnarray}
{\mbox {S}}_{Ga}&=& 80.1 +   6.14\fibecno + 2.43\fib \nonumber \\
{\mbox {S}}_{Cl}&=& 0.248 +   0.236\fibecno + 1.11\fib \nonumber \\
\fib^{Kam}&=& \fib  \quad .
\end{eqnarray}
The result of each experiment can be plotted in the 
($\fib,\fibecno$) plane, as shown in Fig. \ref{intb2}.

With respect to the situation of Sec.~\ref{bounds}, the intersection
moves towards even more negative values of 
$\fibecno$, see Table \ref{limiti},
and the allowed region in the physical part of the plane shrinks.
Whichever experiment is discarded there is no point in the 
physical region within $1\sigma$ from the remaining two results.
The region at $2\sigma$ from each experiment now allows only
$\fibecno<1$.
 This bound is stronger than we found previously,
 {\bf the main reason 
 being that the small, but non negligible, contribution of pep 
 neutrinos to Chlorine and Gallium signals is included from the 
 beginning}.
 
 The bound on the sum clearly holds separately for $^7$Be and CNO 
 neutrinos. For these latter, however, we can get tighter constraints
 by putting $\fibe=0$ in Eqs.~(\ref{sistemone1}--\ref{sistemone4}),
 see again Table \ref{limiti}.
 In this way one finds that in the area
 within two standard deviations from each experiment $\ficno \leq 0.7$.

\subsection{Probabilities, confidence levels and all that}
\label{probabilities}

The intermediate energy neutrino fluxes,  which we derive from the 
experimental signals assuming standard neutrinos, favor 
 negative, \ie~unphysical values. Here we shall discuss 
 the statistical significance
of this information. 
We try to estimate  quantitatively
the probability of compatibility with standard neutrinos
and to determine the significance of upper bounds 
on neutrino fluxes.

  For pedagogical reasons, let us discuss first
a kindred 
 situation where only one variable is present, namely the 
determination of the neutrino mass from $\beta$~decay experiments. 

\subsubsection{Neutrino mass and $\beta$ decay}
   Measurements of the electron energy spectrum from tritium decay
determine a parameter $m^2_{app}$ (apparent squared mass), which
{\em a priori} is just a combination of experimental data and theoretical 
inputs. In principle, $m^2_{app}$ can be positive as well as negative.
This parameter can be identified with the actual $\nu_{e}$ squared mass 
$m^2$ only if the interpretation of the experimental data 
(underlying theory, estimated energy resolution, \ldots) is correct, and
in that case it should be positive. Even so ($m^2_{app}\geq 0$)
experimental results $\hat{m}^2_{app}$ might come out negative due
to statistical fluctuations of the physical quantities entering
the definition of $m^2_{app}$.

   The latest edition of the Particle Data Book~\cite{PDB} considers four
tritium $\beta$ decay experiments, all giving 
$\hat{m}^2_{app}<0$, and presents the weighted average:
\begin{equation}
\hat{m}^2_{app}=(-54\pm 30) \quad \mbox{eV}^2 \, . 
\end{equation}
   This result prompts the following questions:

\begin{itemize}
\item
(a) Is the  interpretation of the data (``working hypothesis'') 
     correct?

\item
(b) Which upper bounds can be set on the actual neutrino mass?

\end{itemize}

   In the discussion we follow --- to some extent --- the approach
of Ref. \cite{PDB}. We assume that the experimental results have
 a Gaussian distribution, centered at an unknown value $m^2_{app}$,
the width being given by the quoted experimental uncertainty
$\Delta=30$~eV$^2$. 

\begin{itemize}
\item
(a)  If the theory is correct ($m^2_{app}=m^2$) then of course
$m^2_{app}\geq 0 $. Evidently, the probability of obtaining a value equal or
smaller than $\hat{m}^2_{app}=-54$~eV$^2$ is maximal when the neutrino
mass is zero ($m^2_{app}=m^2$=0) and in that case it is:
\begin{equation}
   P_1 = { \int^{\hat{m}^2_{app}/\Delta}_{-\infty} dx e^{-x^2/2}   \over 
               \int_{-\infty}^{\infty} dx e^{-x^2/2}       } = 0.036 \, .
\end{equation}
   For $m^2_{app}>0$ the probability is even smaller. Furthermore the
requirement $m^2_{app}\geq 0$ is just a necessary condition for
identifying it with the physical neutrino squared mass.
   Thus our answer to question (a): 
{\bf there is at most a 3.6\% 
probability that the working hypothesis is correct}.

\end{itemize}

    It is somewhat controversial how to set bounds on an observable
when the experimental results lie outside the physical region, and
more than one answer to our question (b)  is possible:

\begin{itemize}
\item
(b1)
   The first approach we consider is similar in spirit to that just
presented. We can determine an upper bound $P_2$ on the probability
that the result is equal to or smaller than $\hat{m}^2_{app}$ if the
apparent squared mass exceeds a value $m^2_0$:
\begin{equation}
   P_2 =  { \int^{(\hat{m}^2_{app}-m^2_0)/\Delta}_{-\infty} dx e^{-x^2/2} 
                \over 
           \int_{-\infty}^{\infty} dx e^{-x^2/2}       }  \, .
\end{equation}
Conversely, by fixing a Confidence Level C.L.= $1-P_2$,  we can determine the 
corresponding value of $m^2_0$ such that the above equation holds.
For instance, if we choose $P_2=1\%$, we find that {\bf at the 99\% C.L. the 
apparent neutrino squared mass is less than $15.8$~eV$^2$}.

   The interpretation of this limit is the following: answering question
(a) we used the fact that only positive values are physical. Now, if we
have some additional information telling us that $m^2>15.8$~eV$^2$,
then there is at most a 1\% chance that the working hypothesis is correct.
\end{itemize}

  Note that the identification $m^2_{app}=m^2$ is not necessary {\em a priori}
and actually it is one of the hypothesis being tested. Also, to reach our
conclusion we  needed only the 
probability distribution function (p.d.f.), $f(\hat{\alpha}|\alpha)$,
giving the probability  of observing an experimental result $\hat{\alpha}$
if the true value is $\alpha$.

  We remark that $f$ is different from the distribution defining
the probability that, given an experimental result $\hat{\alpha}$, 
the true value is $\alpha$. 
This latter p.d.f.,  which we denote by $g(\alpha|\hat{\alpha})$, is the one 
we are going to use in a second approach to question (b), 
the so-called Bayesian approach to confidence limits, 
 described extensively in Ref.~\cite{PDB} and references therein. 

\begin{itemize}
\item
(b2) Here one assumes that the working hypothesis is correct and one seeks
 information on the  mass $m$ from the knowledge
of experimental result $\hat{m}^2$. To be precise, we list the assumptions:

(i) the measured quantity can be identified with the physical neutrino 
    mass, $m^2_{app}=m^2$.

(ii) Bayes' theorem holds:
\begin{equation}
g(m^2|\hat{m}^2) = { f(\hat{m}^2|m^2) \, \pi(m^2)
                                 \over 
          \int f(\hat{m}^2|m^2) \, \pi(m^2)\, dm^2 
                               } \, ,
\end{equation}
where $\pi(m^2)$ is the {\em a priori} p.d.f of the neutrino squared
mass, which we define as:
\begin{equation}
     \pi(m^2)= \left\{
              \begin{array}{ll}
   1        &  {\mbox{if }} m^2>0 \\ 
   0        &  {\mbox{otherwise.}}
              \end{array} 
               \right.
\end{equation}

As we have chosen $f$ to be Gaussian, $g(m^2|\hat{m}^2)$ is also Gaussian, 
but its $m^2$ domain is restricted to the positive axis.
For a confidence level $1-P_3$, the upper limit $m^2_0$ to the 
neutrino squared mass is now given by
\begin{equation}
      P_3={ \int^{(\hat{m}^2-m_0^2)/\Delta}_{-\infty} dx e^{-x^2/2} 
                \over 
           \int_{0}^{\infty} dx e^{-x^2/2}       }  \, .
\end{equation}
For instance, if we choose $P_3=1\%$, this time we find 
$m_0^2=47.4$~eV$^2$ and we can say that {\bf  at the 99\% confidence level
the neutrino squared mass is less than $47.4$~eV$^2$}.
\end{itemize}

   In the present context, for the same $m_0^2$ one has:
\begin{equation}
       P_2 = P_1 \cdot P_3 \quad .
\end{equation}
Thus $P_2$ is always smaller than $P_3$, and correspondingly the bound on
the mass found with the second approach is always weaker than the first
one.

  We may say  in conclusion that the two approaches
 answer two different questions:
 
 \begin{itemize}

\item
(b1) assuming that  the physical mass squared is larger than
     some value $m^2_0$, what is
     the chance that the interpretation of experiments is correct and that
     the result does not exceed $-54$~eV$^2$?

\item
(b2) Assuming that the interpretation is correct, what is the probability
     that the physical mass squared is larger than $m^2_0$ and that 
     experimental result does not exceed  $-54$~eV$^2$?
     
     \end{itemize}

Correspondingly, at a given confidence level, one associates two
different bounds on the squared neutrino mass, which correspond to 
two different physical assumptions and attitudes: the first case (b1) 
is essentially a way of testing the interpretation of the experiment,
if one is confident  in a minimal value for neutrino mass; the second
one (b2), is a way of testing a theoretical prediction on neutrino 
mass, if one believes in the interpretation of the experiment.

\subsubsection{Statistics and the solar neutrino problem}

   The analogy with the previous case should be clear now. 
In the language of the previous section, the right-hand sides of
Eqs.~(\ref{inversione}) define  four apparent neutrino fluxes
$\fip ^{app}$, $\fibe ^{app}$, $\fib ^{app}$, and 
$\ficno ^{app}$, which can be identified with the physical 
fluxes {\bf assuming} standard neutrinos.

Different attitudes are possible and correspondingly different questions
can be raised. If one wants to test the chance of standard neutrinos with
minimal assumptions about solar physics, then the relevant question is:
\begin{itemize}
\item  (a) what is the probability that all the apparent fluxes are non 
negative?
\end{itemize}
If one is confident in some solar models, yielding definite predictions/lower
bounds on  the fluxes, $\Phi_{L,i}$, 
then again as a test of standard neutrinos the question is:
\begin{itemize} 
 \item (b1) what is the probability  that neutrinos are standard if
the true fluxes are at least  $\Phi_{L,i}$, in face of the available results?
\end{itemize}
On the other hand, one who  believes in standard neutrinos and  wants to test
solar models will be interested in  upper bounds $\Phi_{U,i}$on the fluxes.
His question is now:
\begin{itemize}
 \item  (b2) assuming standard neutrinos, what is the probability that the 
       true fluxes do not exceed $\Phi_{U,i}$, in view of the 
       experimental results?
\end{itemize}

As compared to the $\beta$ decay, the only complication is that 
we deal now with several 
variables (fluxes) instead of just one (the neutrino squared mass).

For this reason we resorted to Monte Carlo techniques. We
generated a large ensemble of sets of four simulated signals (Gallium,
Chlorine, KAMIOKANDE and the solar luminosity); each simulated signal was
extracted by a Gaussian distribution with mean and width equal to its
actual experimental central value and error. When appropriate we only
considered the subsets of three simulated signals where one of the
neutrino experiments (Gallium, Chlorine and KAMIOKANDE) was in turn
excluded. Given any set of four (or three) simulated signals out of the
ensemble, we derived three apparent fluxes: $\fip$, $\fib$ and
either $\fibe$ or $\ficno$ assuming the other one ($\ficno$ 
or $\fibe$) equal to zero. We have chosen not to derive
from the simulated signals both $\fibe$ and $\ficno$, since
the possibility of separating the two signals is critically dependent
on the difference between the ratio of the average cross sections for
Be and CNO neutrinos in the Chlorine experiment and the same ratio in the
Gallium experiments. Moreover, the limit obtained by assuming either
$\fibe$ or $\ficno$ to be zero is more conservative.
Note that when we assume $\ficno=0$, the upper
limit on $\fibe$ is in fact an upper limit for $\fibe$ and for
$\fibecno$ being the cross section for CNO neutrinos larger
than the one for Be neutrinos. In practice the system of equations giving
the signals as functions of the fluxes was inverted with the techniques of
singular value decomposition that automatically take care both of the
case of almost degenerate equations (if one decides not to combine the
two Gallium data) and of the case of an overdetermined system (when the
apparent fluxes are less than the number of equations), giving in this
last case effectively the best $\chi^2$ fit.
In summary, we have generated an ensemble of a milion apparent fluxes
from the ensemble of a milion simulated experiments. If 
  $\fibe$ is positive in this ensemble 60000 times out of a milion, we
say that the probability of standard neutrinos is 6\%. The other limits
are derived similarly considering different experiments and/or fluxes.

   Alternatively, one could reduce the problem to a single variable,
by considering linear combinations of the form
\begin{equation}
\Phi(x) = x \fibe ^{app} + (1-x) \ficno ^{app} \, .
\end{equation}
For standard neutrinos and $0\leq x\leq 1$ this combination should be
positive. By requiring that the physical constraints are satisfied for
any $x$ one can determine the required probabilities/bounds.
Both methods give the same results, which,
taking into account the errors on neutrino cross sections, can be 
summarized as follows:

\begin{itemize}

\item (a) 
the probability $P_1$ for {\bf both } $\fibe ^{app}$ and $\ficno ^{app}$ 
to be positive is less than about 2\%. Should we disregard arbitrarily 
one of the experiments, one still has $P_1\leq 6\%,7\%$ or $9\%$ neglecting, 
respectively, the results of Chlorine, Gallium or KAMIOKANDE experiment. This 
indicates that standard neutrinos ($\fibe ^{app}>0$ and 
$\ficno ^{app}>0$) are unlikely;

\item (b1) 
to the 99.5\% C.L., , without any {\em a priori} knowledge,
$\fibe ^{app}+\ficno ^{app}$ should not exceed 
0.7$\cdot10^{9}$~cm$^{-2}$~s$^{-1}$;

\item (b2) 
to the same confidence level, if one assumes {\em a priori}
standard neutrinos, the combined flux of Be and CNO neutrinos does 
not exceed 2$\cdot 10^{9}$~cm$^{-2}$~s$^{-1}$.
\end{itemize}

Similar statements hold for \sBe +CNO neutrino fluxes separately, see
Table~\ref{confidence}.

  The main message can roughly be summarized by saying that
  {\bf the 
probability of standard neutrinos are low, not much more than 2\%. However, 
the precise magnitude of the probability  should be taken with caution},
at least for the following reasons:
 i) 
statistical and systematic errors have been combined together and
ii)
 the assumption of Gaussian fluctuations might underestimate the
probability, especially considering that we are dealing with the tail of 
the distribution.

%%%

\subsection{Experimental results and standard solar models}
\label{standard}
Let us insist on the hypothesis of standard neutrinos and compare 
experimental information with theoretical estimates.

We  report in Fig.~\ref{intb2} 
the results of several recent solar model calculations 
 \cite{BP95,P94,TCL,CL,SDF,CESAM,DS96,RCVD96,Ciacio} 
together with the  experimental results. 
Some of the models predict a \oB neutrino flux  close to the KAMIOKANDE value;
 however
no model is capable of reproducing the low Be+CNO flux implied 
by the experiments.

In Table~\ref{confidence}, we have  considered {\bf only standard  solar 
models}~\cite{P94,BP95,RCVD96,Ciacio}.
For standard neutrinos, the experimental information is also presented in
the same  Table.
 The discrepancy between theory and 
experiment is about a factor two for the boron flux. 
{\bf The discrepancy on $\fibecno$, where the predicted values exceed the 
experimental upper bounds (99.5\% C. L.)
 by a factor three,  appears more important
to us}.

The problem is mostly with beryllium neutrinos and let us examine
it in some detail. The extraction of $\fibe$
from experimental data (with the requirement $\ficno \geq 0$) yields an
unphysically negative Be flux. Without any prior knowledge, $\fibe$ cannot
exceed 1/10 of the RSM prediction at the 99.5\% C.L. If we {\em a priori}
force it to be non-negative, then the upper bound is 1/5 of the RSM at the
95\% C.L.; a value as high as 1/3 of the RSM prediction is only allowed
at the 99.5\% C.L. All this indicates that \sBe neutrino suppression
is much stronger than that of \oB neutrinos.

%%%
\subsection{The beryllium-boron relationship} 
\label{bebo}
 Additional insight on neutrino fluxes can be obtained
by considering the  physical 
connections among them.
The relationship between 
 between \sBe and \oB neutrinos, particularly emphasized  by 
Berezinsky \cite{Bere}, is most interesting.

Both neutrinos  are ``daughters" of the $^7$Be nuclei, see Fig. 
\ref{padre}.
For this nuclide, electron capture (rate $\lambda _{e7}$) 
is  favoured over proton 
capture  (rate $\lambda_{17}$), due to the absence of the Coulomb barrier.
 Thus  the value 
of $\fibe$ is a clear indicator of the parent  $^7$Be concentration, 
$n_7$:
\begin{equation}
n_7 \propto \fibe /  \lambda_{e7} \, .
\end{equation}
Since $\fibe$ comes out to be reduced by a (large) factor with respect to 
the SSM prediction, the same reduction has to occur for the $^7$Be
 equilibrium abundance
($\lambda_{e7}$ is weakly dependent on temperature, and 
it is essentially scaled from measurements in the laboratory~\cite{Rolfs}). 
The puzzle is thus with \oB neutrinos, since:
\begin{equation}
\fib \propto n_7 \lambda_{17} \, .
\end{equation}
The observed (KAMIOKANDE) value of $\fib$ being just a factor 
two below  the SSM prediction, it looks that experiments are 
observing too high $\fib$! To put it in another way, one cannot kill the 
father/mother before the baby is conceived.

Should we insist on this approach, then we need to enhance $\lambda_{17} / 
\lambda_{e7}$. {\bf Any attempt to reduce
the discrepancy between
the  KAMIOKANDE and Chlorine experiments with
respect to SSM by lowering 
 $S_{17}$ } (the zero energy 
astrophysical factor for the p+\sBe$\rightarrow$ \oB $+\gamma$ reaction)
{\bf goes into the wrong direction}.

To make this argument more quantitative, let us define the reduction
factors $R_i$ with respect to the prediction of the Reference Solar Model
($R_i=\fii/\fii ^{RSM}$) \cite{Bere}.
 From the Chlorine and KAMIOKANDE data one gets:
\begin{equation}
\label{cernzero}
\frac{R_{Be}}{R_B} \leq -1 \pm 0.8
\end{equation}
which is another way of presenting the 
``inconsistency'' between the Chlorine and KAMIOKANDE data, see
Sec. \ref{clorokam}.

\subsection{What if the sun were  burning less now?}
\label{luminosity}

One might speculate that
the {\bf present} luminosity L$_{\odot}$ does not correspond to the
{\bf present} nuclear energy production rate  L$_{nuc}$ in the sun.
Actually it takes about eight minutes for neutrinos produced in the solar core
to reach earth, whereas the time for electromagnetic energy to reach 
the solar surface is more than 10$^4$ years,
and one might imagine that L$_{nuc}$ is different from  L$_{\odot}$.
Short time scale fluctuations  in the nuclear energy
production might not alter the photospheric temperature,
the relevant time scale  being of the order
of $10^7$ yr because of the enormous amount
of gravitational energy stored in the solar structure. 
Although this hypothesis  of a thermal instability is rather 
extreme, let us consider it as a way of exploiting the full potential of the 
``solar model independent'' approach.

First of all, we remind that the result of combining Chlorine and
KAMIOKANDE experiments (see Sect. \ref{clorokam}):
 \begin{equation}
\label{eq28bis}
 \Sclint = -0.48\pm 0.49 \quad {\mbox{SNU}}
 \end{equation}
was independent of the solar luminosity constraint. The probability
of finding $\Sclint$ about $1\sigma$ below its physical limit is 16\%.
By exploiting Gallium result, we can find that the present situation
is even less probable. If we insist that Gallium result is consistent
with the other experiments, then the signal originates essentially
from p (=pp+pep) neutrinos:
\begin{equation}
\label{cern1}
\fip=\frac{S_{Ga}}{\sgap} = (60.2 \pm 6.5 ) {\mbox \unitan}
\end{equation}
\ie~p neutrinos are essentially as many as calculated in standard solar
models.
From solar models we only assume $\fipep/\fip=\fipep^{RSM}/\fip^{RSM},
\ie~$\footnote{We 
verified that $\csi$ varies by 2\% when L$_\odot$ is 
changed by 10\%.}:
\begin{equation}
\label{csi2}
\fipep=2.36 \cdot 10^{-3} \fip \quad .
\end{equation}
One can thus
calculate  the contribution of pep neutrinos to the Chlorine signal:
\begin{equation}
\label{cern2}
\Sclpep=\sclpep \fipep= 0.23 \pm 0.025 \quad {\mbox{SNU}} \, .
\end{equation}
After subtracting this contribution from Eq. (\ref{eq28bis}) we are left
with:
\begin{equation}
\label{cern3}
\Sclbecno= -0.71 \pm  0.49 \quad {\mbox{SNU}} \, .
\end{equation}
By requiring  $\Sclbecno \geq 0$, one immediately derives that the 
probability of the present situation is at most 8\%.

Essentially the same argument can be presented in a more precise
form. We abandon the luminosity contraint Eq. (\ref{lum2}) or (\ref{sistemone1})
 and still we 
have the three experimental signals, Eqs. (\ref{sga2} -- \ref{ska}) or
Eqs. (\ref{sistemone2} -- \ref{sistemone4}), in face of four
unknowns: $\fip, \fibe, \ficno$ and $\fib$. One can thus
express one flux (\eg~$\fibe$) in terms of another flux
(\eg~$\ficno$) and of the experimental signals. In this way one finds:
\begin{eqnarray}
\label{cern4}
 \fibe +1.77\ficno & = & 
		4.59 {\mbox{S}}_{Cl} - 1.29\cdot 10^{-2} {\mbox{ S}}_{Ga} -
    5.04 \Phi_{B,Ka} \nonumber \\			
			&= & -3.0 \pm 2.2  \quad ,
\end{eqnarray}
which is essentially equivalent to Eq. (\ref{cern3}). 
This equation also shows {\bf  the relevance of direct \sBe neutrino
detection}. As an example, a measurement of $\fibe$ in excess of 3.3 (\unitanb)
would imply $\ficno < 0 $ at $3\sigma$ level, and thus
would be a proof of non-standard neutrinos, even 
{\bf  without} the stationary sun hypothesis.

\subsection{Universal neutrino oscillations}
\label{universal}

In the case of neutrino oscillations, similar arguments can be used 
when the averaged survival probability 
(P$_{\nu_e \rightarrow \nu_e}$) is the same
for all the components (pp, pep, Be, \ldots) of the neutrino flux.
This situation is realized for $\Delta$m$^2 > 10^{-3}$eV$^2$
(in this case coherent matter effect are negligible at any point in the sun,
for any component of the solar neutrino flux).
It is important to observe that in this situation
the cross sections $\sigma_i$ averaged over the neutrino
energy spectra are the same as for standard neutrinos.

Thus by interpreting again $\fibe,\ficno$ and $\fib$ as the electron
neutrino fluxes at earth, the case of {\bf sterile
neutrinos} is exactly equivalent to the one just discussed in 
Sec. \ref{luminosity}, and we get again Eq. (\ref{cern4}). This
means that the probability of universal oscillations into sterile
neutrinos is less than 8\%.

We remark that we used really minimal information from solar
models, essentially Eq. (\ref{csi2}). 
Clearly any additional
hypothesis on the sun (\eg~a  minimal non vanishing \sBe flux)
will essentially exclude this scenario, see also Ref.~\cite{Petcov}
and Ref. \cite{Futuro}.

\subsection{Concluding remarks}

\begin{itemize}
 \item All in all, a solar model independent evidence 
for non-standard neutrinos exists. It is however not overwhelming:
the probability of standard neutrinos is less than a few percent
(see Sec.~\ref{probabilities}), \ie~the indication is at the 2$\sigma$ level
(we point out that this conclusion is reached when giving up
all our understanding  of stellar physics).
 \item If one insists on standard neutrinos, 
then the $^7$Be neutrino flux has to be  drastically suppressed 
with respect to the  prediction  of SSMs and this suppression
is stronger than that for \oB neutrinos. 
Non-standard solar models 
attempting to solve the solar neutrino puzzle have to account for 
{\bf both} these reductions.
 \item All this does not imply that the $^7$Be  flux on earth
has to be small.
That conclusion holds  for standard neutrinos only.
 There are neutrino oscillation schemes which can account
for all available  data and, at the same time, predict  a $^7$Be signal
quite consistent with the SSM prediction.
(These models exploit the possibility of deforming energy spectra
and/or transforming $\nu_e$ into $\nu_\mu$, which are active 
in the KAMIOKANDE detector, see Ref.~\cite{justso}).
In conclusion {\bf the direct detection of  $^7$Be neutrinos is crucial}.
\end{itemize}

In this respect, one has to avoid the temptations illustrated by the following 
story~\cite{Telegdi}:
 The owner of a villa in Rome was explaining the civilization
 level of  ancient Romans: ``They even had  telegraphs or telephones.
When digging
 in my garden,  I discovered ancient copper wires''.  His friend 
 immediately went digging in his own garden  and returned
with the comment ``You were right about their incredible 
civilization. They even had wireless communication. Indeed I found no
cables upon digging''.
\newpage
%%%%%%%%%%%%%%%  SECTION  THREE %%%%%%%%%%%%%%%%%
%
\section{Non-standard solar models: why and how?}
\label{cap3}

\subsection{Introduction}
\label{introcap3}

For standard neutrinos, one finds that 
the fluxes of intermediate energy neutrinos (\sBe and CNO) are strongly
reduced with respect to the SSM expectations, so that  
the nuclear energy production chain appears  strongly 
shifted  towards the pp-I termination.

The question addressed in the next sections is the  following:
is it possible to build  {\bf non-standard} solar models in 
agreement with  available experimental data? In other words, if we insist
on standard neutrinos, is the solar neutrino puzzle restricted to
the results of SSMs, or is the problem more general?

In order to enhance the pp-I termination, it is necessary
that the ratio between the rates for the \treHe+\treHe and  \treHe+\qHe
reactions,
\begin{equation}
\label{rapporto}
 \Rapporate=\overline{\lambda_{34}} /\overline{ \lambda_{33}}
\end{equation}
is drastically decreased with respect to the SSM prediction
(here and in the following \laij
is the fusion rate of two nuclides with mass numbers $i$ and $j$
 and the average is meant over 
the energy production
region).
In the (realistic) approximation that the total flux is given by the 
sum of \sBe and pp components, 
\ie~only ppI and ppII contribute, the first termination has probability
P=$\overline{\lambda_{33}}/ (\overline{\lambda_{33}}+\overline{\lambda_{34}})$
and of course the other one has 1$-$P. In ppI two pp neutrinos
are emitted, against one pp plus one \sBe neutrino in ppII, so that:
\begin{equation}
\frac{\fibe}{\fipp} = \frac{\Rapporate} {2+ \Rapporate} \quad .
\end{equation}
In the RSM one has  $\frac{\fibe}{\fipp}=0.09$
so that
$\Rapporate_{RSM}= 0.2$;
if the \sBe neutrino flux has to be reduced by --- say --- a factor three,
 then one needs
$\Rapporate \approx 0.07$.

One finds just two ways for decreasing $\Rapporate$:

\noindent a) 
Adjusting the parameters
so as to lower the inner  temperature.

\noindent b) 
Adjusting the \treHe nuclear cross sections, so 
as to make the \treHe+\treHe 
reaction  even more favoured with respect to the \treHe+\qHe reaction.

We note, by the way, that the p+\sBe $\rightarrow $\oB+$\gamma$ cross section
is not relevant for enhancing the pp-I branch.

In this section we begin the discussion of case a), by  identifying
the  parameters which could affect the inner solar temperature.
Non-standard solar models with reduced  central temperature will
be explicitly presented  in the next section, whereas the effect
of varying the \treHe nuclear  cross sections will be discussed in 
section~\ref{cap5}. As a way of characterizing the non-standard solar 
models which we shall discuss later, we introduce  in the final part of 
this section some algorithms, essentially based on ``homology''
(scaling) concepts.

\subsection{Cooler solar models: why and how?}
\label{sec32}

By lowering the temperature and therefore
the collision energies, the tunnelling probabilities are decreased 
for both the \treHe+\treHe and \treHe+\qHe branches, the latter being
more suppressed as heavier nuclei are involved, see Eq.~(\ref{Ptunnel});
in conclusion, the \treHe+\treHe branch gets favoured.

How can one decrease the inner solar temperature?
As we shall see, there
are a few analogies between the solar core and the human body.
For this latter the following statements clearly hold:

\noindent  i) 
the (absolute) temperature is fixed to the level of $\pm$ 1\%;

\noindent ii) 
alteration of the temperature is a symptom, and not an illness in itself;

\noindent iii) 
once you measure the temperature somewhere, you know it everywhere.

As regards the inner solar temperature, a comparison of the results
of various SSM calculations  (see Table~\ref{Modelli}) immediately shows
the first point.
The stability of the internal temperature for any given input 
physics will be further analysed in Sec.~\ref{sec4accuracy}.
 Point iii) will be discussed more extensively at the end of this section.
  Let us now concentrate on ii). It  means
that one cannot treat --- in principle --- the central  temperature $T_c$
as an independent  parameter. In fact, $T_c$ cannot be decoupled 
from the solar structure, since severe constraints arise from the stability
criterion. 
One has thus  to study how $T_c$ is altered when  physical/chemical
parameters are varied, while the basic  equilibrium conditions
are still satisfied. In other words, one has to study 
``different-but-still-reasonable'' suns,
\ie~pseudo-solar structures where stellar matter behaves differently
 from
the predictions of standard physics,
 the basic equations for  stellar structure and
evolution still being satisfied.

In order to reduce $T_c$ one can resort to several manipulations,
playing on the physical and chemical inputs which determine the structure
of the star, see Sec.~\ref{structure}:

{\em a) A larger pp cross section.}

\noindent Increasing \spp~(the astrophysical zero energy S-factor for the
$p+p \rightarrow d+ e^+ +\nu_e$ reaction) implies a lower central 
temperature, since fusion gets easier while the solar luminosity has to 
be kept constant. Although \spp~is theoretically well determined 
(see Sec.~\ref{subsec4spp}), one can introduce an artificial variation
just  to get cooler pseudo-suns. In the language of 
Sec.~\ref{structure}, one is essentially changing the energy production 
rate per unit mass.
 
{\em b) A less opaque sun.}

\noindent This is another way to get a cooler solar interior, as a smaller 
temperature gradient enlarges the region of nuclear burning, with less
 energy needed from the innermost core. In practice, this can be 
accomplished by using different {\em ad-hoc} assumptions:

   \noindent {\em b1}) 
   the metal fraction Z/X is significantly smaller than that indicated by 
   the photospheric and/or meteoritic composition;

   \noindent {\em b2}) the radiative opacity is smaller than that computed 
   by several authors~\cite{HU77,IGL90,IGL92};

   \noindent {\em b3})
   some new mechanism could contribute to energy transport through the
   sun. In this connection some years years ago (see \eg~\cite{SperPress},
   \cite{Kaplan}) the possibility of Weakly Interacting Massive
   Particles (WIMPS) captured by and trapped inside the sun was discussed. 
We do not consider this hypothesis in view of the negative results of direct
searches for such  particles and for the lack of any observational
evidence in later stellar evolution~\cite{WIMPS}.

According to Sec.~\ref{structure}, one could also 
study modification of the equation of state. We do not consider
this possibility as in the region where neutrinos are produced
the deviation from a perfect gas law, although relevant for a detailed
evaluation of the solar structure, cannot deeply modify the present
sun.

On the other hand, there is an additional, still hypothetical possibility:

{\em c) A younger sun.}

\noindent Should the sun be younger, the hydrogen mass fraction in the 
 center would be higher and the same  nuclear energy output could be 
 produced at lower temperatures.

\subsection{Homology relationships.}

In the next sections we shall present many solar models, where some
physical input $X$ (solar age, chemical composition...) 
is varied from the starting value used in the calculations,
$X^*$, by a scale factor $x$:
\begin{equation}
		x = X/X^*
\end{equation}
As a result, one finds  profiles for the physical quantities
$\O$ (\eg~pressure, temperature, density...) 
which are different from the profiles $\O^*$
of  the starting model,
at any point in the solar interior.

As we shall see, we find that in many cases the distributions 
of physical quantities follow with a 
good accuracy an ``homologous'' scaling relation
\begin{equation}
\label{omol}
	\O(x, m)= f_\O(x) \O^*(m)\, ,
\end{equation}
(where $m$ represents the mass coordinate $m={\mbox{M}}(r)/{\mbox{M}}_\odot$,
and ${\mbox{M}}(r)$
is the mass within the current values of the radius $r$)
\ie~for any physical quantity, the profile along the mass coordinate
has  the same shape  as in the starting model; the difference with respect to
it is just  in a scale factor, which depends on  the amount $x$
of the variation of the parameter $X$  and which can be different for different
physical quantities (however $f_\O=1$ for any $\O$ and $X$, when $X=X^*$).

This is not in principle an unexpected behaviour, since we know from
the theory of stellar structures that similar relations holds
for ``homologous'' stellar model. To understand the point, let
us recall that a stellar model is governed by a set of differential
equations whose solutions give the correct distribution of the physical
quantities $\O$ throughout the structure. According to the linear
structure of differential relations, given one solution
(\eg~the starting model) further solutions can be generated by a 
scaling transformation
of the physical quantities, as in Eq.~(\ref{omol}), the different scaling
factors being related through algebraic constraints.

Equation~(\ref{omol}) is thus essentially a way of transforming solutions
into solutions. Note however that not any solution is obtained just by
these transformations, much in the same way that not all stars are 
homologous to each other. As an example, we shall see that
homology is violated when considering huge solar age variations,
as the age alters the profile of helium abundance.
  
 More generally, bearing in mind that  R$_\odot$ has
to be matched by tuning the mixing length,
 one finds that non-standard 
structures tend to have an homologous internal structure, with  a strong 
departure from homology just in the more external layers.

We recall that our main interest is in the resulting  
neutrino flux, so that two points have to be borne in mind:

\noindent i) 
We are primarily concerned with the energy production region  
 ($m\le$ 0.3 or R/R$_\odot$ $\le$ 0.2).

\noindent ii) 
We are mainly interested in the  temperature profile,
as neutrino production depends crucially on temperature.

All this means that we do not bother if Eqs.~(\ref{omol}) are badly violated
outside the central core, and that  for physical quantities other
than T we shall be satisfied  if Eq.~(\ref{omol}) holds  only to a 
fair approximation.
What really does matter is the temperature profile in the energy
production region.

Clearly, if Eq.~(\ref{omol}) holds for the temperature profile,
\begin{equation}
\label{omol2}
	T(x, m)= f_T(x)T^*(m)  \quad ,
\end{equation}
then from the knowledge of the temperature at a point
 say the center,   we are able to compute it at any other point,
the only  relevant parameter being the scaling factor:
\begin{equation}
\label{Tc}
	\tau = T_c/T_c^*
\end{equation}
In other words,
the test of homology for the temperature profiles correspond to check  the 
third 
 statement  mentioned in the previous
section.

It is useful to specify, in preparation for the next sections, the
algorithms we use to test Eqs.~(\ref{omol}) and to extract the dependence
of the physical quantities on the inputs which will be varied when building
non-standard solar models.

If Eq.~(\ref{omol}) holds for a physical quantity $\O$, then the ratio
\begin{equation}
\label{ratioomol}
	\RO =\frac {\O(x,m)}{\O^*(m)}
\end{equation}
is independent on the mass coordinate $m$ and is purely determined by
the physical input which has been varied. Qualitatively, this can  be
seen by looking at a graph where $\RO$ is plotted versus $m$. For quantitative
statements, it is useful to compute the average value of $\RO$ over the cells
of our solar model and  its variance  $\Delta \RO$ 
\begin{equation}
\label{ratioomol2} 
 {\overline {\RO}} =\frac{1}{N_m}\sum_{k}\RO (m_k)
\end{equation}
\begin{equation}
\label{ratioomol3} 
  (\Delta \RO)^2 =\frac{1}{N_m}\sum_{k}[\RO (m_k) -{\overline {\RO}}]^2
\end{equation}
where we assume to divide (a portion of) the solar profile into 
$N_m$ cells, the $k$-th one being centered  at $m=m_k$. By definition,
for a perfect homology $\Delta \RO=0$. The ratio
\begin{equation}
\label{ratioomol4} 
	\delta_\O= \Delta \RO/{\overline {\RO}}
\end{equation}
is an indicator of the validity of the homology relationship for the quantity
$\O$, and typically we consider two regions:

\noindent a) 
the energy production region ($m\le$ 0.3 or R/R$_\odot$ $\le$ 0.2)
which is of primary interest to us, as already remarked.

\noindent b) 
The full radiative interior, ($m \le$ 0.98 or $R/R_\odot$ $\le$ 0.7).
Although such an extended area is not important for neutrino production, 
nevertheless it is interesting to study the behaviour of physical 
quantities up to the bottom of the convective layer, where useful 
constraints arise from helioseismological measurements.

\subsection{Scaling laws for the physical quantities}
\label{seciiie}

Once homology has been tested, one still has to study the function 
$f_\O(x)$, \ie~the dependence on the input parameter $X$ which is being 
varied.

A natural parameterization, again reminiscent of those encountered in the
study of the homology relationship is of the form:
\begin{equation}
\label{alfa0}
\O (x,m)= x^{\alphaO} \O^*(m)   \quad .
\end{equation}
The coefficient $\alphaO$ depends on the physical quantity $\O$
(as well as on $X$). For 
simplicity of notation we will understand this dependence.
Note that for $x \rightarrow 1, \O \rightarrow \O^*$.

From Eq. (\ref{alfa0}),  the power law coefficient $\alphaO$
is determined as:
\begin{equation}
\label{alfa01}
\alphaO =\frac{ \log( \O / \O^* )}{\log x}
\end{equation}

If we have built $N_x$ models, labelled by an index $j$ specifying
the value of the input parameter ($x=x_j$) and if each
model contains $N_m$ cells, labelled by an index $k$
indicating the value of the mass coordinate ($m=m_k$) then one has
$N_x \times N_m$ independent determinations of the coefficient $\alphaO$:
\begin{equation}
\label{alfa02}
\alpha_{jk}=\frac{\log \frac{\O(x_j,m_k)}{\O^*(m_k)}}{\log x_j} \quad .
\end{equation}

In order to extract a suitable average value for $\alpha$, we
proceed in the following way:

\noindent a) 
for the $j$-th  model, we perform an average over the cells
\begin{equation}
\label{alfa03}
{\overline {\alpha_j}} =\frac{1}{N_m} \, \sum_k \alpha_{jk}
\end{equation}
and evaluate the corresponding variance $\Delta \alpha_j$ from:
\begin{equation}
\label{alfa04}
(\Delta \alpha_j) ^2= \frac{1}{N_m} \,
 \sum_k (\alpha_{jk}-{\overline {\alpha_j}})^2  \, .
\end{equation}

\noindent b) 
We take then a weighted average over the different models, using
$1/(\Delta \alpha_j) ^2$ as a weighting factor:
\begin{equation}
\label{alfa05}
\alpha= \frac{ \sum_j {\overline {\alpha_j}}/ (\Delta \alpha_j) ^2 }
{\sum_j 1/(\Delta \alpha_j) ^2 }
\end{equation}
We remind that homology is well verified when $\Delta \alpha_j / \alpha_j$
is small, and this provides a justification for the choice of the weighting
factor.

\noindent c) 
We can also define a variance $\Delta \alpha$, specifying in some sense
the global uncertainty on the coefficient just determined:
\begin{equation}
\label{alfa06}
(\Delta \alpha) ^2 = \frac{1}{N_x} \sum_j (\alpha_j -\alpha)^2
\end{equation}
Note that $\Delta \alpha / \alpha$ essentially
 estimates the accuracy of the
power law approximation (whereas $\Delta \alpha_j / \alpha_j$ indicates the 
accuracy of the homology relationships).

Generally we consider two kinds of variations of the input parameter,
for the calculation of the power law coefficients $\alpha$ and of their
variances:

\noindent i) 
small variations: $\mid x-1 \mid \leq$ 10\%. 
 This procedure, which was pioneered
in \cite{BU89}, is useful to study the effect of changing an input 
parameter of the SSM within its estimated uncertainty (which is generally 
of the order of few percent, see section~\ref{cap4}).

\noindent ii) 
Large variations: typically by an order of magnitude, \ie~for $x$ well outside 
the range allowed for the SSMs.  It is in this way that one is really 
building and 
testing non-standard solar models. Such models are actually called for if one 
wants to  effectively suppress \sBe 
neutrinos.

Clearly the coefficients found in i) and ii) should be equal ---- within 
numerical uncertainties --- if the power laws were exact. Actually
there is no deep reason for these laws to hold. They are 
just  parameterizations of data and one should not be astonished to get 
different 
numbers for cases i) and ii). It is more of  a surprise if the values 
are found to be close, indicating that the same simple parameterization holds 
over a wide range for the input parameter.

\subsection{Scaling laws for neutrino fluxes}

We assume again  a power-law behaviour:
\begin{equation}
\label{lawf}
	\Phi _\ii= \ x^{\alpha _\ii}  \cdot \Phi _\ii ^* 
\end{equation}

In order to determine the coefficients $\alpha_\ii$, we use algorithms 
similar to those of the previous section.
As the fluxes are already summed  over the cells we can skip point 
a) of the previous section and 
we  construct directly the average over the models. Again
omitting for simplicity the index specifying  the flux component,
one has:
\begin{equation}
\label{alfa1}
 \alphai = \frac{1}{ N_x} \sum_j \frac{\log \frac
{\Phii(x_j)}{\Phii^*}}{\log (x_j)}
\end{equation}
and the corresponding variance:
\begin{equation}
\label{alfa11}
	(\Delta \alphai)^2 =\frac {1}{N_x} \sum_{j} \left[ \frac{\log \frac
{\Phii(x_j)}{\Phii^*}}{\log (x_j)} - \alphai \right]^2
\end{equation}
Since all pseudo-suns have the same luminosity, they will give 
(approximately) the same 
total neutrino flux:
\begin{equation}
\label{sommafi}
\sum_\ii \qi \Phi_\ii(x) = {\mathrm const} =  \sum_\ii \qi \Phi_\ii ^* \, ,
\end{equation}
so that equations of the form of Eq.~(\ref{lawf}) cannot hold exactly 
for all the components
of the neutrino flux and for arbitrary variations of the parameters. 
For small variations one gets, 
by differentiating Eq.~(\ref{sommafi})
 with respect
to $x$,  the constraint:
\begin{equation}
\label{alfa13}
		\sum_\ii \alpha_\ii \qi \Phi_\ii ^* =0
\end{equation}
which can be used
as a check of the calculations.
For the case of large  variations it is convenient to use Eq.~(\ref{sommafi})
as a way of expressing one of the fluxes in terms of the others, so as to 
maintain the luminosity constraint.
 This is best done for the case
of pp+pep neutrinos, as these are the least sensitive to variations of
the physical inputs. For these latter, thus, instead of Eq.~(\ref{lawf}) 
we will generally use the expression:
\begin{equation}
\label{alfa14}
\fip=\fip ^* - \frac{\qbe}{\qp}(\fibe - \fibe ^*)
     -\frac{\qcno}{\qp}(\ficno - \ficno ^*)
\end{equation}
where $\fibe$ and $\ficno$ are given through Eq.~(\ref{lawf}).

\subsection{Dependence on the central temperature}

All in all, we are mainly interested in the change of neutrino fluxes
(and other physical quantities characterizing the stellar interior) on
the inner solar temperature, when some input parameter is varied.

This question can be easily answered on the basis of the above discussion,
if we find --- and we shall --- that the temperature 
profiles satisfy the homology relationship:
\begin{equation}
\label{omoT}
T(x,m) = \frac{T_c(x)}{T_c^*}\cdot T^*(m) \quad .
\end{equation}
Neutrino fluxes, as well as the other quantities, can then be parameterized
in terms of the scale factor $ T_c/T_c^*$, see Eq.~(\ref{Tc}):
\begin{eqnarray}
\label{FiTc}
\Phi_\ii&=& \Phi_\ii ^* \cdot (T_c/T_c^*)^{\beta_\ii} \\ \nonumber
\O_i&=&\O_i ^*  \cdot (T_c/T_c^*)^{\beta_\ii} \,\, .
\end{eqnarray}

The $\beta_\ii$ coefficients can be determined directly  from the calculated
fluxes for several models with a procedure similar to that presented
in the previous section (this is the way we shall use for neutrino
fluxes). 
Alternatively, one can profit from the  previous results. If one has 
determined the  dependence of the central temperature on
the parameter $x$:
\begin{equation}
\label{alfa15}
T_c =  x^{\alpha_T} \cdot  T_c^* 
\end{equation}
and the dependence 
of the fluxes on the same parameter, see Eq.~(\ref{lawf}), then one
has:
\begin{equation}
\label{beta}
\beta_\ii = \frac{\alpha_\ii}{\alpha_T}
\end{equation}
For neutrino fluxes, the
most interesting  comparison will be that of the  $\beta_\ii$ values 
 obtained by varying different physical inputs. In principle
they do not need to be the same. If we find that they are close,
no matter  which parameter is varied, then this will be a confirmation
of our expectation that $T_c$ is the quantity controlling
the neutrino fluxes, independently of how that particular value of $T_c$ is
achieved.

\newpage
%%%%%%%%%%%%%% SECTION FOUR  %%%%%%%%%%%%%%%%%%
%
\section{Low central temperature solar models}
\label{cap4}

\subsection{Introduction}

Solar models builders claim that the central temperature of the sun is
known with an accuracy of  one percent or better. This claim
is often questioned by other physicists, who feel such an accuracy
as too high on a matter where 
 no direct observational data are available. Independently of personal
feelings, a few pertinent questions are the following.
\begin{itemize}
\item
If we insist that a low temperature solar model yields a 
drastically  (say a factor three) reduced  \sBe  neutrino flux,
how much should the physical and/or chemical inputs of solar models be varied?
Is it enough to go slightly beyond  estimated uncertainties,
or are wild changes actually needed?
\item
Is it possible to get fluxes of both \sBe and \oB neutrinos consistent
with available experimental information?
\end{itemize}

Following the lines sketched in section~\ref{cap3}, we now
construct and discuss low inner temperature solar models.
As the temperature is not an independent variable, we shall
construct our pseudo-suns by acting on different 
inputs of the solar model: the p+p $\rightarrow$ d + e$^+$ + $\nu$ 
cross section, the metal content of the  sun, the adopted values for 
the radiative opacity and the solar age.

These inputs will be varied
well beyond their estimated uncertainty, so as to build  
non-standard solar models.
Generally we will attempt to vary the input parameters as long as we get
a \sBe neutrino flux reduced to one third of the RSM prediction.

To give and estimate of the sensitivity of central temperature
to the chosen input parameter, we will define $x(0.1)$ as the value of the
scaling factor such that the central temperature is reduced by 10\%.

A few common features of all the computed models
will be summarized in section~\ref{para}, namely: 

\noindent i) 
the temperature  profiles appear, to quite a good approximation, 
homologous among the different models, in the sense specified in the 
foregoing section:
\begin{equation}
T(m) = \frac{T_c}{T_c ^*} \cdot T^*(m)
\end{equation}

\noindent ii) 
no matter how the temperature variation is obtained, the neutrino fluxes
are essentially determined by the  scale factor
\begin{equation}
\label{Tc2}
	\tau = T_c/T_c^*
\end{equation}
In the same section the temperature dependence
of the main components of the neutrino fluxes, as obtained by
numerical simulations, will be also demonstrated 
analytically.

In section~\ref{still} we will speculate on the possibility of  
getting information on the central 
solar temperature through homology and of testing the homology relationship
itself
with next generation experiments, elaborating 
 an observation by  Bahcall \cite{Bahcall93,Bahcall94a},
 see also Ref.~\cite{PRD}.
The final section contains our answer to the  questions raised  at the 
beginning.

In  this  entire section~\ref{cap4},  we shall  for clarity divide the neutrino 
flux into the four  components already introduced, 
$\fip$, $\fibe$, $\ficno$ and $\fib$. In the appendix B
we will briefly discuss the ratios $\fipep$/$\fip$ and $\fin$/$\ficno$,
mainly for substantiating our assertion about the stability of these quantities,
among standard and non-standard models.

Clearly our interest is on the changes of physical quantities with 
respect to the Reference Solar Model, BP95, when some input parameters
are varied.

Actually, following Refs.~\cite{Future,PRD} all throughout this section and
 section \ref{cap5} we report results obtained by modifying the 
inputs of a  starting  solar model
(CDF94) described in Ref.~\cite{PRD}: the equation of state was 
taken 
from Ref.~\cite{Oscar88}, internal opacity tables 
from Ref.~\cite{IGL92}, corresponding
to the chemical composition of Ref.~\cite{G91} and diffusion was
 neglected\footnote{CDF94 cannot be considered anymore as a standard 
 solar model, since it does not satisfy the helioseismological
 constraints Eqs. (\ref{ranges}) and (\ref{yrange}).}.

{\bf All quantities corresponding to the CDF94 model will be labelled 
with the index (*) here and in Sect. \ref{cap5}. }

\subsection{The p+p $\rightarrow$ d + e$^+$ + $\nu$ reaction rate}
\label{subsec4spp}

%\subsection{The uncertainty on $S_{pp}$}

The rate of the initial reaction in the pp chain is too low to be
directly  measured in
the laboratory (even  in the sun's center this rate is extremely slow,
of the order of 10$^{-10}$ yr$^{-1}$
consistently  with the  solar age) 
and it can be determined only by using the theory
of low energy weak interactions, together with the measured
properties of both the proton proton scattering and  the deuteron.
In terms of the astrophysical factor, $S_{11}$(E)
what really matters is its zero energy value, which for brevity,
and following the usual notation,
 will be indicated simply as $\Sppzero$. 
While we refer to Refs.~\cite{BP92,KamB} for an updated review, we remark that 
the calculated values \cite{KamB,GGue}
 are all in the range  (3.89--4.21) 10$^{-25}$ MeV b,
\ie~they differ from their mean 
 by no more than 3\%.
Kamionkowski and Bahcall \cite{KamB} give an useful parameterization,
in terms of the three quantities of
physical interest for the determination of $\Sppzero$:
 the squared overlap integral $\lambda^2(0)$, the
ratio G$_A$/G$_V$ of the axial to vector coupling constants and the
fractional correction $\delta$ to the nuclear matrix element due to
exchange currents
\begin {equation}
\label{SPP}
\Sppzero[10^{-25}{\mathrm {MeV b}}]
= 3.89 [\lambda^{2}(0)/6.92] [(G_{A}/G_{V})/1.2573]^{2} 
[(1+\delta)/1.01]^{2} \quad .
\end {equation}
The most recent evaluation of $\lambda^2(0)$
 is from \cite{KamB}, obtained by using 
 improved data for pp scattering and for the deuteron wave function
and also including the effect of vacuum polarization.
The estimated uncertainty is
 about $\pm 1\%$. The ratio $G_A/G_V$ can be obtained with an
accuracy of about 0.3\% \cite{PDB}
%\cite{Hik92} 
 from a weighted average of 
five precise modern experiments.
The contribution of the exchange currents is $\delta \approx $1\%, with a 
comparable uncertainty.

In summary, one has \cite{KamB}
\begin{equation}
\label{provo}
\Sppzero=3.89 \cdot 10^{-25} (1 \pm 0.01) \, {\mbox{ MeV b}}  \quad .
\end{equation}

Although some warning is  
in order as to the meaning
of the quoted (1$\sigma$) error, one may  conclude that well 
 known physics 
determines $\Sppzero$ to the level of few per cent or even better.
Nevertheless, as  explained in the previous section,  the variation
of \spp~well beyond its  estimated uncertainty provides  a good theoretical 
laboratory for investigating alternative solar-like structures.

In the RSM as well as in CDF94, $S_{pp}$ corresponds to the central value 
of Eq. (\ref{provo}).

When  varying \spp ($S_{pp} \rightarrow s_{pp} \, S_{pp}$),
we considered both the case of small variations ($s_{pp}$
in the range 0.9 to 1.1 ) and large variations, up to  $s_{pp}=3.5$ 
which corresponds to a 
\sBe neutrino flux reduced by a factor three. For drastically  reductions
of $\fib$, 
 unreasonable variations of $S_{pp}$ are needed, orders
of magnitude  larger than compatible with  the estimated uncertainty.
Briefly, we remark the following occurrences (see Ref.~\cite{CDF93} for 
details):

\noindent i) 
in the energy production region, temperature, density, pressure
and radius all satisfy the homology relationship Eq.~(\ref{ratioomol4})
 to better than 1\%, and the same holds throughout all the radiative 
interior (see Fig.~\ref{omospp}). The same holds  over the entire explored 
range in $s_{pp}$.

\noindent ii) 
The hydrogen mass fraction, as a function af the mass 
coordinate, is essentially unchanged with respect to the CDF94 estimate, 
as it is constrained by the solar age.

\noindent iii) 
For the power law coefficients $\alpha$ and $\beta$
[see Eqs.~(\ref{lawf}) and (\ref{FiTc})]  of the 
quantities characterizing the physical interior we found the values 
 in Table~\ref{talpp}.
The power laws look  accurate ($\Delta \alpha / \alpha$ and
$\Delta \beta / \beta$ being just a few percent)  and the coefficients 
for small and large variation of $S_{pp}$ are consistent, within
their estimated uncertainties. In addition $s_{pp}(0.1)$=2.5, \ie~one has
to multiply  $S_{pp}$ by this huge factor to reduce the central
temperature by 10\%.

\noindent iv) 
The dependence of neutrino fluxes is also shown in Table~\ref{talpp}.
For the case of small variations,  already  investigated by Bahcall
in Ref.~\cite{BU89,Bahcall1989,BahUlm}, we essentially  agree with his 
results. Even for 
large variations, power laws are very  accurate for $\fibe$ and $\fib$. 
For $\fipp$ the accuracy is smaller,
for the reasons outlined in section \ref{seciiie}.
This holds  for $\ficno$ as well, as this flux is the sum of two terms
($\fin$ and $\fio$) which depend differently on temperature.

We stress that these features (points iii and iv ) can  be well understood
analytically, assuming that the starting model and the pseudo-suns are
connected by a homology transformation and that the hydrogen mass fraction
profiles are the same, see Ref.~\cite{CDF93}.

\subsection{Radiative opacity}

An extensive and critical discussion of the uncertainties
on radiative opacities is given in Ref.~\cite{BP92}.
In the energy production region 
the typical difference between the 
outputs of the Livermore and the Los Alamos 
code is about (2--5)\%~\cite{RI91}.
At least half of the  opacity in the central region
 is due to scattering on  electrons
and  inverse bremsstrahlung in the
field of H and He nuclei,
 processes which can be calculated
with an accuracy of about 10\%, or better.
 Bahcall and Pinsonneault \cite{BP92} estimate a 1$\sigma$ uncertainty
of about 2.5\%.
On the other hand Turck-Chi\`eze \etal~\cite{TC88}
claim the uncertainty to exceed 5\%, a point criticized
in \cite{BP92}.
Recently Tsytovitch \etal~ \cite{TSY} argued that some plasma
physics effect have not been included in the calculations of
the Livermore group, so that opacity at the solar center might
be overestimated by 9\%.

Although it is hard to make a definitive statement on such a 
complex matter,
we shall conservatively take 5\% as a  1$\sigma$ uncertainty.

Since we aim to lower the internal temperatures, we
investigated the effect of reducing the opacity.
We scaled it uniformly along the solar profile
by a factor $opa$ with respect to CDF94.
We consider $opa$ as low as 0.6.
 This corresponds to a temperature reduction
of 6\% and $\fibe \approx 0.5 \fibe^*$.
For even smaller opacity, the resulting pseudo-sun would have an
original helium abundance well below the cosmological value.
 By extrapolating, one finds $opa(0.1)$=0.42.

The following points are to be noted:

\noindent i)
The homology relationship is accurate to at least  1\%  for 
any  variable characterizing the internal structure, over the entire radiative
region, see Fig~\ref{omopa}.
\noindent ii) 
Density, pressure and radius are essentially insensitive to opacity  variations,
 see Table~\ref{talopa}. On the other hand, when the 
opacity decreases the hydrogen mass
fraction increases. This can be understood by observing that as the 
star gets less opaque, the interior  becomes cooler. At the lower temperature,
the pressure gradient needed to sustain gravity is then  maintained
with a larger hydrogen abundance.

\noindent iii) 
The dependence of neutrino fluxes on the opacity parameter and the 
connection with the central temperature are also shown in Table~\ref{talopa}.
Again our results for small variations are well in agreement with those 
of~\cite{BU89,Bahcall1989,BahUlm}. 

\subsection {The ``metallicity''  of the sun}

Let us remind that the metallicity\footnote{According to the 
astrophysical jargon, any element heavier than He is termed
``metal'' or ``heavy element''}  is an input parameter
 constrained by  photospheric observations.
 For a comprehensive review about the metal 
content of the solar interior  we refer again to~\cite{BP95}
 and we mention just the main points:
 spectroscopic observations
 give the   mass abundance (Z) of the heavy elements, relative 
 to the hydrogen mass fraction (X),
  in the atmosphere of the present sun.
About 75\% 
of the heavy elements is accounted by carbon, nitrogen and  oxygen.
 For these elements, the analysis of
vibration-rotation and pure rotation lines of molecules as CO, CH,
OH, NH in the infrared, from space experiments \cite{G90,G91,Bie91} 
provide accurate information.
The abundance of elements heavier than oxygen is usually determined
by absorption lines in the optical range; these clearly show the predominance
of Iron. The relative distribution of these elements can be usefully 
tested to the distribution of CI carbonaceous chondrite meteorites, which 
should keep
the composition of these not volatile elements in the original solar
nebula~\cite{AG89,GN93}.
It is important to note that the recent photospheric 
Iron abundances 
\cite{GN93} agree 
 now well with the meteoritic values.

According to the most recent evaluations \cite{GN93}, one has
for the photosphere of the present sun:
\begin{equation}
({\mathrm Z/X})_{photo}= 0.0245
\end{equation}
with an accuracy better than 10\%.

 We remark that, due to diffusion 
towards the solar center, the original
heavy elements abundance in the sun should be higher, by 10-15\%.

When building solar models, 
this time we keep $z$ = (Z/X)/ (Z/X)$^*$ as a free variable.
We considered small variations 
($z$ = 0.9--1.1) and  large variations, with $z$ as small as 0.1.
With such a small value the central temperature is decreased by about
10\% and $\fibe$ is  about 1/3 of the initial prediction.
In this case $z$(0.1)=0.17.
While we refer for details to~\cite{PRD,Future}, we summarize here the
main results, see Fig.~\ref{omoz}. 

\noindent i)
The temperature profile satisfies the homology relationship with an
accuracy better than 1\% throughout  the entire radiative interior. 
 Homology holds also for R$(m)$, but with less 
accuracy. On the other hand,  even in the energy production region, homology
is only a fair approximation for pressure and density: for large 
variation of Z/X even in this restricted  region $\delta \omega / \omega 
\approx 5\%$.

\noindent ii)
The picture is somehow similar to that of the opacity variations:  R
remains essentially unchanged on the average, whereas X grows as metallicity
is diminished, see Table~\ref{talz}.

\noindent iii)
This similarity also  reflects on the neutrino fluxes, the temperature 
dependence of these latter are essentially the same as in the previous 
subsection a part from $\ficno$, see below.

As already mentioned, the similarity with the variations of the opacity can be 
readily understood since  metallicity affects mainly the stellar opacity.

\subsection{Rejuvenated suns}

Unlike the other  parameters used to constrain the SSM, 
%(M$_\odot$, R$_\odot$, L$_\odot$ and Z/X)
 the age of the sun 
is not an observable.
It is inferred from the dating of the oldest meteorites, 
provided that a connection between the   formation  time
 of the meteorites and the birth of the sun 
(\ie~the ignition of H burning) is achieved. 
A recent discussion is provided by Wasserburg in \cite{BP95}. 
On this basis Bahcall and Pinsonneault estimate (at $1\sigma$) \cite{BP95}
\begin{equation}
{\mbox{t}}_\odot= 4.57 \pm 0.01 \, \mbox{Gyr}
\end{equation}

Since the main difficulty is in establishing the evolution phase of the sun
 at the formation time of meteorites,  we shall take as a conservative estimate
of the uncertainty  the duration of the pre-main sequence
phase, $\Delta t_\odot \approx 30$ Myr.

The solar luminosity and age fix the amount
of H which is burned into He in the solar core.
In younger suns more H
is available, so that fusion reactions are more
likely  and the observed luminosity  can be reached 
at smaller $T_c$.

Correspondingly, a 
younger sun  has a different
composition in the energy production region,
this difference becoming more and more marked as t$_\odot$ is shortened,
and this drives the structure progressively away from  homology
(see Fig.~\ref{tempeta}).

In CDF94 we started with t$_\odot^*$= 4.6 Gyr.

We reduced the solar age down to t$_\odot=0.1$ t$_\odot^*$, this extreme 
corresponding
to a central temperature reduction of about 6\% and 
$\fibe \approx 1/2 \, \fibe^*$.
For even shorter ages, the structure of the pseudo-sun would be deeply 
modified by
the occurrence of a central (\tHe driven) convective core. We found the
following results:

\noindent i) 
homology is now just a fair approximation  (see Fig.~\ref{omoeta}). 
Even restricting oneself only to the energy production region, the accuracy
in temperature is about 2\% or less; that in  density and pressure
about  10\% and for the hydrogen abundance it is merely about  25\%.

\noindent ii) 
Correspondingly, the power laws for the variables 
characterizing the  interior are only rather approximate
($\Delta \beta / \beta \approx 20\%$, see Table~\ref{talt}).

\noindent iii) 
As regards neutrino fluxes, we are in fair agreement 
with the results given in \cite{BU89,Bahcall1989,BahUlm} for small variations.
The dependences on temperature are in any case similar to those previously 
found.

\subsection{Low temperature models and neutrino fluxes:  summary and 
explanation}
\label{para}

An important general feature of  the models discussed above 
is the approximate homology of the 
temperature profiles
\begin{equation}
\label{homo}
  T(m)= \tau T^*(m) \quad ,
\end{equation}
where $m={\mbox{M}}/{\mbox{M}}_\odot$ is the mass coordinate,
 and the factor $\tau$ depends on the 
parameter which is varied but does not on $m$.

We have verified that Eq.~(\ref{homo}) holds to an accuracy better than 1\% 
in the entire  radiative zone (M/M$_\odot<0.98$ or 
R/R$_\odot<0.7$) for all the 
models considered, except for huge (and really unreasonable) 
variations of t$_\odot$  (see Fig.~\ref{fig5}).
% and Table~\ref{thomo}.
It is worth 
noting that $T(m)/T^*(m)$ stays constant throughout a region where 
$T(m)$ change by a factor five.
% see Fig.~\ref{fig6}.
The scaling factor $\tau$ may be taken as a 
the ratio of 
 the central temperature $T_c$ to that of the starting model:
\begin{equation}
\tau=T_c/T_c^* \quad .
\end{equation}
The coefficients $\beta_i$ for the power laws of neutrino fluxes {\em vs}
temperature
\begin{equation}
\label{ftdep}
  \Phi_i = \Phi_i^*\,
            \left(\frac{T_c}{T_c^*}\right)^{\beta_i}  \quad.
\end{equation}
are collected in Table~\ref{betas}, where we include for completeness
all flux  components.

 One notes that 
$\beta_{p}$, $\beta_{Be}$, and $\beta_{B}$
are largely independent of the parameter 
which is being varied (see also Fig.~\ref{fig7}).
In other words,  {\bf these fluxes are mainly determined by the central 
temperature, almost independently of the way the temperature 
variation is imposed}.

The dependences of the  fluxes  can be understood
semi-quantitatively  by  simple analytical arguments.
As a  zeroth order  approximation, let us assume that 
energy production occurs entirely through the ppI termination, and that
the chain is fully equilibrated.
Requiring that the rate of \tHe burning 
($n_3^2\langle\sigma v\rangle _{33}$)
corresponds to the fusion rate ($ n_1^2 \langle \sigma v \rangle_{11} /2$)
at each point in the stellar core,
the equilibrium \tHe density $n_3$ is given by \cite{CDF93}:
\begin{equation}
\label{equi3he}
n_3^2=
\frac{1}{2} n_1^2 \frac{\langle \sigma v \rangle_{11} }
                 {\langle \sigma v \rangle_{33}}
=\frac{2 \epsilon \rho } {\qpp \langle \sigma v \rangle_{33}} \quad ,
\end{equation}
where $\epsilon$ is the energy production rate per unit mass.

The production of \sBe nuclei,  and consequently of \sBe neutrinos, can
be treated as a perturbation to the ppI termination. Practically
every \sBe nucleus  produced is destroyed through electron 
capture, with emission of a \sBe neutrino. The production rate
per unit mass of the latter, $w_{^7Be}$, is thus equal
to the production rate of \sBe nuclei, again per unit mass:
\begin{equation}
\label{rate7be}
w_{^7Be}=n_3 n_4 \langle \sigma v \rangle_{34} / \rho
\end{equation}
With Eq.~(\ref{equi3he}) and $n_4=\rho Y N_{A}$ one gets:
\begin{equation}
\label{equepsi}
w_{^7Be}= \frac{N_{A} \sqrt{2} }{4 \sqrt{\qpp}}
\frac{\langle \sigma v \rangle_{34}}{\sqrt{\langle \sigma v \rangle_{33}}}
 Y \sqrt{\epsilon \rho}
\end{equation}
The nuclear reaction rates
$\langle \sigma v \rangle_{ij}$ are strongly temperature dependent.
This dependence is usually parameterized by power laws \cite{Rolfs}:
\begin{equation}
\label{paramrate}
\langle \sigma v \rangle_{ij} \propto T^{\gamma_{ij}}
\end{equation}
and the coefficients $\gamma_{ij}$ 
are given in Table~\ref{gammas}.

Clearly quantities like Y, $\epsilon$ and $\rho$ are also connected with 
temperature, but the dependence is much weaker than for the reaction
rates; as a further approximation, we assume they are the
same as in the starting model.
On the other hand, we use  homology for the temperature
profiles. 
In this way, from Eq.~(\ref{equepsi}),
the neutrino production at any mass coordinate $m$
is related to that of the starting model:
\begin{equation}
w_{^7Be}(m)=w_{^7Be}^*(m)
\tau^{\gamma_{34}-\frac{1}{2} \gamma_{33}}
\end{equation}

The same equation obviously holds for the fluxes on earth,
$\fii= \frac{1}{4\pi R_{ES}^2} \int dm \, w_i(m)$:
\begin{equation}
\label{flussobeana}
\fibe=\fibe ^* \tau^{\gamma_{34}-\frac{1}{2} \gamma_{33}}
	=\fibe ^* \tau^{8}
\end{equation}

One can study the production of \oB neutrinos similarly. Their
production rate per unit mass is:
\begin{equation}
\label{rate8b}
w_{^8B}=n_1 n_7 \langle \sigma v \rangle_{17} / \rho \quad ,
\end{equation}
whereas for \sBe neutrinos one has:
\begin{equation}
\label{rate7be2}
w_{^7Be}=n_e n_7 \langle \sigma v \rangle_{e7} / \rho  \quad . 
\end{equation}
By eliminating the equilibrium \sBe nuclei density $n_7$
one has:
\begin{equation}
\label{rate8be2}
w_{^8B}= \frac{\langle \sigma v \rangle_{17}}
               {\langle \sigma v \rangle_{e7}}
		\frac{n_1}{n_e} w_{^7Be}
\end{equation}
Again assuming $n_1/n_e$ to be essentially that of the  starting model,
 one obtains after integrating 
over the mass coordinate:	
\begin{equation}
\label{flussobana}
\fib=\fib ^* \tau^{\gamma_{17}+\gamma_{34}-\frac{1}{2} \gamma_{33}
-\gamma_{e7}} =
    \fib ^* \tau^{21.5}
\end{equation}

We are now also able to estimate the  temperature  dependence of $\fip$.
 We recall that the two main components 
are $\fip$ and $\fibe$, and that their sum is fixed by the luminosity 
constraint.
By differentiating with respect to
$T_c$, this implies approximatively:
\begin{equation}
\beta_p= - \beta_{Be} \frac{\fibe ^*}{\fip ^*}=-0.6
\end{equation}

The values we find are in agreement with the
numerical estimates  of Table~\ref{betas}. It is
not surprising that the analytical values
 are quite close to the coefficients obtained by varying
$S_{pp}$, since  scaling works
best   in this case. 

Let us remark a few relevant points:

\begin{itemize}

\item
the temperature dependences of
$\fip,\fibe$ and $\fib$  are 
well under control. One sees from the foregoing discussion that
they are
essentially determined by the behaviour of 
$\langle \sigma v \rangle _{ij}$ as a function of temperature;
the latter behaviour is 
fixed mainly by  the Coulomb barrier
\cite{Rolfs}.
Solar physics only enters through the homology relationships
of the  temperature profiles. 

\item
The flux $\fibe$
can been determined independently of the value of the \sBe lifetime.
No matter what  the value of this latter is, practically  all \sBe nuclei
 produced will emit a \sBe neutrino.

\item
The ratio of $\fib$ to $\fibe$
is essentially governed 
by nuclear physics, its temperature dependence being that of
$\langle \sigma v \rangle_{17}/ \langle \sigma v \rangle_{e7}$.

\item
For any temperature one has
\begin{equation}
\label{berbor}
\frac{\fib}{\fib ^*}\approx (\frac{\fibe}{\fibe^*})^{2} 
\end{equation}
\ie~{\bf the suppression/enhancement
of $\fib$ is much stronger than that of $\fibe$.
A reduction of $\fibe$ to 1/3 of the RSM prediction
 implies a reduction of $\fib$ by an order of magnitude.
This essentially  illustrates the failure of low temperature models
when compared with experimental data. }

\end{itemize}

The CNO flux appears sensitive not only
to the temperature, but also to other parameters  characterizing the solar 
interior. The drastically different exponents, found when 
varying the metallicity, can be understood by noting that the efficiency 
of the CN cycle is affected not only by temperature but also by the number
densities 
of nuclei which  act as catalyzers of the chain.

\subsection{More about  homology:  additional consequences and a possible test}
\label{still}

One generally thinks that the neutrino production zone is so well  hidden 
below the solar surface that it can  hardly be 
studied experimentally other than with neutrinos.

Actually, the accurate  homology of temperature 
profiles, valid up to the border
of the radiative interior, indicates  a strict connection between the properties
of the energy production region and of more external layers of the sun.
If one is confident in homology, then a measurement of temperature at, say,
the bottom of the convective zone immediately gives
the temperature of the solar center.

A new generation of  experiments is being planned for detecting 
monochromatic  neutrinos produced in electron capture
(\sBe$+ e^- \rightarrow {}^7$Li$+ \nu$)
and in the $pep$ ($ p + e^- + p \rightarrow  d + \nu$)
reactions~\cite{borex,Alessa,Ragha}.
Bahcall~\cite{Bahcall93,Bahcall94a} has 
pointed out that one can, from  measurements of the average energy difference
between neutrinos emitted in solar and laboratory decay, infer
the temperature of the production zone. 
The possibility of measuring inner solar temperatures through thermal
effects
on monochromatic neutrino lines is extremely
fascinating (although remote). In this respect the homology
relationship, Eq.~(\ref{homo}),
is particularly interesting, see Fig.~\ref{fig13}.
If homology holds, then a measurement of the solar temperature  in the 
 \sBe production zone gives the value of $T_c$.

In addition,
the homology relation itself is testable -- in principle -- by comparing
the temperatures at two different points, as can be done by looking
at the shapes of both the
$\nu_{Be}$ and $\nu_{pep}$ lines. We remark that this
would be a test of the energy transport  mechanism in the
inner sun. 

\subsection{On the accuracy of the central solar temperature}
\label{sec4accuracy}

A rough estimate of $T_c$ can be obtained by equating the thermal 
energy of a hydrogen nucleus to its gravitational energy 
($k T_c \approx G{\mbox{M}}_\odot m_p /{\mbox{R}}_\odot$). In this way 
one finds 
$T_c \approx 2 \cdot 10^7$ K,  in  good agreement with the 
much more refined SSM
estimates (see Table \ref{Modelli}). Different standard solar models
give  the same value of $T_c$ within 1\% 
and solar model builders 
claim that the present accuracy on $T_c$ is of the same order.

From the preceding discussion, the reader can derive his own
opinion about this claim. We have seen that the main regulators of the
central solar temperature are $S_{pp}$, the metallicity,
the solar age and the opacity of the solar interior. In terms of the  
uncertainties in these quantities, one has:
\begin{equation}
(\frac {\Delta T_c} {T_c} )^2=
 			(\alpha_{T,S_{pp}}\frac{\Delta S_{pp}}{S_{pp}})^2
			+(\alpha_{T,Opa}\frac{\Delta Opa}{Opa})^2
			+(\alpha_{T,Z/X}\frac{\Delta (Z/X)}{(Z/X)})^2
			+(\alpha_{T,Age}\frac{\Delta Age}{Age})^2
 \end{equation}
Using the power law coefficients $\alpha_{T,i}$ and the estimated
 uncertainties summarized in Table \ref{tempera}, one actually
gets $\Delta {T_c}/T_c \approx 1\%$. One also sees that 
most of the error results from the uncertainties 
in  opacity  and in metallicity (note
that with the uncertainties
as estimated  in \cite{BP92,BP95} the global uncertainty on $T_c$ would
be halved). 

The next question is how the main parameters should be varied 
in order to get  a solar model with drastically reduced 
\oB and/or \sBe neutrino fluxes. For instance if one requires
$\fib/\fib ^* \approx 1/2$ a reduction of $T_c$ by 3\%
is needed. Although this  does not look terribly outside
the allowed $T_c$ range, really huge variations of the physical
inputs are needed  (see again Table \ref{tempera}, 4th column).

The situation is even more desperate if one tries
to reduce the \sBe flux by a factor two (three). In this
case temperature has to be reduced  by at least 7\% (13\%), 
which  requires the really unreasonable variations given in the last columns
of Table \ref{tempera}.
The input parameters are to be varied by an order of magnitude (or more)
with respect to their estimated uncertainties. Even a 40\% 
reduction of opacity  can only reduce the \sBe neutrino flux
by at most a factor two.
Last but non least, as is clear from Eq.~(\ref{berbor}), it is essentially 
impossible to account for the reductions of both $\fibe$ and $\fib$.

{\bf One concludes that a solution to the solar neutrino problem cannot be found
merely by reducing  the central temperature }.
\newpage
%%%%%%%%%%%%%% SECTION  FIVE %%%%%%%%%%%%%%%%%
%

\section{Nuclear reactions in the sun}
\label{cap5}

One may suspect  that the solar 
neutrino problem is due to some inadequacy in  our understanding 
of nuclear reactions in the solar interior.
  What could be wrong with the nuclear burning rates used in standard
  solar model
  calculations? There are (at least) three sides to this question, {\em viz.}:
  
 \begin{itemize}
 \item  
  Nuclear physics: as repeatedly stated, the astrophysical S-factors 
  used in 
  stellar model calculations are 
  generally obtained by  extrapolating 
  experimental data taken at energies higher than those relevant for 
  the solar interior. Although the underlying theory is robust,
  one can be suspicious  of extrapolations.

  \item
   Atomic/molecular physics: experiments in the laboratory use atomic or 
molecular targets. At the lowest measurable energies, electron screening is
relevant and its effect has to be subtracted when deriving
cross sections for bare nuclei. Indeed, the effect of elctron screening
has been detected \cite{Rolfsschermo}, however theory and experiments seem to 
disagree \cite{Rolfsschermo,bracci}.
   
  \item  
  Plasma physics: the burning rates for bare nuclei are then to be corrected
  for the screening of nuclear charges  by the solar plasma.
  In any calculation, the predicted effects are small; however,  the theory 
  is not completely satisfactory \cite{schermonoi} and there are no direct 
  experimental tests.
   The disagreement between theory and experiment for electron screening in 
atomic/molecular targets provides some warnings:
   although that is a different context,
    one has to keep an open mind about the possible effect of plasma screening
    on neutrino production.
  \end{itemize}

As shown in Sec. \ref{cap2}, the main problem is with the \sBe 
neutrinos, so that particular attention has to be given to
 the  reactions in the pp chain preceding
the formation of \sBe nuclei; in other words, those reactions which 
determine the branching between the  ppI and ppII chains
(see Fig.~\ref{catenapp}). 
The r\^{o}le of the p+p cross section was already 
discussed in Sec.~\ref{subsec4spp}, as it is one of central temperature 
regulators.
The value of the p+d $\rightarrow$\treHe +$\gamma$ cross section 
is unimportant. It is orders of magnitude larger than that of the weak 
interaction p+p$\rightarrow$ d+$e^+ +\nu_e$ process, so that on a very 
short time scale equilibrium is reached and any practically pp reaction is 
followed by the production of one \treHe nucleus. 
 
 The cross sections for reactions  between He isotopes,
 \begin{equation}
 {\mbox {\treHe+\treHe}} \rightarrow {\mbox {\qHe}} +2p 
 {\mbox{~~and~~\treHe+\qHe}}\rightarrow {\mbox{\sBe}}+\gamma \quad ,
 \end{equation}
  are clearly crucial and we shall pay particular attention to them. On 
  the other hand,
  the proton capture by \sBe:
 \begin{equation} 
  p+{\mbox{\sBe}} \rightarrow{\mbox{\oB}} +\gamma
  \end{equation}
  is of minor  relevance for the flux of \sBe neutrinos
 and governs essentially only the production of \oB neutrinos,
see Sec.~\ref{nuova}. Similarly, the raction:
\begin{equation}
p+^{14}{\mbox{N}} \rightarrow ^{15}{\mbox{O}} +\gamma \quad ,
\end{equation}
which will be briefly discussed in Sect. \ref{subsecn}, essentially determines
the production of CNO neutrinos only.
 
The goal of this section is to study the changes induced  in neutrino fluxes 
when the  nuclear burning rates  are changed with respect to RSM inputs.
The comparison of these results with  neutrino experiments will be 
discussed in the section~\ref{cap6}.
As mentioned in Sec.~\ref{cap4}, we recall that our starting  standard
solar model  here is CDF94 \cite{Future,PRD}.

  \subsection{The status of \treHe+\treHe $\rightarrow $ \qHe + 2p 
 reaction}
 \label{sec533}
  
    The relevant energy range is  determined by the  energy
    $E_o$ and the 
    halfwidth,  $\Delta /2$ of  the Gamow
     peak\footnote{$E_o=1.22 (Z_1^2 Z_2^2 \mu  T_6^2 )^{1/3}$ keV and
    $\Delta = 0.749 (Z_1^2 Z_2^2 \mu T_6^5)^{1/6}$ keV,
    see \eg~\cite{Rolfs}.}.
 For the central solar temperature one has
    $E_o$=22 keV and $\Delta$=12 keV.
    The  available experimental data 
\cite{GO54,KR87,DW74,DW71,WA66,BA67,BR87} are shown in  
Fig.~\ref{datis33}. Data  below
25 keV do not exist
and at the lowest measured energies errors are of the order of 20\% or 
greater, so that some extrapolation  is necessary in 
order to reach the relevant energy range.

The  astrophysical S-factors are usually  
 parameterized with a Taylor expansion:
 \begin{equation}
 \label{polinomio}
S_{ij}(E)=S_{ij}(0) + S_{ij}'(0) E+ \frac{1}{2} S_{ij}''(0) E^2 \, ,
\end{equation}
where  mainly the coefficient $S_{ij}(0)$ matters for the sun,
as $E_o$ is much smaller than the nuclear energy scale.
 In Table~\ref{S033} we present some estimates of $S_{33}(0)$:
the value  used in the Caughlan and Fowler compilation \cite{CF88} 
corresponds to  
the experimental result of \cite{KR87}; 
Parker and Rolfs  \cite{PA91} give a weighted  average
of several experimental results.

We derived a  new estimate of $S_{33}(0)$  by reanalyzing all available data,
except for those of the pioneer experiment by Good 
\etal~\cite{GO54} which
is systematically a factor 2--3 below the 
others  and carries no estimated error. A fit to the data using 
Eq.~(\ref{polinomio}) gives $S(0)$= (5.3$\pm0.1)$ MeV b. 
The data at low energies are however affected by electron screening
\cite{Rolfs}.
This effect is negligible at $E \geq 100$ keV and data above this threshold 
give the value of the  4th row in 
Table \ref{S033}. A slightly smaller result is obtained if one considers all 
energies and corrects for electron screening effect in the adiabatic 
approximation \cite{bracci}, as shown 
in the last row. We consider this last value as the best estimate of $S_{33}(0)$
for bare nuclei. 
Correspondingly we find:
\begin{equation}
\label{poli33}
S_{33}(E)=(5.1 \pm 0.2)  +  (-3.0\pm 0.4 ) E  
           + \frac{1}{2} (3.0 \pm 1.0 ) E^2 \, .
\end{equation}
This expression (energies in MeV and $S$ in MeV b)
 gives a good fit to all data ($\chi^2_{d.o.f.}=0.8$).

At the Gamow peak near the solar center, one thus  finds
 $S_{33}=5.0$ MeV b, slightly higher than, but still 
consistent with the value
used in the RSM, $S_{33}=4.8$ MeV b.
In CDF94 we used $S_{33}^*=4.98$ MeV b.

\subsection{The status of \treHe + \qHe $\rightarrow$ \sBe + $\gamma$
reaction}
\label{sec534}

The range of  relevant energies  is  essentially the same as above
($E_o= 23$ keV, $\Delta= 12$ keV), 
whereas data are available only at $E>100$ keV. Concerning the experimental 
results collected in Fig.~\ref{datis34}, the following comments are 
needed:
 %\noindent
i) the original data of \cite{PA63} have been corrected, following the comment
in \cite{NA69}; 
%\noindent
ii) data from \cite{KR82} have been multiplied by a factor 1.4 
according to\cite{HI88};
% \noindent
iii) errors quoted in \cite{OS84} have been doubled
so that  fluctuations among data points become  statistically consistent;
iv)at all 
 the energies where data are available electron screening is 
irrelevant.

Again what matters is  the astrophysical S-factor
at zero energy. In Table~\ref{S034} we report 
some different determinations of this quantity. 
The value used in   \cite{CF88}  coincides with that
of the review paper \cite{PA85}, which was obtained as a weighted average
of the extrapolations provided  by different experiments. A similar, more 
recent analysis 
 \cite{PA91} gave a slightly smaller value.
It has however to be remarked  that the extrapolations were performed
 using different theoretical models, so that combination of extrapolated 
values is dubious.
We performed  a new analysis of all experimental data
\cite{HO59,PA63,NA69,KR82,OS84,AL84,HI88,RO83,VO83}.
For a quadratic expansion of the astrophysical S-factor we find
\begin{equation}
\label{poli34}
S_{34}(E)= [(4.8 \pm 0.1) + (-2.9\pm 0.2)E  + (0.9\pm0.1)E^2 ] \cdot 
10^{-4} 
\end{equation}
(again $E$ in MeV and $S$ in MeV b). Alternatively, 
by using an exponential parameterization, as frequently adopted
 in the literature and supported by theoretical
models (see \eg~\cite{WI81}), we get:
\begin{equation}
\label{expo34} 
 S_{34}(E)= (5.1 \pm 0.1) \cdot 
10^{-4}  \exp [(-0.83\pm0.07)E + (0.25\pm0.03)E^2] \, .
 \end{equation} 

By using these two parametrizations, at the Gamow peak near 
the solar center one finds respectively
$S_{34}(E_o)=4.74$ and 5.01 in $10^{-4}$ MeV b.
One sees that uncertainties due to the extrapolation procedure
are at least comparable to the quoted statistical
error, so that the global error  is about $\pm 2\cdot 10^{-5}$ MeV b.

The value of $S_{34}(E_o)$ used in the RSM is based on
\cite{PA91}, after introducing a 1.6\% decrease due to the
vacuum polarization effect 
\cite{Kamion}, yielding $S_{34}(E_o)=5.17 \cdot 10^{-4}$ MeV b,
slightly larger than that given by 
our preferred expression Eq. (\ref{expo34}), but still
consistent with it within uncertainties. In CDF94 we had
$S_{34}^*(E_o)=5.26 \cdot 10^{-4}$ MeV b.

 \subsection{Neutrino fluxes and helium reactions}
 \label{sec5neutrini}
  
  In order to estimate the dependence of \sBe and \oB neutrino fluxes
  on the nuclear cross sections,
essentially we repeat here  the argument of Sec.~\ref{para}.
 Let us first consider the 
  local equilibrium concentration of the parent \sBe nuclei, $n_7$.
\sBe is created in the \treHe+\qHe reaction and destroyed, essentially via 
electron capture, with a lifetime $\tau_{e7}$. Thus at equilibrium:
\begin{equation}
n_7=\tau_{e7} n_3 n_4 \langle \sigma v \rangle _{34} \quad ,
\end{equation}
where $n_i$ is the number density of the nuclei with mass number equal to 
$i$. The \treHe equilibrium density is obtained by equating 
its creation rate
(\ie~the rate of the p+d$\rightarrow$ \treHe +$ \gamma $ reaction,
which equals the rate of p+p$\rightarrow$ d +e$^+ +\nu_e$)
 to the burning rate, which is dominated by the \treHe + \treHe reaction.
With this approximation, one gets:
\begin{equation}
n_3^2= \frac{1}{2} n_1^2 \frac{\langle \sigma v \rangle _{11}} 
                           {\langle \sigma v \rangle _{33}}\quad .
\end{equation}
By using the above equations  one can derive:
\begin{equation}
n_7 = \frac{1}{\sqrt{2}} \tau_{e7} n_1 n_4 
\sqrt{\langle \sigma v \rangle _{11}}
                    \frac{\langle \sigma v \rangle _{34}}
                  { \sqrt{\langle \sigma v \rangle _{33}}}
\end{equation}

If  the  rates \svtt~and \svtq~differ from
 the starting inputs,  one
expects that only the probabilities of the 
 pp terminations are  varied whereas the densities
$n_1$, $n_4$ and the temperature  are essentially unchanged with respect to
the starting predictions, since the hydrogen burning rate and the 
helium abundance in the solar interior are determined by the
present luminosity and the present age of the sun. This expectation is 
confirmed by numerical experiments, see for example Fig. \ref{temps33}.
One thus gets that
 only the following combination of \svtt~and \svtq~matters:
\begin{equation}
\chi=\frac{\langle \sigma v \rangle _{34}}
                  { \sqrt{\langle \sigma v \rangle _{33}}}
\end{equation} 
In terms of $\chi$ one has:
                \begin{equation}
\label{n7x}
n_7= n_7^* \frac{\chi}{\chi^*}
\end{equation}
We remark that Eq.~(\ref{n7x}) holds at each point  in the solar interior.

Assuming for the moment that there are no resonances in the energy range of 
interest and  that the low-energy astrophysical S-factors differ from 
those  used in the starting model by a constant factor $s_{ij}$
 ($S_{ij}(E)=s_{ij} S_{ij}^*(E)$),
then at any point in the sun:
\begin{equation}
\frac{\chi}{\chi^*} = \frac{s_{34}} {\sqrt{s_{33}}}
\end{equation}
and consequently:
\begin{equation}
n_7= n_7^{SSM}\frac{s_{34}} {\sqrt{s_{33}}}
\end{equation}
The \sBe and \oB neutrino fluxes, which are obviously both proportional to 
$n_7$, scale then in the same way
\begin{equation}
\label{scala3He}
\fibe=\fibe ^* \quad s_{34} \quad s_{33}^{-0.5}\quad ; \quad
\fib=\fib ^* \quad s_{34} \quad s_{33}^{-0.5}\quad
\end{equation}
The p (= pp+pep) neutrino flux can be best derived by the 
luminosity constraint, Eq. (\ref{lum2}). 
 
 Numerical experiments confirm these analytical estimates.
In Fig. \ref{flussis33} one sees that for each component the flux is 
actually determined by the variable $\chi$ and that Eq.~(\ref{scala3He})
provides a good approximation. 
In Table \ref{tabalfas33} we show the power law coefficients for the 
main components of neutrino flux.
Our results for small variations agree with those which can be derived
from \cite{Bahcall1989}.
The last column presents the case of large variations.
One notes that the numerical values are well in agreement with the analytical
estimates presented above.

Note also that pp+pep and CNO neutrinos are essentially insensitive 
to variations of $\chi$ as expected since the temperature is unchanged.

Large variations
are really necessary if one wants a
 reduction of \oB and \sBe neutrino flux by a factor 2--3.
  As an example, if $S_{33}$ is kept fixed then $S_{34}$ has to be 
reduced by a factor 2--3, 
or  $S_{33}$ has to be enhanced by a factor 4--9 if $S_{34}$ is unchanged.
 This  is clearly in conflict with the experimental
  situation discussed in the previous 
 sections.
 
  The only way out,
  a desperate one, is to invoke a resonance
 \cite{Fowres} in the 
 \treHe+\treHe reaction, the resonance energy $E_r$
 being below the  experimentally explored region,
 \ie~$E_r \leq 25$ keV.
 Such a resonance is not predicted theoretically; furthermore 
 experimental searches for excited $^6$Be states  in reactions like
 $^6$Li(p,n)$^6$Be and others failed \cite{Rolfs}. One can however not
 definitely exclude this possibility \cite{Rolfs}.
 
A resonance would  affect 
 the various components of the neutrino flux differently. 
Qualitatively, a very low 
 energy resonance will be more effective in the more external 
 (cooler) solar regions, so that the \sBe neutrino flux can be more 
 suppressed  than the \oB flux. 
 The converse is true for a higher energy 
 resonance, the border between the two regimes being provided 
 by the Gamow energy for \treHe+\treHe reaction, about 22 keV
 near the solar center.

 The best case for our purpose \cite{AA} is that of $E_r$=0 which corresponds,
for a strong resonance, to 
 \begin{equation}
 \frac{\fibe/\fibe ^*}{\fib/\fib ^*} =0.75
\end{equation}

In practice, the effect of such a zero energy resonance can be mimicked by
introducing a resonance S-factor, S$_{res}$ and parameterizing the fluxes
as:
\begin{equation}
 \label{fibetesi}
\fibe=\fibe^* \, \left ( 1+ \frac{16}{9} \frac {S_{res}}{S_{33}^*} \right ) 
^{-1/2}
   \end{equation}
   \begin{equation}
   \label{fibotesi}
  \fib=\fib^* \, \left ( 1+  \frac {S_{res}}{S_{33}^*} \right ) ^{-1/2}
 \end{equation}
All other fluxes are unchanged,
except for the pp+pep flux which can be derived by the luminosity constraint,
Eq.~(\ref{lum2}).
In the Sec.~\ref{cap6} we compare
the results of solar models with such a hypothetical resonance
with experiments.

Before concluding this section, we remark that, in principle, also
a narrow resonance below threshold (\ie~$E_r <0$) could work, since
again it is more efficient in the \sBe production region than in
the solar center, where \oB is produced. Note however that the  resonance
has to be very close to the threshold, otherwise it is uninfluential.

The experimental situation on such hypothetical resonances should be clarified
by the ongoing LUNA experiment \cite{Luna} at the underground Gran Sasso
National Laboratory.

\subsection{The p+\sBe$\rightarrow$ \oB +$\gamma$ reaction}
\label{nuova}

The determination  of $S_{17}$, 
the astrophysical factor for the  p+\sBe$\rightarrow$ \oB +$\gamma$ reaction,
involves several complications:
 i)
extrapolations are needed 
to reach the relevant energies in the solar interior
($E_o =19 $keV) as measurements have been performed only
at $E_{CM}>100$ keV \cite{KAV60,PAR68,KAV69,VAU70,WIE77,FIL83},
 see Johnson \etal \cite{koonin2} for a detailed discussion about the 
extrapolation procedure;
ii)
the number of \sBe nuclei present in the target is most accurately
determined by monitoring the build-up of $^7$Li (the \sBe decay product)
as a function of time, using the $^7$Li(d,p)$^8$Li reaction, see \cite{FIL86}. 
 For this technique to be useful, the cross section for this latter 
reaction must be known and there has been considerable work on this point
since 1978, resulting in a determination with an accuracy of 6\% \cite{FIL86}.
Note that in the past different values of this  normalization cross section
have been  used by different authors. 

The available data from different experiments, all normalized to the same
value  $\sigma_{dp}=157\pm10$~mb and extrapolated according to \cite{koonin2},
are summarized in Table \ref{tabs17}. For deriving an average
value, the experimental results of \cite{KAV69} and \cite{FIL83} are 
particularly important since they correspond to measurements 
performed over a wide energy interval, reaching the lowest 
energies. One has to remark that, at each energy, the values
of \cite{FIL83} are systematically lower than those of \cite{KAV69}.
Keeping into account this uncertainty, Johnson 
\etal~derived\footnote{Recently, the value  $\sigma_{dp}=146$~mb has been 
recommended in Ref. \cite{ST96}. A reanalysis of all data by 
U. Greife and M. Junker within the NACRE collaboration 
yields $S_{17}(0)= 19.9\, \mbox{eV b}$\cite{Greife}.}:
\begin{equation}
\label{S17}
S_{17}(0)= 22.4\pm 2.1\, \mbox{eV b}
\end{equation}

This is the value adopted in the Reference Solar Model (Table \ref{cross}).

The Coulomb dissociation process has recently attracted
a great deal of attention as an alternative method
to study radiative capture reactions of astrophysical interest
at low energies. The process can be treated as the absorption of
a virtual photon, essentially the inverse of the radioactive
capture process, see Fig.\ref{diagrammi}. The cross section for
 Coulomb dissociation of \oB --- the
$^{208}$Pb(\oB,\sBe p)$^{208}$P reaction --- was measured  with a radioactive
\oB beam at RIKEN \cite{MOTO} and a preliminary value 
$S_{17}(0)= 16.7 \pm 3.2$ eV b was deduced from the data.
A theoretical analysis of  the same data by Langanke and Shoppa
\cite{LS} yielded  an even smaller value, $S_{17}(0)= 12 \pm 3$ eV b,
which was however criticized  in \cite{GAIBERT}. 

All in all, this indirect approach looks very nice, but the extraction
of $S_{17}$ is experimentally  difficult and 
theoretically complex, and we agree with the authors of Ref.~\cite{LS} that
``{\em The recently developed technique of Coulomb
dissociation  might prove itself as a very
useful tool in nuclear astrophysics \ldots
The \oB Coulomb dissociation experiment at RIKEN 
takes a first step in this direction and should be continued
and refined. However a reliable determination of  the 
astrophysically important cross section for the \sBe(p,$\gamma$)\oB
reaction from \oB Coulomb dissociation experiments has to 
wait  until improved data become available}''~\cite{LSreply}.

We thus consider Eq.~(\ref{S17}) as our reference
value, and we will  study the effects of varying $S_{17}$.
It is clear that acting on a very minor termination of the fusion chain,
the properties of the solar interior (profiles of temperature,
density, \ldots) will be unchanged. This expectation is confirmed
by numerical experiments, see for instance Fig. \ref{TCS17}. Thus, only
the relative intensity of the ppII and ppIII terminations are affected, and
consequently only the \sBe and \oB neutrino fluxes will change, their sum being
fixed at the SSM value. If the astrophysical factor is scaled
by an amount $s_{17}$,
one has:
\begin{eqnarray}
&\frac{\fib}{\fibe}= \mbox{$s_{17}$} \frac{\fib ^*}{\fibe ^*}\\ \nonumber
& \fib+\fibe =\fib ^* +\fibe ^* 
\end{eqnarray}
This clearly implies:
\begin{eqnarray}
&\fib=\fibssm \, \mbox{$s_{17}$} \, 
\frac{1+\fib ^*/\fibe ^*}{1+ \mbox{$s_{17}$} \fib ^*/\fibe ^*} \\ \nonumber
&\fibe=\fibessm  \, 
\frac{1+\fib ^*/\fibe ^*}{1+ \mbox{$s_{17}$} \fib ^*/\fibe ^*}
\end{eqnarray}
We remind that the ppIII termination is very disfavored with respect
to the ppII,
so that, to a very
good approximation:
\begin{eqnarray}
&\fib=\mbox{$s_{17}$} \, \fib ^* \\ \nonumber
& \fibe=\fibe ^*
\end{eqnarray}
These analytical estimates are (obviously) confirmed by numerical 
experiments, see Table~\ref{tabalfas17}. Note that the slight 
dependences of $\fipp$ and $\ficno$ are 
not significant, being at the level of the numerical accuracy.

{\bf In conclusion, only the flux of \oB neutrinos is significantly affected
when $S_{17}$ is changed, the dependence being linear to a 
very good approximation.
Clearly, playing with $S_{17}$ cannot be the solution of the solar neutrino
puzzle, as the production of intermediate energy neutrinos is unchanged.}

\subsection{The p+$^{14}$N $\rightarrow ^{15}$O +$\gamma$ reaction }
\label{subsecn}

We will briefly review the status of this reaction 
since it is the slowest process
in the main CN cycle, and thus it controls the production of CNO neutrinos.

The Gamow peak in the solar center corresponds to $E_o = 27$ keV, whereas
laboratory measurements have been performed  at energy in excess of 100 keV, 
so that extrapolations are necessary.

The data of earlier investigations
\cite{Wood49,Dun51,Lam57,Pix57,Bai63,Hen67}
were extrapolated by various groups, yielding different values
for $S_{1,14}(0)$ ranging from about 2 to 10 keV b \cite{Rolfs}. This
discrepancy illustrates the difficulty in determining absolute
cross sections.
Fowler \etal~in 1975 \cite{CFZ75} recommended $S_{1,14}(0)=3.32$ keV b, 
by means of
some unspecified, judicious treatment of experimental data and/or theoretical
input. Schr\"{o}der \etal~\cite{Sch87} performed a comprehensive
experimental study, the only one covering continuously the energy range
$E= 200 - 3600$ keV and measuring absolute cross sections, $\gamma$-ray angular
distribution and excitation functions. This work removed most
of the appearent discrepancies among previous experiments. The extrapolated 
result, $S_{1,14}(0)=3.20 \pm 0.54$ keV b, is ``{\em {essentially identical
with the previously recommended value of 3.32 keV b}}'' \cite{PA91,Sch87}.
In his latest compilation Fowler \cite{CF88} adopted the value found
by  Schr\"{o}der \etal~On the other hand, Bahcall and Pinsonneault \cite{BP92}
quoted:
\begin{equation}
S_{1,14}(0)= 3.32 \pm 0.40 \quad {\mbox{keV b}}
\end{equation}
and later reduced the central value by 1\%, after including the effect
of vacuum polarization \cite{BP95}.

Concerning solar neutrinos, the CNO  neutrino flux scales linearly with
$s_{1,14}$:
\begin{equation}
\ficno= \ficno ^* \,  s_{1,14}
\end{equation}
In order to keep the same luminosity, also pp and pep neutrino fluxes 
are slightly changed:
\begin{equation}
\fip=\fip ^* \, s_{1,14}^{-0.02} \quad ,
\end{equation}
all other components being essentially 
unsensitive to $s_{1,14}$.

 \subsection{Plasma screening of nuclear charges}
 
\subsubsection{The results of different models}
\label{sec5screen}
 
The burning rates  of bare nuclei derived from experiments need to be corrected 
 to take into account
  the screening  provided by the 
 stellar plasma.
 The study of screened nuclear 
reaction rates was started with the pioneer work of Salpeter \cite{salp}; 
it  has been addressed by several authors, 
see \eg~\cite{svh,gdgc,koonin,mitler}, 
and recently reviewed in \cite{dzitko,schermonoi}. In the sun 
the screening effects are small, however different calculations 
yield relatively different  nuclear reaction rates.

As a starting point let us
neglect  any screening effect, \ie~reactions 
 take place for bare  ions  with  rates $\lambda_{NOS}$. The results of the 
corresponding solar model are shown in Table \ref{solischermo} (NOS).
Due to the screening, the actual rates $\lambda$ will be 
larger :
\begin{equation} 
\lambda=\lambda_{NOS} f
\end{equation} 
where the enhancement factors $f$ depend on the reaction and on the 
plasma properties. Various approaches have been developed for evaluating
these factors.

In the weak-screening approximation (WES), originally introduced by Salpeter 
\cite{salp}, one has for a  Debye plasma wherein partial electron 
degeneracy is included:
\begin{equation}
\label{fweak} 
{\mbox{ln}} f^{WES} = Z_1 Z_2 e^{2}  / (R_D \, kT)
\end{equation}
where  $Z_{1,2}$ are the charges of the reacting nuclei, T is the 
temperature and  
$R_D$ is the Debye radius, see ref. \cite{Clayton}.
In this scheme, the reacting particles are assumed to move slowly in 
comparison to the plasma particles (adiabatic approximation). Also 
modifications of the Coulomb potential are assumed to be sufficiently 
weak so that the 
linear approximation holds  (see \cite{koonin2} for a more extensive discussion 
about the validity of this approach).
In the sun, the weak-screening approximation is
justified 
(to some extent) for the $pp$-reaction, whereas the other nuclear 
reactions  occur in the so-called intermediate screening regime.

 Graboske \etal~\cite{gdgc} (GDGC) used  Eq.~(\ref{fweak}) when 
 $f^{WES}< 1.1$; for 
larger values (up to 2) they derived the enhancement factors by using general
thermodynamic arguments  and interpolation of  
Monte Carlo calculations. The explicit expressions can be found
in ref. \cite{gdgc}.

 Mitler \cite{mitler} (MIT) developed an analytical method  
which goes beyond the linearized approach and which correctly reproduces 
both the limits of weak and strong screening. Neglecting 
the small effects of a
radial dependence in the effective potential, see \cite{dzitko}, the 
enhancement 
factors are given then by: 
\begin{equation}
{\mbox{ln}} f^{MIT}=-\frac{8}{5} (\pi e n_e)^2 R_D^5 \, 
[(\zeta_1+\zeta_2 +1)^{5/3} - (\zeta_{1} +1)^{5/3} - (\zeta_{2}+1)^{5/3} ] 
/ (k T)
\end{equation}
where $\zeta_{1,2}=3 Z_{1,2}/(4\pi n_e R_D^3)$ and $n_e$ is 
the average electron density.

In the Salpeter approach one assumes  the reacting nuclei
to be so slow that the plasma can fully rearrange itself while the
nuclei are moving (as in the Born-Oppenheimer approximation used in
molecular physics) and dynamical effects (corrections to the 
Born-Oppenheimer approximation) are completely neglected.
 Carraro \etal~\cite{koonin}  (CSK) observed that  the reacting nuclei 
actually  move 
faster than most of the plasma ions (the Gamow peak energy is generally larger 
than thermal energy), so that ionic screening plays  a smaller r\^{o}le under 
this condition. They calculated the 
dynamic response of the plasma in the framework of the
linearized theory. The resulting enhancement factors are expressed in 
terms of 
those of the weak screening: 
\begin{equation}
{\mbox{ln}} f^{CSK}= C\quad {\mbox{ln}} f^{WES}  \quad .
\end{equation}
The coefficients $C$ essentially specify the corrections to 
the adiabatic limit, which obviously corresponds to $C=1$.
At the center of the sun the correction factors are \cite{koonin}: 
$C_{p+p}=0.76$, 
 $ C_{^3He+^3He}=0.75$,    
$ C_{^3He+^4He}=0.76$,  $ C_{p+^7Be}=0.80$,
 $ C_{p+^{14}N}=0.82 $.
 Note that the isotopic dependence is rather weak (in all previous models
 there was no isotopic dependence).

We shall not discuss the strong screening limit  
which is  definitely  too far from solar conditions (see 
\eg~\cite{Clayton}).

From Table \ref{solischermo}, where we report the results 
of solar models corresponding to the different approaches,
 one notes the following features:

i)
 The largest differences arise between the no screening  model and the 
 weak screening  model.  
$\fib$ can vary   by at most 15\%, the 
Chlorine  
signal is stable  to within 13\% and the Gallium signal at the level of 3\%. 

ii)
The GDGC model,  extensively used in stellar evolution codes, 
yields values very close to the no screening model; the difference between 
the two is
at the level of 1\% for $\fibe$ and $\fib$, as well as 
for the Chlorine and Gallium signals.

In Fig. \ref{figffactors}  we show the enhancement factors  along 
the solar profile, calculated by using 
the different prescriptions outlined above, for the  reactions relevant to 
hydrogen burning in the sun; as concerns 
the CN cycle, we pay attention only to the slowest reaction:  
$p+{} ^{14}${N} $\rightarrow ^{15}$O+$\gamma$.
All the enhancement factors depend very weakly 
on the mass coordinate, at least in the energy production region 
(M/M$_{\odot} <0.3$). This is clear in the weak screening 
regime, since the dependence on the solar structure parameters
 is just of 
the form $\rho/T^3$
(see Eq. (\ref{fweak}) and remind $R_D \propto \sqrt{T/\rho}$)
 and this quantity is 
approximately constant along the solar profile. The same holds in the 
strong  regime \cite{Clayton}, and thus 
 the approximate constancy in the 
intermediate regime is not a surprise. For these reasons, in the 
following we shall 
 concentrate on the enhancement factors calculated at the 
solar center (see Table \ref{ffactors}).

The weak-screening approximation, 
Eq.~(\ref{fweak}), always yields the largest enhancement factors, 
as it is physically clear since  electrons and ions are assumed 
to be free and capable of following 
the reacting  nuclei and in addition the electron cloud is allowed to strongly 
condense around the nuclei (in the linear approximation the electron density 
becomes infinite at  the nuclear site).
By using the Mitler model, 
where electron density at the nuclear site is fixed at 
$n_e$, one obtains smaller enhancement factors. The same holds 
for the model where the limited mobility of ions and thus
their partial screening capability is taken into account.
The GDGC  enhancement factors 
are systematically smaller than the others 
(except for the $pp$ reactions where, by definition, they are equal to 
the  weak screening prescription). It is thus clear that the 
corresponding neutrino fluxes and 
experimental signals  are the closest ones to those of 
the no-screening models. 
One notes that the enhancement factors for He+He and p+$^7$Be are very 
close: actually, in the weak screening approximation only the product of 
nuclear charges enters.

All in all, the enhancement factors are relatively close to unity, however 
none of the approaches to screening discussed above is completely 
satisfactory.
 The weak screening approximation is not justified for reactions other than
  the $pp$, since   $Z_1 Z_2 e^2 /( R_D \, kT)$ is not  small.
   The GDGC results stems from an interpolation of numerical computations 
and  the prescription of the authors yields an unphysical discontinuity at 
the border between the weak and intermediate regimes
\cite{dzitko}. The CSK result,
which incorporates  dynamic effects of finite nuclear velocity, is 
however  derived in the framework of a linear theory, \ie~the weak screening 
approximation.
 The Mitler approach goes beyond the weak screening approximation;
  on the other hand, the partial mobility of ions due to ion-interaction effects
   and/or to the finite thermal velocity is not taken into
    account. Also, the value of the  electron density at the nucleus
     is somehow artificially kept equal to the average electron density 
     $n_e$.

\subsubsection{A model independent analysis}
\label{sec5model}

The r\^{o}le of  screening  
on solar neutrinos can be investigated,  more generally, in a model 
independent way \cite{schermonoi}.
In the previous section we saw that a few features are common to any
screening model:

i) 
the enhancement factors $f$ can be taken constant in the energy production 
region, so that one needs to specify  the values at the solar center only.

ii) 
They are almost insensitive to isotopic effects, \ie~approximately
\begin{equation}
\ftt=\ftq
\end{equation}
and we will refer generically to
an enhancement factor for helium-helium reactions, $\fhh$.

iii) 
To a good approximation, the enhancement factor is determined by the 
product of the electric charges of the reacting nuclei, so that we can take:
\begin{equation}
\fpbe=\fhh
\end{equation}
 
In this case one is left with just three numbers,
$\fpp$, $\fhh$ and  $\fpn$,
{\bf which we shall consider as free parameters}. 
We recall that $f\geq 1$ since in the plasma  the Coulomb repulsion 
between the reacting nuclei is diminished.

The introduction of (spatially constant) enhancement factors is equivalent
to an overall change of the astrophysical S-factors:
\begin{equation}
S_{ij}\to S_{ij} \quad f_{i+j}
\end{equation}
and we can exploit the results for the variations of the astrophysical 
S-factors presented previously.

Concerning the r\^{o}le of $\fpp$,  we recall that
 an increase of $S_{pp}$ immediately implies a reduction of the 
central 
temperature $T_c$. From Table \ref{talpp}: 
\begin{equation}
T_c/T_c^{NOS}=(\fpp)^{-1/9}
\end{equation}
where  the superscript NOS refers to the 
no-screening solar model. 
From the same table, by using the dependence of neutrino fluxes
on \spp, one immediately gets the effect of the screening factor: 
\begin{equation}
\fibe=\fibe^{NOS} 
\, (\fpp)^{-1.1} 
\end{equation}
\begin{equation}
\fib=\fib^{NOS}
\, (\fpp)^{-2.7}
\end{equation} 
\begin{equation}
\ficno=\ficno^{NOS} 
 \, (\fpp)^{-2.2} \quad.
\end{equation}

Concerning the \treHe+\treHe and \treHe+\qHe 
reactions, we recall  from the foregoing section that the 
\sBe equilibrium concentration scales as: 
\begin{equation}
n_7 \propto S_{34}/ 
\sqrt{S_{33}}
\propto \sqrt{\fhh}
\end{equation}
This is clearly the dependence of \sBe neutrino flux; for $\fib$
one has to remind that it is proportional to $S_{17}$ and thus an extra
power of $\fhh$ occurs:
\begin{equation}
\fibe=\fibe^{NOS}\quad 
(\fhh)^{1/2} 
\end{equation}
\begin{equation}
\fib=\fib^{NOS}
\quad (\fhh)^{3/2} 
\end{equation}
The enhancement factor for p+$^{14}$N only matters for the CN cycle:
\begin{equation}
\ficno=\ficno^{NOS} \quad \fpn
\end{equation}
We can put together all the previous results in the following way:
\begin{equation}
\fibe=\fibe^{NOS} \quad (\fhh)^{1/2} \quad (\fpp)^{-1.1}
\end{equation}
\begin{equation}
\fib=\fib^{NOS} \quad (\fhh)^{3/2} \quad (\fpp)^{-2.7} 
\end{equation}
\begin{equation}
\ficno=\ficno^{NOS} \quad \fpn \quad (\fpp)^{-2.2} \quad .
\end{equation}

The behaviour of pp-neutrinos, as usual, can be best derived best by using the 
conservation 
of luminosity.

By using the above equations with the enhancement factors given in 
Table \ref{ffactors}, one can quantitatively reproduce, to a large extent, the 
numerical results  presented in Table \ref{solischermo}.
Furthermore,  we have verified that these analytical results are quite accurate
 for a wide range  of the $f$ factors, by using our stellar evolution code.

Note also that although \treHe+\treHe and \treHe+\qHe 
have the same enhancement factor, the equilibrium concentration of ${}^7$Be, 
and thus $\fibe$ and $\fib$, are changed when screening is introduced.
\newpage
%%%%%%%%%%%%%%%%%%SECTION SIX  %%%%%%%%%%%%
%
  \section{Non-standard solar models and experimental results}
  \label{cap6}
  \subsection{Introduction}
    
  We compare now the predictions of non-standard solar models with the 
experimental
  results on solar neutrinos. The basic questions are:
  
  \begin{itemize}
  
  \item
  Is there a solar model that accounts for all available results, 
    {\bf assuming standard neutrinos}?
  
  \item
  What would change if one of the experiments were wrong?
  
  \item
  If --- as it happens --- no model is successful, what is the reason of
    the failure?
   
  \end{itemize}
  
  For this purpose, we consider solar models characterized by three  
    parameters:

   (1) the central solar temperature, $T_c$, which accounts for (most of) 
       the effect of changing the astrophysical factor $S_{pp}$ for the
       p+p$\to ^2\mbox{H}$+e$^{+}$+$\nu_e$  cross section, the solar 
       opacity or the age of the sun; neutrino fluxes are determined
       essentially by $T_c$, independently of the way in which that particular
       temperature is achieved (see Fig.~\ref{fig7} and
       Refs.~\cite{PRD,Hata3,Hata94b});

   (2) the astrophysical factor $S_{33}$, which can be used 
       as an effective parameter controlling both cross sections for the 
       He+He reactions: by varying $S_{33}$, at fixed $S_{34}$, one can tune
       the parameter $\chi= S_{34}/\sqrt{S_{33}}$, which determines the neutrino
       fluxes when the cross sections for the $^4$He + $^3$He and/or
       $^3$He + $^3$He channels are altered (see Sec.~\ref{sec5neutrini});

   (3) the  astrophysical factor $S_{17}$ for the 
       p$ + {}^7\mbox{Be}\to {}^8\mbox{B} + \gamma$ reaction.
  
  These three parameters are really independent of each other; we recall 
  that $T_c$ is essentially unaffected by variations of
  $S_{33}$ and/or $S_{17}$.
  
  Possible variations of screening factors will not be discussed 
  extensively since, as shown in the previous section, they can be rephrased
  in terms of variations of astrophysical S-factors (see however
  Ref.~\cite{schermonoi}).

  The dependence of the neutrino fluxes on the three parameters is shown
  in Fig.~\ref{dependence}.
  For $T_c$ we use the exponents corresponding to (large) variations 
  of $S_{pp}$ (see Table~\ref{talpp}). For the dependence on the cross sections
  of the two He + He reactions, we explicitly consider the case of a 
  zero-energy resonance in the $^3$He + $^3$He channel, as given by 
  Eqs.~(\ref{fibetesi}) and (\ref{fibotesi}) and use the resonant contribution 
  $S_{33}^{res}$ as a free parameter.
  This possibility has the best chance of producing agreement with the data,
  since 
  it suppresses more strongly $\fibe$ than  $\fib$: a nonresonant 
  variation is even less effective~\cite{Fowres,AA,PRD}. 
  The pp-neutrino flux, $\fipp$,
  is determined by imposing the luminosity constraint, see Eq.~(\ref{alfa14}).
  Note that $S_{17}$ affects only the $^8$B neutrino flux significantly.
  As reference fluxes $\Phi_i^{RSM}$, we use the ones of the reference
  standard model (RSM) of Bahcall and Pinsonneault (``best model with
  diffusion''), which are given in Table~\ref{Modellibis} (BP95).
  
  We remark that the precise values of the exponents and coefficients
  in Fig.~\ref{dependence} are not  important for our
  discussion; somewhat different exponents could origin, if
  the change in $T_c$ was induced by means other than changing
  the astrophysical factor $S_{pp}$, or if these power laws were fitted 
  on different ranges of parameters.  Such different choices 
  do not affect the essence of our conclusions.
  
  In the following sections, we first vary one single parameter at a time,
  and then  analyze all possible combined variations. In
  addition, we repeat the analysis after having arbitrarily discarded
  any one of the experiments and, finally, we discuss the reasons of the 
  failure of all our attempts. This section updates and extends the analyses
  of Refs.~\cite{PRD,Future,Degli94,Bere94b,Bere96}.
  
\subsection{Fitting all experimental results}
  
    Let us consider the ensemble of solar models originating fluxes 
  $\Phi_i$ as parameterized in Fig.~\ref{dependence}. We performed a 
  $\chi^2$ analysis to establish quantitatively how well these models 
  compare with  experimental data. A given solar model predicts
  theoretical signals $S^{\mbox{\footnotesize th}}_X$ 
  ($X=$  Gallium, Chlorine and KAMIOKANDE) according to the formula
  \begin{equation}
  S^{\mbox{\footnotesize th}}_X = \sum_{i} \sigma_{X,i} \,\Phi_i \quad ,
  \end{equation}
  with the averaged neutrino cross sections $\sigma_{X,i}$ given in 
  Table~\ref{SIGMA}. These theoretical signals are to be compared with
  the experimental signals $S^{\mbox{\footnotesize ex}}_X$ reported in 
  Table \ref{EXPE}. 
  GALLEX and SAGE results have been combined into a single Gallium signal
  as in Eq.~(\ref{Sga}).
  As usual, we define the likelihood function as:
  \begin{equation}
  \chi^2 (T_c,S_{33}^{res}, S_{17}) =\sum_{XY} ( 
  S^{\mbox{\footnotesize ex}}_X -
                          S^{\mbox{\footnotesize th}}_X ) V^{-1}_{XY}
                        ( S^{\mbox{\footnotesize ex}}_Y -
                          S^{\mbox{\footnotesize th}}_Y ) \quad .
  \end{equation}
  The covariance matrix $V_{XY}$ takes into account both  
  experimental and  theoretical uncertainties. These latter
  always include the errors on the averaged neutrino cross sections 
  $\sigma_{X,i}$. Theoretical uncertainties on neutrino fluxes reflect
  uncertainties on $T_c$, $S_{33}^{res}$ and $S_{17}$. When some of these 
  variables are used as free parameters only the uncertainties corresponding 
  to the remaining ones are included. The propagation of these latter 
  uncertainties to the fluxes is established by means of power laws similar 
  to those of Fig.~\ref{dependence}, optimized for the case of small variations.
  The matrix $V_{XY}$ is not diagonal because the same parameter can affect more
  than one flux and the same flux contributes  in general to more than 
  one signal.  
  The use of the full covariance matrix is really necessary, since otherwise 
  apparently good fits can be achieved in an unphysical 
  way~\cite{PRD,Hata94b,lisierr,Gates}.
  Error correlation means, for instance, that
  we cannot use the uncertainty in  $\fib$ to strongly reduce 
  its contribution to the Davis experiment, while  having at the same time a 
  smaller reduction in the KAMIOKANDE experiment. More details on the
  calculation of the covariance matrix can be found in Ref.~\cite{PRD}.
  
  We explored a wide region for the 
  three parameters: $T_c$ down to 8\% of the RSM value, $S_{17}$ from
  zero up to five times the RSM value and $S_{33}^{res}$ from zero up
  to 200 times the value of $S_{33}^{RSM}$. With  three experimental
  results, one has two degrees of freedom (d.o.f.) when the three 
  parameters are varied one at the time, and no d.o.f. when all three
  are varied simultaneously. 
  The results of this analysis are summarized in Table~\ref{tbl1}
  which needs the following comments.

   (1) None of the attempts succeeds in giving an acceptable $\chi^2$ 
       even when we use as many parameters as the experimental data.

   (2) The smallest $\chi^2$, obviously obtained when all parameters are left
        free, corresponds anyhow to quite unphysical  variations: 
        $T_c$ comes out 7\% smaller than the RSM value and the $S_{17}$ 
       astrophysical factor is 5 times larger.

   (3) All the best fits give too high signals for the Gallium and Chlorine 
       experiments and too low a  $^8$B flux as compared to KAMIOKANDE; in 
       other words, {\bf the KAMIOKANDE signal looks  too high for
       an astrophysical solution}.

   (4) When varying  only one parameter at a time, a resonant increase of the 
       \treHe + \treHe cross section is the most effective way, as it allows 
       a higher suppression of \sBe relative to \oB neutrino flux. The central
       temperature is less effective, since it gives a too large suppression of 
       the $^8$B neutrino flux.  $S_{17}$ only affects the $^8$B neutrino
       flux and it is totally useless for the present solar neutrino
       problem,{\bf  which is mostly a problem of the intermediate energy 
       neutrinos}.

   (5) A significantly reduced \chiq\ is obtained by changing $S_{17}$
       together with $T_c$ or $S_{33}$. Note, however, that the smallest
       $\chi^2$ is obtained at the border of the parameter space, 
       \ie~in an extremely unphysical region. On the other hand, there is 
       no 
       significant gain when varying at the same time $T_c$ and $S_{33}^{res}$ 
       as they both mainly shift the balance towards the ppI chain.
  
  \subsection{What if one experiment were wrong?}
  
     There is no rational reason for doubting any of the
  experimental data.
   Nevertheless in the same spirit as discussed in Sec.~\ref{cap2},
   we have repeated the analysis of the previous section
  excluding in turn the KAMIOKANDE, Chlorine and Gallium result. 
  (Note  that, excluding the Gallium result, we are giving up two experiments,
  GALLEX and SAGE.)
  
    The results of this exercise, reported in Table~\ref{tbl2}, show that the 
  situation is essentially unchanged:

   (1) Again none of the attempts succeeds in giving an acceptable \chiq, 
       even when we use more parameters than experimental data.

   (2) As previously, the best fit corresponds anyhow to extremely unphysical 
       values of the input parameters.

   (3) The smallest $\chi^2$ are obtained disregarding either the KAMIOKANDE or
      Gallium data. In other words, these two results taken together are
    the hardest to reconcile with astrophysics. {\bf The point is that Gallium 
      result implies that only the ppI chain is effective, whereas KAMIOKANDE
      shows that ppIII,  and hence ppII, are operational}.
   
  \subsection{What went wrong with solar models?}
     Tables~\ref{tbl1} and \ref{tbl2} present the quantitative evidence that
   any solar model (which we are able to parametrize) cannot account for the 
  experimental data, even when one of these data is arbitrarily removed. 
  In this section we discuss the physical motivations of such failure,
  giving through Figs.~\ref{fig2}, \ref{fig3} and \ref{fig4} a graphical 
  illustration of the results of Tables~\ref{tbl1} and \ref{tbl2}.
  
     The starting point is that any combination of two experimental data
  requires, with respect to the RSM, a strong suppression of the 
  intermediate-energy--neutrino flux ($^7$Be and CNO), much stronger 
  than that of the $^8$B flux, as was extensively discussed in Sec.~\ref{cap2}.
  
  In the ($\fib$, $\fibecno$) plane, assuming  standard neutrinos, 
  the regions allowed at the $2\sigma$ level by each experiment and their
  intersection are indicated in Fig.~\ref{fig2}.
  
  The aim of non-standard solar models is to push the predictions close to 
  this area allowed by  experiments.
  
   In Fig.~\ref{fig2} we show the effect of changing  one of the 
  three parameters in turn. A reduction of $S_{17}$ only affects the
  $^8$B flux and we are left with a much too high $\fibecno$. Reducing 
  the temperature decreases both $\fib$ and $\fibe$, however $\fib$ is 
  suppressed more strongly than $\fibe$ contrary to experimental evidence. 
  The resonant increase of the \treHe + \treHe cross section succeeds in 
  reducing $\fibe$  almost to zero while leaving part of $\fib$; however,
  $\fibecno$ is still too high since $\ficno=\ficno^{RSM}$ is by itself 
  sufficient to spoil the agreement with the data.
  
     In Fig.~\ref{fig3} we show the effect of changing two parameters at
  the same time. The best case is when $T_c$ and $S_{17}$
  are changed. In fact, only $T_c$ is able to cut both $\fibe$
  and $\ficno$, and $S_{17}$ helps to bring $\fib$ back up.
   We stopped at $T_c=0.88$ and $S_{17}=5S_{17}^{RSM}$  (which are already
  completely unphysical values!) and yet  have not reached the allowed region 
  (which of course could be attained by even smaller $T_c$ together with 
  larger $S_{17}$).
  
      Finally, we show the effect of changing all three parameters in
  Fig.~\ref{fig4}. This case does not differ from the previous one; the
  main difference is that since we use $S_{33}^{res}$ to cut $\fibe$ and
  use $T_c$ mostly to cut $\ficno$ we can reach even a lower value of
  $\fibecno$ with a relatively higher temperature. However, within the
  framework of present physical knowledge, one cannot find a justification
  to accept a non-standard solar model with the required decrease of
  temperature (7\%).
  
  For the same reasons, playing with screening factors does not help;
  see Fig.~\ref{figschermo} where the three screening factors $\fpp$,
  $\fhh$ and $\fpn$ (see section~\ref{sec5model}) are arbitrarily
  varied by huge amounts with respect to the RSM estimates. Again the
  suppression of intermediate energy neutrinos is too weak in comparison
  with experimental results.
  
  All in all, the fits we found are not acceptable,
  \ie~the chances of an astrophysical solution look very weak even if
  we are extremely generous on the region spanned by the
  physical inputs.
  In Sec.~\ref{cap2} we found that the probability of standard neutrinos is 
  at most 2\% when all fluxes are left as absolutely free parameters.
  It is thus not a surprise to reach an even stronger conclusion when
  some astrophysical input is used.
\newpage
%%%%%%%%CONCLUSION
%
\section{The never  ending  story?}

The sun or the neutrino, who is  at fault? To express our opinion on such 
a matter, let us summarize the main points of this review. \\

{\centerline{\bf Assuming  standard neutrinos:}}
\begin{itemize}
\item
the available experimental results appear mutually inconsistent, 
even if some of the 
four experiments was wrong. 
 This is a solar-model independent evidence for non-standard neutrinos.
 This evidence, however, is not an  overwhelming one, since there is still a
 few percent (at most) probability of neutrinos being standard, \ie~we
  have a ``2-$\sigma$ level effect'' (see Sec.~\ref{cap2}).
We remark that this conclusion is reached at the cost of giving up all our 
understanding of stellar physics.
\item
Even neglecting these inconsistencies, the flux of intermediate
energy neutrinos (\sBe + CNO), as derived from experiments, is 
significantly smaller than the predictions of SSMs. The main 
puzzle is with $^{7}$Be neutrinos, for which the theoretical predictions 
are really robust. If one insists on standard neutrinos, one has definitely 
to abandon Standard Solar Models, all 
predicting  a much too high $^{7}$Be neutrino flux.
\item
In addition, the suppression of $^{7}$Be neutrinos is much stronger 
than that of $^{8}$B neutrinos. Non-standard solar models fail to 
account for {\bf both} these reductions.
\item
Even extremely non-standard models cannot account for  
data, even 
when one of the experimental data is arbitrarily omitted. All the fits 
that we attempted 
are unacceptable (see Sec.~\ref{cap6}), \ie~the chances of a nuclear or 
astrophysical solution look to us very weak, although we have been
extremely generous in the region spanned by the physical inputs. 
Since we have found that the probability of standard neutrinos is about 2\% or 
less when all fluxes are left as free parameters, it is not a surprise
to reach an even stronger limit when some astrophysical input is used.

\end{itemize}

Let us  wait (and hope) for  future experiments, bringing 
{\bf direct},  decisive evidence of 
some non-standard neutrino property. 
%
%%%%%%%%%%%%%%%%%%
%
\acknowledgments

We are grateful to J.N. Bahcall, V. Berezinsky, S. D'Angelo,
S. Gershtein, E. Lisi and  C. Rolfs for providing us with extremely
 useful comments and suggestions.

We wish to express our deep gratitude to V. Telegdi for a painfully critical 
 reading of the draft text.

The analysis of \treHe+\treHe and \treHe+\qHe data was performed
with the NACRE ({\underline N}uclear {\underline A}strophysics
{\underline C}ompilation of {\underline {Re}}action rates),
collaboration, supported by the European Union through the Human
Capital and Mobility programme. We are grateful to the NACRE colleagues for
the scientific collaboration and to the European Union for financial
support.

This work was supported by MURST, the Italian Ministry for University and 
Research.

\newpage
%%%%%%%%%%%%%%%% APPENDIX A  %%%%%%%%%%
\appendix

\section{Our standard solar model }

~\\
 We present the  solar 
 model
 resulting from an updated version of FRANEC\footnote{Based on 
 work of Ciacio \etal~\cite{Ciacio}.}
 (Frascati Raphson Newton Evolutionary Code),
 where helium and heavy elements diffusion is included and the
 OPAL equation of state (EOS) \cite{refEOS,RSI96} 
 is used. The EOS is consistent with the adopted opacity tables,
 \ie~the most recent evaluation from the same Livermore group \cite{RI95}
 and this should further enhance the reliability of the model.
 In addition, updated values of the relevant nuclear cross sections are
 used and more refined values of the solar constant
 and age are adopted.

We discuss the effects of each of these improvements, showing that
they are essential in order to get agreement with the
helioseismological information about the bottom of the convective zone. 
We  also calculate neutrino fluxes and
the expected signals in ongoing experiments.

FRANEC has been described in previous papers 
(\eg~see Ref.~\cite{Cieffi89,Castel92}). Recent determinations of the
the solar luminosity (L$_\odot =3.844 \cdot 10^{33}$ erg/s)
and of the solar age (t$_\odot =4.57 \cdot 10^9$ yr)
are used ~\cite{BP95}. The present ratio 
of the solar metallicity to solar hydrogen abundance by mass 
corresponds to  the most recent value of Ref.~\cite{GN93}: 
(Z/X)$_{photo}$~=~0.0245.

Following the standard procedure,
for each set of assumed physical inputs, the initial Y, Z and  the
mixing length parameter $\alpha$ were varied until the radius,
luminosity and (Z/X)$_{photo}$ at the solar age matched the observed values
within a tenth of percent or better. 
We considered the following steps:

 a) As a starting point we used the Straniero equation 
of state~\cite{Oscar88}, the OPAL opacity tables 
which were available in 1993~\cite{RG92,IGL92} for the
solar metallicity ratio of Ref.~\cite{GN93},
combined with the molecular opacities of Ref.~\cite{Alexander}; diffusion was 
ignored. This model may be useful for a comparison with
the Bahcall and Pinsonneault model without diffusion, described
in ref.~\cite{BP95},
as the chemical composition is the same,
although it  is not the most updated one.

b) Next, we introduced the OPAL equation of state. With respect to 
other commonly
used EOS, this one avoids an ad hoc treatment of the pressure
ionization and it provides a systematic expansion in the Coulomb
coupling parameter that includes various quantum effects generally
not included in other computations (see refs.~\cite{refEOS} and 
\cite{RSI96} for more details).

c) 
We used the latest OPAL opacity tables~\cite{RI95}, solar
 metallicity ratio as in \cite{GN93} and again the molecular opacities
 of Ref.~\cite{Alexander} for temperature below 10$^4$ K.
 With respect to Ref.~\cite{RG92} the new OPAL tables include
the effects on the opacity of seven additional elements and some minor
physics changes; moreover the temperature 
grid has been made denser.

d) 
We included the diffusion of helium and heavy
elements. The diffusion coefficients have been calculated using the
subroutine developed by Thoul, see Ref.~\cite{Thoule}.
The diffusion equations were integrated numerically. At any time step,
after updating the physical and chemical quantities as usually done in FRANEC,
we took into account the effect of diffusion. Our time steps are about 
3 $\cdot$ 10 $^7$ yr.
The variations of the abundance of  H, He , C, N, O and Fe are followed;
all these elements are treated as fully ionized.
According to Ref.~\cite{Thoule}
all other elements are assumed
to diffuse at the same rate as the fully-ionized iron.
To account for the effect of heavy element diffusion 
on the opacity coefficients, we calculated the total heavy elements
abundance in each spherical shell in which the solar model
is divided and we interpolated (by a cubic spline interpolation)
between opacity tables with different total metallicity
(Z = 0.01, 0.02, 0.03, 0.04).

e) 
Finally, we investigated the effect of updating the nuclear cross sections 
for $^3$He+$^3$He and $^3$He+$^4$He reactions, following a recent 
new analysis of all available data, described in Sec.~\ref{cap5}. 
For $S_{34}$, we used the exponential parametrization of Eq. (\ref{expo34}).

 The resulting solar models are summarized in Table~\ref{modellinoi},
 and deserve the following comments.

(a $\rightarrow$ b):
   The introduction of the new OPAL EOS reduces appreciably
the initial helium abundance.
The Straniero (1988) EOS underestimates the
Coulomb effects neglecting the contribution due to the electrons,
 which are considered as completely degenerate, whereas the OPAL EOS includes
corrections for Coulomb forces which are correctly treated.
The models
with Straniero EOS have a higher central pressure and a higher central
temperature, and correspondingly a higher initial helium abundance. 
The effect of an underestimated Coulomb correction was discussed in 
Ref.~\cite{TCL}.
Note that 
the transition between radiative and convective
regions is not correctly predicted by the model, 
the convective region 
being definitely too shallow.

(b $\rightarrow$ c): 
The  updating of the radiative opacity coefficients has minor effects.
The convective zone is again too
shallow.

(c $\rightarrow$ d):
This step shows the effects of diffusion. Helium and heavy elements sink 
relative to hydrogen in the radiative interior of the star because of the
combined effect of gravitational settling and of thermal diffusion.
This increases the molecular weight in the core
and thus the central temperature raises.
 The surface abundances of hydrogen, helium and  heavy elements
are appreciably affected by diffusion. For example, the initial value
Y$_{in}$~=~0.269 is reduced to the present photospheric value 
Y$_{photo}$~=~0.238,
The predicted depth of the convective zone and the
sound speed are now in good agreement with helioseismological
values. It is natural that when diffusion is included
the radiative region reduces its size. With respect to models without
diffusion, in the external regions the present helium fraction is reduced
while the metal fraction stays at the observed photospheric value. Thus the
opacity increases and convection starts deeper in the sun.

(d $\rightarrow$ e):
The modifications of our solar model arising from the new values of 
the nuclear cross sections are negligible with respect to the other 
improvements just presented.

The predicted neutrino fluxes and signals are summarized in 
Table~\ref{flussinoi} for our models. All in all, the results 
are quite stable with respect to the changes we have introduced 
as long as diffusion is neglected. On the other hand, due to the higher central
temperature, model d) has significantly higher $^8$B and
CNO neutrino fluxes.
It is essentially the increase  of $\fib$  which enhances the  
predicted Chlorine signal. 
The slight change in the nuclear cross sections weakly
affect neutrino fluxes and signals. The $^7$Be and $^8$B fluxes are 
reduced by about 5\% 
as a consequence of the correspondingly smaller value of
$S_{34}$ (see Table~\ref{flussinoi}). Should we use the 
polynomial expansion of Eq.~(\ref{poli34}) one would get a further 5\% decrease.
\newpage

%%%%%%%%%%%%%%%%APPENDIX  B %%%%%%%
%
\section{The stability of pep and CNO neutrino fluxes} 
\label{appendicsi}

We already compared the predictions of Standard Solar Models by different 
authors in Sec.~\ref{cap1}, see also Table \ref{Modellibis}.
We discuss here the stability of the ratios $\csi=\fipep/(\fipp+\fipep)$
and $\eta=\fin/(\fin+\fio)$ among the non-standard solar
models which we have built in section \ref{cap4}.

Figs. \ref{figapp1} and \ref{figapp2} show the situations as
some input parameter ($S_{pp}$, Z/X, \ldots) is varied by a scaling 
factor $X/X^*$. In all the presented models, the central temperatures
differ by that of the starting solar model by no
more than 5\%. The following features are to be remarked:

i) variations of $\csi$ with respect to the starting  model prediction
do not exceed ten per cent;

ii) $\eta$ is stable to within 20\% level or better;

iii) as in all models we considered
the central temperature is smaller or equal to that of 
the starting solar model, the rate
of the key reaction p+$^{14}$N$\rightarrow \gamma +^{15}$O is reduced
and the CN chain is less equilibrated, resulting in a smaller 
$^{15}$O neutrino production, which accounts for $\eta > \eta ^*$
systematically.
 On the other hand the behaviour of $\csi$
looks erratical: it may increase as well as decrease when the
central temperature is reduced, presumably since $\csi$ is sensitive
to the electron density more than to the temperature.
\newpage

%
%%%%%%%%%%%%%%%%%%%  REFERENCES %%%%%%%%%%%%%

%
\newpage

%%%%%%%%%%%%% TABLES
%
%                          TABELLA 3
%
\begin{table}
%\footnotesize
%\centering
\caption[bb]{Comparison among several solar models, all calculated 
including diffusion
of helium and heavy elements. The ``best model
with diffusion'' of \cite{BP95} will be used as the Reference 
Solar  Model  (RSM)
in this paper. FRANEC96 indicates our best model with diffusion, model 
(e) of  Appendix A. The dag (\dag) indicates values of c$_b$ calculated 
by us, assuming fully ionized perfect gas EOS.
The last row indicates the consistency with helioseismology, see
section \ref{structure}.}
\begin{tabular}{lr@{}lr@{}lr@{}lr@{}lr@{}lr@{}lr@l}
%\hline
%\hline
 &\multicolumn{2}{c}{CGK89}
                          &\multicolumn{2}{c}{P94}
                                    &\multicolumn{2}{c}{DS96}
                                           &\multicolumn{2}{c}{RVCD96}
                                                  &\multicolumn{2}{c}{RSM}
                                               &\multicolumn{2}{c}{FRANEC96}\\
Ref. &\multicolumn{2}{c}{~\cite{COX}}
                           &\multicolumn{2}{c}{~\cite{P94}}
                                      &\multicolumn{2}{c}{~\cite{DS96}}
                                           &\multicolumn{2}{c}{~\cite{RCVD96}}
                                          &\multicolumn{2}{c}{~\cite{BP95}}
                                           & \multicolumn{2}{c}{} \\
\hline
t$_\odot$[Gyr]          &    4.&54              &  
4.&60   &    4.&57    &  4.&60    &  4.&57    & 4.&57\\
L$_\odot$[\unitaerg]    &    3.&828             &  
3.&846  &    3.&844  &  3.&851  &  3.&844   & 3.&844\\
R$_\odot$[10$^{10}$ cm] &    6.&9599            &  
6.&9599 &     6.&960  &  6.&959   &  6.&9599   & 6.&960\\
(Z/X)$_{photo}$         &    0.&02464            &  
0.&02694&     0.&02263&  0.&0263  &  0.&02446  & 0.&0245\\
\hline
X$_{in}$                &    0.&691             &  
0.&6984  &    0.&7295 &  0.&7012  & 0.&70247  & 0.&711\\
Y$_{in}$                &    0.&289             &  
0.&2803  &    0.&2509  &  0.&2793 &  0.&27753 & 0.&269\\
Z$_{in}$                &    0.&02              &  
0.&02127&    0.&01833  &  0.&0195 &  0.&02    & 0.&0198\\
\hline
X$_{photo}$             &    0.&7265            &  
0.&7290 &    0.&7512  &  0.&7226 &  0.&73507 & 0.&744\\
Y$_{photo}$             &    0.&2556            &  
0.&2514 &    0.&2308  &  0.&2584 &  0.&24695 & 0.&238\\
Z$_{photo}$             &    0.&0179            &  
0.&01964&    0.&0170 &  0.&0190 &  0.&01798 & 0.&0182\\
\hline
R$_b$/R$_\odot$           &    0.&721             &  
0.&7115 &    0.&7301  &  0.&716  &  0.&712   & 0.&716\\
$T_b$[10$^6$ K]         &    2.&142             &    &     &    
2.&105  &  2.&175  &  2.&204   & 2.&17\\
$c_b$[10$^7$ cm s$^{-1}$] &  2.&21$^{\mbox{\dag}}$       &    &     &    
2.&21$^{\mbox{\dag}}$  
 &  2.&22$^{\mbox{\dag}}$  &  2.&25$^{\mbox{\dag}}$   & 2.&22\\
\hline
$T_c$[10$^7$ K]         &    1.&573             &  
1.&581  &    1.&561   &  1.&567  &  1.&5843  & 1.&569\\
$\rho_c$[100 gr cm$^-3$]&    1.&633             &  
1.&559  &    1.&554   &  1.&545  &  1.&562   & 1.&518\\
%Y$_c$                   &    0.&6580           &  
%0.&6488 &      &      &  0.&646  &  0.&64564 & 0.&63\\
\hline   
Helioseismology          &      &No             &    &Yes  &      & 
No   &    &Yes  &    &Yes  &   &Yes  \\
%\hline
\end{tabular}
\label{Modelli}
\end{table}
\begin{table}
%\footnotesize
%\centering
\caption[bb]{Physical and chemical inputs of the solar models 
in Table \ref{Modelli}.
The correspondence between acronyms and references is as follows:
C\&S70=\cite{Coxsteward}, MHD~=~\cite{MHD}, CEFF~=~\cite{Egg73}
with the Coulomb correction added (see \cite{CDD}),
EFF~=~\cite{Egg73}, BP92~=~\cite{BP92}, G\&N93~=~\cite{GN93},
G91~=~\cite{G91}, F75~=~\cite{CFZ75}, C\&F88~=~\cite{CF88}. Note that
RVCD96 also includes rotational mixing.}
\begin{tabular}{c c c c c c c}
%\hline
%\hline
 &\multicolumn{1}{c}{CGK89}
                         &\multicolumn{1}{c}{P94}
                                    &\multicolumn{1}{c}{DS96}
                                             &\multicolumn{1}{c}{RVCD96}
                                                  &\multicolumn{1}{c}{RSM}
                                           &\multicolumn{1}{c}{FRANEC96}\\
Ref. &~\cite{COX}&~\cite{P94}&~\cite{DS}&~\cite{RCVD96}&~\cite{BP95}&\\
\hline
%\hline
 OPACITY &C\&S70& OPAL     &OPAL   &OPAL  & OPAL&   OPAL\\
EOS &MHD  & CEFF&DS96  &      MHD &     BP92&      OPAL\\
 MIXTURE & G\&N93     & G91  &    G\&N93&   G\&N93&  G\&N93    & G\&N93\\
 CROSS SECTIONS    &F75 & BP92 &DS96 &C\&F88 & 
Table \ref{cross}& Table \ref{cross}
%\hline
%\hline
\label{modelliinput}
\end{tabular}
\end{table}
\begin{table}
%\footnotesize
%\centering
\caption[bb]{Comparison among the neutrino fluxes of the solar models of 
Table~\ref{Modelli}. All of them, except for DS96, are SSMs 
according to our definition.}
\begin{tabular}{lr@{}lr@{}lr@{}lr@{}lr@{}lr@{}l}
%\hline
                               &\multicolumn{2}{c}{P94}
                                         &\multicolumn{2}{c}{DS96}
                                                 &\multicolumn{2}{c}{RVCD96}
                                                       &\multicolumn{2}{c}{RSM}
                                              &\multicolumn{2}{c}{FRANEC96} \\
Ref.
                              &\multicolumn{2}{c}{~\cite{P94}}
                                         &\multicolumn{2}{c}{~\cite{DS96}}
                                           &\multicolumn{2}{c}{~\cite{RCVD96}}
                                          &\multicolumn{2}{c}{~\cite{BP95}}
                                            & \multicolumn{2}{c}{}  \\
\hline
$\fipp$ [\unitanb]     & 59.&1   &    61.&0     &  59.&4    & 59.&1     & 
59.&92\\
$\fipep$[\unitanb]     &  0.&139 &     0.&143   &   0.&138  &  0.&140   &
0.&14\\
$\fibe$ [\unitanb]     &  5.&18  &     3.&71    &   4.&8    &  5.&15    &
4.&49 \\
$\fin $  ~[\unitanb]   &  0.&64  &     0.&382  &   0.&559  &  0.&618   &
0.&53\\
$\fio $  ~[\unitanb]   &  0.&557 &     0.&374  &   0.&481  &  0.&545   &
0.&45\\
$\fib$  ~[\unitasb]    &  6.&48  &     2.&49    &   6.&33   &  6.&62    &
5.&16\\
\hline
S$_{Ga}$[SNU]          &136.&9   &   115&     & 132.&77   &137.&0     &
 128\\
S$_{Cl}$[SNU]          &  9.&02  &     4.&1     &   8.&49   &  9.&3     &
7.&4\\
%\hline
%\hline
\end{tabular} 
\label{Modellibis}
\end{table}

\begin{table}
\caption[cc]{Helioseismological determinations of the present surface
He abundance, Y$_{photo}$. In RVCD96 uncertainties depending on the inversion
method and on the EOS are included.}
\begin{tabular}{ll}
%\hline
%\hline
  Reference         &   Y$_{photo}$  \\
\hline
Dappen (1988) \cite{Dappen88} &    Y=0.233 $\pm$ 0.003 \\
Dappen (1991) \cite{Dappen91}  &    Y=0.268 $\pm$ 0.002\\
Dziembowski (1991) \cite{DJ91}  &  Y=0.234 $\pm$ 0.005 \\
Vorontsov (1991) \cite{Vor91}   &  Y=0.250 $\pm$ 0.010 \\
Dziembowski (1994) \cite{DJ}  &   Y=0.24295 $\pm$ 0.0005\\
Hernandez (1994) \cite{hern}  &   Y=0.242 $\pm$ 0.003  \\
 RVCD96 \cite{RCVD96}            &   Y=0.250 $\pm$ 0.005     
%\hline
%\hline
\label{elios}
\end{tabular}
\end{table}
\begin{table}
%\centering
\caption[pesi]{For each component of the  neutrino flux, we show the 
average neutrino energy $\langle E_\nu\rangle$ and the averaged neutrino 
capture cross sections (10$^{-9}$~SNU~cm$^2$~s = 10$^{-45}$ cm$^2$)
for Chlorine  and Gallium, with errors at 1$\sigma$ level .
When averaging the pp and pep components to get p, we use the 
relative weights of 
the RSM; similarly for $^{13}$N and $^{15}$O to get CNO.
All data from \cite{Bahcall1989} but for $\sigma_{Cl,B}$ from \cite{sfiga}.
}
\begin{tabular}{lr@{}lr@{}l@{}lr@{}l@{}l}
%\hline
%\hline
         &\multicolumn{2}{c}{$\langle E_\nu \rangle_i$}
                    &\multicolumn{3}{c}{$\sigma_{Cl,i}$}
                                       &\multicolumn{3}{c}{$\sigma_{Ga,i}$} \\
         &\multicolumn{2}{c}{[MeV]}
                    &\multicolumn{3}{c}{[10$^{-9}$SNU cm$^2$s]}
                         &\multicolumn{3}{c}{[10$^{-9}$SNU cm$^2$s]} \\
\hline
pp       & 0. & 265 &    0. &    &               &    1. & 18 & $(1\pm 0.02)$\\
pep      & 1. & 442 &    1. & 6  & $(1\pm 0.02)$ &   21. & 5  & $(1\pm 0.07)$\\
p=pp+pep & 0. & 268 &    0. & 0038& $(1\pm 0.02)$ &    1. & 23 & $(1\pm 0.02)$\\
$^7$Be   & 0. & 814 &    0. & 24 & $(1\pm 0.02)$ &    7. & 32 & $(1\pm 0.03)$\\
$^{13}$N & 0. & 707 &    0. & 17 & $(1\pm 0.02)$ &    6. & 18 & $(1\pm 0.03)$\\
$^{15}$O & 0. & 996 &    0. & 68 & $(1\pm 0.02)$ &   11. & 6  & $(1\pm 0.06)$\\
CNO=$^{13}$N + $^{15}$O
         & 0. & 842 &    0. & 41 & $(1\pm 0.02)$ &    8. & 72 & $(1\pm 0.05)$\\
$^8$B    & 6. & 71  & 1110. &    & $(1\pm 0.03)$ & 2430. &    & $(1\pm 0.25)$\\
%\hline
%\hline
\end{tabular}
\label{SIGMA}
\end{table}
\begin{table}
%\centering
\caption[exper]{
 The main characteristic of each neutrino experiment:
 type, detection reaction, energy 
threshold $E_{th}$, experimental results with  statistical and
systematical errors. In the last column the Reference 
Solar Model   \cite{BP95} predictions 
are presented. Errors are at $1\sigma$ level.}
\begin{tabular}{ccccc}
%\hline
%\hline
%Experiment& type & $E_{th}^{a)}$& result $^{b)}$ &BP95$^{b)}$\\
Experiment& type & $E_{th}$%
\tablenote{Energy in Mev}
& result%
\tablenote{in SNU for radiochemical experiments; in 10$^6$ cm$^{-2}$s$^{-1}$ 
 for KAMIOKANDE.}
&BP95$^{b}$\\
\hline
 &&&&\\
Homestake & radiochemical & 0.814& 2.55 $\pm0.17\pm0.18$&9.3$^{+1.2}_{-1.4}$\\
          &$\nu+^{37}$Cl$\rightarrow$ e$^- + ^{37}$Ar&&&\\
           \hline
            &&&&\\
KAMIOKANDE& scattering& 7&$(2.73\pm0.17 \pm0.34)$&6.62$(1.00^{+0.14}_{-0.17})$\\
          &$\nu+ e^{-}\rightarrow \nu+ e^{-}$&&&\\ 
          \hline
           &&&&\\
GALLEX    & radiochemical &0.233&$77.1\pm8.5^{+4.4}_{-5.4}$&137$^{+8}_{-7}$\\
           &$\nu+^{71}$Ga$\rightarrow$ e$^- + ^{71}$Ge&&&\\ 
           \hline
            &&&&\\
SAGE      & radiochemical &0.233&$69\pm10^{+5}_{-7}$&137$^{+8}_{-7}$\\
           &$\nu+^{71}$Ga$\rightarrow$ e$^- + ^{71}$Ge&&&\\
%           \hline
%           \hline
% \multicolumn{5}{l}{\scriptsize $^{a)}$ Energy in Mev}\\         
%\multicolumn{5}{l}{\scriptsize $^{b)}$ in SNU for radiochemical exp.; 
%in 10$^6$ cm$^{-2}$s$^{-1}$ KAMIOKANDE.}\\          
\end{tabular}          
\label{EXPE}
\end{table}
\begin{table}%[htb]
%\centering
\caption[bb]{Contribution  from the main components of neutrino flux
to the signals (SNU) in $^{71}$Ga  and 
  $^{37}$Cl detectors according to  the RSM, 
from~\cite{BP95}.}
\begin{tabular}{lr@{}lr@{}lr@{}l}
%\hline
%\hline
         &\multicolumn{2}{c}{$^{71}$Ga} & \multicolumn{2}{c}{$^{37}$Cl}\\
\hline
pp       &  69.&7  &    0.&0 \\
pep      &   3.&0  &    0.&22 \\
$^7$Be   &  37.&7  &    1.&24 \\
$^{13}$N &   3.&8  &    0.&11 \\
$^{15}$O &   6.&3  &    0.&37  \\
$^8$B    &  16.&1  &    7.&36  \\
\hline
Total    & 136.&6  &    9.&30 \\
%\hline
%\hline                 
\end{tabular}
\label{contribution}
\end{table}
\begin{table}
\caption[bb]{
Experimental information on the fluxes of intermediate energy neutrinos
(units of $10^9$cm$^{-2}$s$^{-1}$). In (a) only the luminosity constraint 
is assumed. In (b) we assume also $\csi$ and 
$\eta$ as given by the RSM \cite{BP95}, see Sec.~\ref{bounds}. The 
best fit points
and upper limits, within 2$\sigma$ from each experimental results, are 
presented.}
\begin{tabular}{lcccccc}
%\hline
%\hline
&\multicolumn{4}{c}{(a)}& \multicolumn{2}{c}{(b)} \\
\cline{2-5} \cline{6-7}
&  $^{13}$N & $^7$Be  &  $^{15}$O & pep& $^7$Be+CNO & CNO \\ 
\hline
best fit & -2.2 & -1.6 & -0.8 & -0.2 & -2.2 & -1.7 \\
upper limit (2$\sigma$) &1.8 & 1.5 & 0.8 & 0.5 & 0.9 & 0.7 \\
%\hline
%\hline
\end{tabular}
\label{limiti}
\end{table}
\begin{table}
\caption[model]{Information on neutrino fluxes.
$\fibe$  and $\ficno$ ($\fib$) in units of $10^9$cm$^{-2}$ s$^{-1}$ 
($10^6$cm$^{-2}$ s$^{-1}$).
All bounds are at the 99.5\% C.L. Direct information (exp) is only available 
for \oB neutrinos, from Ref.~\cite{Kamioka}, 
the indicated error here corresponds to $3\sigma$.
 The bounds in (b1) correspond to
no prior knowledge on the unknown variables. In (b2) we assume {\em a
priori} $\Phi_i \geq 0$.
The results of SSMs are also shown (same notation
as in Table \ref{Modelli}).
}
\label{confidence}
\begin{tabular}{lccccccc}
%\hline
     &(exp)&(b1)&(b2)&P94&
%KS94&
RVCD96&RSM&FRANEC96\\
Ref.     &~\cite{Kamioka}& & &\cite{P94}&
%\cite{DS}&
~\cite{RCVD96}&\cite{BP95}& \\
\hline 
$\fib$&2.73$\pm$1.14&&&6.48&
%5.83&
6.33&6.62&5.16\\
$\fibecno$&&$\leq$0.7&$\leq$2.0&6.38&
%4.99&
5.9&6.31&5.47\\
$\fibe$&&$\leq$0.7&$\leq$2.0&5.18&
%4.91&
4.8&5.15&4.49\\
$\ficno$&&$\leq$0.5&$\leq$1.5&1.20&
%0.08&
6.33&1.16&0.98\\
%\hline
\end{tabular}
\end{table}
%
%\begin{table}
%\caption[cfactor]
%{Analytical expression and numerical values for the $C_i$ and $G_i$ 
%coefficients, see text.}
%\label{cgcoeff}
%\begin{tabular}{lcc}
%\hline
%\hline
%$C_0$     &  $\sclp \ksole /\qp$          & 0.246 \\
%$C_{Be}$  &  $\sclbe - \sclp \qbe/\qp$    & 0.236 \\
%$C_{CNO}$ &  $\sclcno - \sclp \qcno/\qp$  & 0.405 \\
%$C_B$     &  $\sclb - \sclp \qb/\qp$      & 1.110 \\
%\hline
%$G_0$     &  $\sgap \ksole /\qp$          & 80.01 \\
%$G_{Be}$  &  $\sgabe - \sgap \qbe/\qp$    & 6.143 \\
%$G_{CNO}$ &  $\sgacno - \sgap \qcno/\qp$  & 7.546 \\
%$G_B$     &  $\sclb - \sgap \qb/\qp$      & 2.430 \\
%\hline
%\end{tabular}
%\end{table}
%
\begin{table}
\caption[alpp]{Results of variations of \spp, for quantities 
characterizing the solar interior
and for the neutrino fluxes. We present the calculated values of 
$\alpha$, see Eqs.~(\ref{alfa0}) and (\ref{lawf}), for the case of small 
and large variations. For this latter case also the variances
$\Delta \alpha$ are shown, Eq.~(\ref{alfa06}). For the dependence 
on the central temperature, the corresponding 
$\beta$ coefficients, Eq.~(\ref{FiTc}), are also presented.}
\begin{tabular}{lcccc}
%\hline
%\hline
\multicolumn{1}{l}{\spp/\spp$^*$}
&\multicolumn{1}{c}{0.9--1.1} & \multicolumn{1}{c}{1--3.5}
&\multicolumn{1}{c}{0.9--1.1} & \multicolumn{1}{c}{1--3.5}\\
\multicolumn{1}{l}{coefficient}
&\multicolumn{1}{c}{$\alpha$} & \multicolumn{1}{c}{$\alpha \pm \Delta \alpha$}
&\multicolumn{1}{c}{$\beta$} & \multicolumn{1}{c}{$\beta \pm \Delta \beta$}\\
\hline
T  				& 
-0.11 & -0.11 $\pm$ 0.001 &  -  &     -         \\
$\rho$ & -0.37 & -0.37 $\pm$ 0.01  & 3.3 & 3.3 $\pm$ 0.05 \\ 
P      & -0.49 & -0.51 $\pm$ 0.01  & 4.4 & 4.6 $\pm$ 0.1  \\
R      & +0.12 & +0.13 $\pm$ 0.003 & -1.1&-1.1 $\pm$ 0.03 \\ 
X      & -0.01 & -0.03 $\pm$ 0.004 & 0.1 & 0.3 $\pm$ 0.04\\
\hline
$\fip$ & +0.11 & +0.07 $\pm$ 0.01  & -0.9&-0.6 $\pm$ 0.1   \\
$\fibe$& -1.02 &- 1.1 $\pm$ 0.04  &  8  &  9  $\pm$ 0.3   \\
$\ficno$&-2.7  & -2.2  $\pm$ 0.3   & 21  & 18  $\pm$ 2     \\
$\fib$ & -2.7  & -2.7  $\pm$ 0.1   & 21  & 22  $\pm$ 1     \\
%\hline 
%\hline   
\end{tabular}          
\label{talpp}
\end{table}
\begin{table}
\caption[alpp]{Variations of opacity. Same notation as in Table~\ref{talpp}.}
\begin{tabular}{lcccc}
%\hline
%\hline
\multicolumn{1}{l}{$opa$}
&\multicolumn{1}{c}{0.9--1.1} & \multicolumn{1}{c}{0.6--1}
&\multicolumn{1}{c}{0.9--1.1} & \multicolumn{1}{c}{0.6--1}\\
\multicolumn{1}{l}{coefficient}
&\multicolumn{1}{c}{$\alpha$} & \multicolumn{1}{c}{$\alpha \pm \Delta \alpha$}
&\multicolumn{1}{c}{$\beta$} & \multicolumn{1}{c}{$\beta \pm \Delta \beta$}\\
\hline
T  				& 
+0.12& +0.12 $\pm$ 0.003 &  -  &     -         \\
$\rho$ & -0.06 & -0.05 $\pm$ 0.006 &-0.5 &-0.4 $\pm$ 0.05 \\ 
P      & -0.08 & -0.07 $\pm$ 0.003 &-0.7 &-0.6 $\pm$ 0.03 \\
R      &   0   & +0.01 $\pm$ 0.01  & 0   &+0.1 $\pm$ 0.1 \\ 
X      & -0.28 & -0.28 $\pm$ 0.01  &-2.3 &-2.3 $\pm$ 0.1 \\
\hline
$\fip$ &-0.11  & -0.09 $\pm$ 0.01  & -0.9&-0.68 $\pm$ 0.1\\
$\fibe$&+1.1   & +1.2  $\pm$ 0.1   & +8.5&+9.1 $\pm$ 1     \\
$\ficno$&+1.7  & +1.7  $\pm$ 0.1   &13.4 &+13  $\pm$ 1     \\
$\fib$ & +2.4  & +2.6  $\pm$ 0.2   & 19  & 20.1  $\pm$ 1     \\
%\hline 
%\hline   
\end{tabular}          
\label{talopa}
\end{table}
\begin{table}
\caption[alpp]{Variations of Z/X. Same notation as in Table~\ref{talpp}}
\begin{tabular}{lcccc}
%\hline
%\hline
\multicolumn{1}{l}{(Z/X)/(Z/X)$^*$}
&\multicolumn{1}{c}{0.9--1.1} & \multicolumn{1}{c}{0.1--1}
&\multicolumn{1}{c}{0.9--1.1} & \multicolumn{1}{c}{0.1--1}\\
\multicolumn{1}{l}{coefficient}
&\multicolumn{1}{c}{$\alpha$} & \multicolumn{1}{c}{$\alpha \pm \Delta \alpha$}
&\multicolumn{1}{c}{$\beta$} & \multicolumn{1}{c}{$\beta \pm \Delta \beta$}\\
\hline
T  				& 
+0.06 & 0.05  $\pm$ 0.005 &  -  &     -         \\
$\rho$ & +0.03 & +0.02 $\pm$ 0.003 & 0.5 & 0.4 $\pm$ 0.06 \\ 
P      & 0.006 & +0.001$\pm$ 0.001 & 0.1&0.02 $\pm$ 0.02 \\
R      & -0.04 & -0.01 $\pm$ 0.004 &-0.7 &-0.2 $\pm$ 0.1 \\ 
X      & -0.2  & -0.13 $\pm$ 0.01  &-3   &-2.6 $\pm$ 0.2 \\
\hline
$\fip$ &-0.06  & -0.04 $\pm$ 0.01  & -0.9&-0.7 $\pm$ 0.1   \\
$\fibe$&+0.62  &+0.54  $\pm$ 0.03  &  9.9&10.6 $\pm$ 0.4  \\
$\ficno$&+2    & +1.7  $\pm$ 0.1   &  31 & 33  $\pm$ 2     \\
$\fib$ & +1.3  & +1.1  $\pm$ 0.1   &  21 & 21  $\pm$ 1     \\
%\hline 
%\hline   
\end{tabular}          
\label{talz}
\end{table}
\begin{table}
\caption[alpp]{Variations of solar age. Same notation as in 
Table~\ref{talpp}.
}
\begin{tabular}{lcccc}
%\hline
%\hline
\multicolumn{1}{l}{t$_\odot$/t$_\odot ^*$}
&\multicolumn{1}{c}{0.9--1.1} & \multicolumn{1}{c}{0.1--1}
&\multicolumn{1}{c}{0.9--1.1} & \multicolumn{1}{c}{0.1--1}\\
\multicolumn{1}{l}{coefficient}
&\multicolumn{1}{c}{$\alpha$} & \multicolumn{1}{c}{$\alpha \pm \Delta \alpha$}
&\multicolumn{1}{c}{$\beta$} & \multicolumn{1}{c}{$\beta \pm \Delta \beta$}\\
\hline
T  				& 
+0.03 & +0.02 $\pm$ 0.007 &  -  &     -         \\
$\rho$ & +0.18 & +0.11 $\pm$ 0.04  &+6   & +5.5 $\pm$ 2    \\ 
P      & +0.12 & +0.08 $\pm$ 0.03  &+4   & +4  $\pm$ 2    \\
R      & -0.08 & -0.05 $\pm$ 0.01  &-3   & -2.5 $\pm$ 0.5 \\ 
X      & -0.2  & -0.13 $\pm$ 0.05  &-7   & -6.5 $\pm$ 2.5 \\
\hline
$\fip$ &-0.08  & -0.04 $\pm$ 0.01  & -1.4&-0.8 $\pm$ 0.1   \\
$\fibe$&+0.57  & +0.5  $\pm$ 0.1   & +10 &+11  $\pm$ 1  \\
$\ficno$&+0.9  & +0.5  $\pm$ 0.2   & +16 &+12  $\pm$ 3     \\
$\fib$ & +1.   & +0.8  $\pm$ 0.2   & +18 &+20  $\pm$ 1     \\
%\hline 
%\hline   
\end{tabular}          
\label{talt}
\end{table}
\begin{table}
\caption[gg]{The power law coefficients for the reaction rates as a function
of temperature, Eq.~(\ref{paramrate}), calculated for $T=15.6\cdot 10^6$K.}
\begin{tabular}{lc}
%\hline
%\hline
reaction & $ \gamma =\frac{\d ln \langle \sigma v \rangle}{\d lnT}$ \\
&\\
\hline
p+p       & 4 \\
\tHe+\tHe & 16\\
\tHe+\qHe & 16\\
p+\sBe    & 13\\
e+\sBe    &-0.5\\
p+\quaN     & 20 \\
%\hline
\end{tabular}
\label{gammas}
\end{table}
\begin{table}
\caption[ffff]{The $\beta$ coefficients connecting the neutrino fluxes 
with the temperature, Eq.~(\ref{FiTc}).
The components of neutrino flux are indicated in the first column.
The values presented are the best fit to the 
numerical calculations performed when each input parameter is 
varied in the range specified in the second row. }
\begin{tabular}{lr@{}lr@{}lr@{}lr@{}l}
%\hline
%\hline
Parameter       &\multicolumn{2}{c}{$ S_{pp} $} 
                  &\multicolumn{2}{c}{Opacity} 
                             &\multicolumn{2}{c}{Z/X} 
                                      &\multicolumn{2}{c}{t$_\odot $} \\
Scaling factor      &\multicolumn{2}{c}{ 1 -- 3.5 } 
                  &\multicolumn{2}{c}{0.6 -- 1 } 
                             &\multicolumn{2}{c}{ 0.1 -- 1 } 
                                      &\multicolumn{2}{c}{ 0.1 -- 1 } \\
\hline
$\fipp$   & -0. & 6 & -0. & 7 & -0. & 7 & -0. & 8 \\
$\fipep$  &  2. & 2 & -2. & 3 & -1. & 7 &  0. & 5 \\
$\fip=\fipppep$     & -0. & 6 & -0. & 7  & -0. & 7  & -0. & 8 \\
$\fibe$ &  9 &  &  9 &  & 11 &   & 11 &   \\
$\fin$&15  &   & 12 &   & 31 &   &  9 &  \\
$\fio$&24  &  & 15  &   & 36 &   & 18 &   \\
$\ficno=\fin+\fio$   & 18  &   &  13&   & 33 &   & 12  &   \\
$\fib$  & 22 &   & 21 &  & 20 &   & 20 &   \\
%\hline
%\hline
\end{tabular}
\label{betas}
\end{table}
\begin{table}
\caption[tt]{Values of the coefficients $\alpha_{T,X}$
relating temperature to the input parameter, 
$T_c=T_c^*(X/X^*)^{\alpha_{T,X}}$ see Eq.~(\ref{FiTc}), our estimated 
uncertainties of the  input parameters ($\Delta X/X$), and 
variations  ($\delta X/X$)
required to reduce $T_c$ by 3\%, 7\% and 13\%, respectively.}
\begin{tabular}{cccccc}
%\hline
%\hline
$X$            & $\alpha_{T,X}$ & $\ssm$ &$\tre$   & $\sette$ & $\boh$\\
\hline
S$_{pp}$       & -0.13         & 1\%    & +25\%   &+70\% & +190\%\\
Opacity        & +0.13         & 5\%    & -20\%   &-40\% &  -66\%\\
Z/X            & +0.06         & 10\%    & -40\%   &-70\% &  -90\%\\
T$_\odot$      & +0.05         & 0.6\%    & -45\%   &-75\% &  -94\%\\
%\hline
%\hline
\end{tabular}
\label{tempera}
\end{table}
\begin{table}[htb]
\caption{Estimates for $S_{33}(0)$ [MeVb].}
\begin{tabular}{lc}
%\hline
CF88\cite{CF88} & 5.57\\
PA91 \cite{PA91} &5.0$\pm$0.3\\
All data         &5.3$\pm$0.2\\
$E>100$KeV       &5.2$\pm$0.2\\
adiabatic scr.   &5.1$\pm$0.2\\
%\hline
\end{tabular}
\label{S033}
\end{table}
\begin{table}[htb]
\caption{Estimates for $S_{34}(0)$ [keVb].}
\begin{tabular}{lc}
%\hline
CF88\cite{CF88} & 0.54\\
PA91 \cite{PA91} & 0.533$\pm$0.017\\
quadratic        & 0.48$\pm$0.01\\
exponential      & 0.51$\pm$0.01\\
%\hline
\end{tabular}
\label{S034}
\end{table}
\begin{table}
\caption[chi]{Variation of $\chi$= $S_{34}/ \sqrt{S_{33}}$.
We present the coefficients $\alpha_i$ for the parameterization
$\fii=\fii^* (\chi/\chi^*)^{\alpha_\i}$, 
for $\chi$ in the range indicated in the first row.
In the second column the values of Ref. \cite{Bahcall1989} are shown.
For the case of large variations, the variances $\pm \Delta \alpha$ are also
presented.} 
\begin{tabular}{lccc}
%\hline
%\hline
$\chi / \chi^*$  & \multicolumn{1}{c}{ 0.9--1.1} 
& \multicolumn{1}{c}{ 0.9--1.1}  & \multicolumn{1}{c}{ 0.1--1}\\
reference          & \multicolumn{1}{c}{\cite{Bahcall1989}}
 & \multicolumn{1}{c}{ this work} & \multicolumn{1}{c}{ this work }\\
\hline
$\fip$   & -0.06     &  -0.05   & -0.05 $\pm$ 0.01 \\
$\fibe$  & +0.86     & +0.86    &  +0.92 $\pm$ 0.02 \\
$\ficno$ & -0.05     & -0.04    & -0.02 $\pm$  0.01 \\
$\fib$   &+0.80      & +0.92    & 0.91 $\pm$  0.02 \\
%\hline
%\hline
\end{tabular}
\label{tabalfas33}
\end{table}
\begin{table}
\caption[sss]{Different determinations of $S(0)_{17}$, from \cite{koonin2}}
\begin{tabular}{lc}
%\hline
%\hline
Ref.     &  $S(0)_{17}$ [eV b] \\
\hline
\cite{KAV60} & $15\pm6$ \\
\cite{PAR68} & $27\pm4$ \\
\cite{KAV69} & $25.2\pm2.4$\\
\cite{VAU70} & $19.4\pm2.8$\\
\cite{WIE77} & $41.5\pm9.3$\\
\cite{FIL83} & $20.2\pm2.4$\\
%\hline
%\hline
\end{tabular}
\label{tabs17}
\end{table}
\begin{table}
\caption[sss]{Power law coefficients for variations of S$_{17}$.
Same notation as in Table~\ref{talpp}.}
\begin{tabular}{lc}
%\hline
%\hline
\multicolumn{1}{l}{$S_{17}$/$S_{17}^*$}
&\multicolumn{1}{c}{0.1--10}\\
\hline
$\fip$ & $(+1.1\pm 0.4)10^{-4}$ \\
$\fibe$& $(-2.5\pm2.0)10^{-3}$  \\
$\ficno$&$(-1.7\pm0.7)10^{-4}$ \\
$\fib$  &$(0.996\pm0.003)$ \\
%\hline
%\hline
\end{tabular}
\label{tabalfas17}
\end{table}
\begin{table}%[htb]
\caption[aaaa]{Comparison among solar models with different screening 
predictions: NOS$=$no screening, WES$=$weak screening \cite{salp}, 
MIT$=$Mitler 1977~\cite{mitler}, GDGC$=$Graboske \etal~1973~\cite{gdgc} 
and CSK$=$Carraro \etal~1988~\cite{koonin}. We show the central temperature
$T_c[10^7$K], the helium abundance in mass Y, the metal fraction Z, the values
of each component of the neutrino flux $[10^9$cm$^{-2}$s$^{-1}]$, the 
calculated signals for the Chlorine (Cl) and the Gallium (Ga) experiments 
[SNU], from \cite{schermonoi}.}
\begin{tabular}{lr@{}lr@{}lr@{}lr@{}lr@{}lr@{}}
%\hline
%\hline
&\multicolumn{2}{c}{NOS}
  &\multicolumn{2}{c}{WES}
     &\multicolumn{2}{c}{MIT}
       &\multicolumn{2}{c}{GDGC}
           &\multicolumn{2}{c}{CSK}\\
\hline
$T_c$                               &   1. & 573
   &        1.  & 566               &   1. & 566
   &        1.  & 564               &   1. & 567  \\
Y                                   &   0. & 288
   &        0.  & 289               &   0. & 289
   &        0.  & 289               &   0. & 289 \\
Z ($\times 10^{2}$)                 &   1. & 85
   &        1.  & 85                &   1. & 84
   &        1.  & 84                &   1. & 85 \\
\hline
$pp$                                &  60. & 0
   &       59.  & 6                &  59. & 7
   &       60.  & 0                 &  59. & 7 \\
$pep$                               &   0. & 146
   &        0.  & 142               &   0. & 142
   &        0.  & 143               &   0. & 143 \\
$^7$Be                              &   4. & 82
   &        4.  & 97                &   4. & 93
   &        4.  & 79                &   4. & 94 \\  
$^8$B ($\times 10^{3}$)             &   5. & 51
   &        6. & 36                 &   6. & 13
   &        5. & 59                &   6. & 21  \\
$^{13}$N                            &   0. & 46
   &        0.  & 55                &   0. & 52
   &        0.  & 47                &   0. & 54 \\
$^{15}$O                            &   0. & 39
   &        0.  & 48                &   0. & 45
   &        0.  & 40                &   0. & 47 \\
\hline
Cl                                  &   7. & 7
   &         8. & 8                 &   8. & 5
   &         7. & 8                 &   8. & 6 \\               
Ga                                  & 130 &
   &       134  &                   & 133 &
  &        130  &                   & 134 & \\
%  \hline
%\hline
\end{tabular}
\label{solischermo}
\end{table} 
\begin{table}
\caption[aaaa]{Enhancement factors for different screening prescriptions, 
calculated in central solar conditions. Same notations of 
Table~\ref{solischermo}, from~\cite{schermonoi}.}
\begin{tabular}{lr@{}lr@{}lr@{}lr@{}lr@{}lr@{}}
%&\\
%\hline
&\multicolumn{2}{c}{{}{}{}}
&\multicolumn{2}{c}{WES}
     &\multicolumn{2}{c}{MIT}
       &\multicolumn{2}{c}{GDGC}
           &\multicolumn{2}{c}{CSK}\\
\hline
$p+p$        &~~ &~~~                     &   1. & 049
   &        1. &045                &   1. & 049
   &        1. &038              \\
He+He       &~~ &~~~                       &  1. & 213
   &        1. &176                & 1. &115
   &        1. & 158            \\
$^7$Be+p       &~~ &~~~                       &   1. & 213
   &        1.  &171                &   1. & 112
   &        1.  &169            \\
$^{14}$N+p         &~~ &~~~                        &   1. & 403
   &        1.  & 293               &   1. &192
   &        1.  & 324              \\  
%   \hline
\end{tabular}
\label{ffactors}
\end{table} 
 \begin{table}
 \caption[taa]{
 Best fits to the combined experimental results obtained by varying the 
 seven possible combinations of the parameters $T_c$, 
 $S_{33}^{res}$ and $S_{17}$. For comparison, the first two 
 rows also show the experimental and RSM results. The first column reports 
 the $\chi^2$/d.o.f (just the $\chi^2$ when the number of parameters 
 is greater or equal to the number of data);
 a dag (\dag) indicates that the value is not a local minimum, but it is the 
 lowest value within the explored region ($S_{17}\leq 5 S_{17}^{RSM}$ and 
 $S_{33}^{res}\leq 200 S_{33}^{RSM})$. The second column reports the best-fit 
 values of the parameters in units of the RSM, and the last three columns the 
 corresponding signals for the Gallium, Chlorine and KAMIOKANDE experiments.
 \label{tbl1}
                }
 \begin{tabular}{lccccc}
% \hline
% \hline
               & $\chi^2$/d.o.f.  & best-fit
               & $S_{Ga}$ & $S_{Cl}$ & $\fib^{Ka}$ \\
               &                  & values
               &  \multicolumn{2}{c}{[SNU]} & $10^6$~cm$^{-2}$s$^{-1}$\\
 \hline
 experiment      &  &  & $74\pm8$ & $2.55\pm0.25$  & $2.73\pm0.38$   \\
 \hline
  RSM &  58.5/3 &    & 137.0  & 9.3  & 6.62  \\
 \hline
 \hline
 $S_{17}$ & 50.7/2 & 0.28  & 125 & 4.00  & 1.85  \\
 \hline
 $T_c$ & 22.0/2 & 0.936 & 103  & 2.71  & 1.54  \\
 \hline
 $S_{33}^{res}$ & 18.1/2 & 13.9  & 99  & 2.90  & 1.72  \\
 \hline
 \hline
 $T_c$     & 16.1/1 & 0.973 & 97 & 2.68 & 1.66  \\
 $S_{33}^{res}$ &        & 4   &    &      &       \\
 \hline
 $S_{33}^{res}$       & $13.9^{\mbox{\dag}}/1$ & 200 & 96  & 3.03  & 1.96  \\
 $S_{17}$        &            & 4.2 &   &   &             \\
 \hline
 $T_c$     & $10.7^{\mbox{\dag}}/1$ & 0.879  & 94  & 2.74  & 1.94 \\
 $S_{17}$  &            & 5.0    &     &       &      \\
 \hline
 \hline
 $T_c$      & 7.4$^{\mbox{\dag}} $ & 0.929 & 90 & 2.76 & 2.07 \\
 $S_{33}^{res}$  &      & 9   &    &      &      \\
 $S_{17}$      &      & 5.0   &    &      &      \\
% \hline
% \hline
 \end{tabular}
 \end{table}
 \begin{table}
 \caption[tbb]{
 The best fits when just two experimental results are included.
 Same notations as in Table~\ref{tbl1}.
 \label{tbl2}
                }
 \begin{tabular}{lcccccc}
% \hline
% \hline
         & \multicolumn{2}{c}{Ga + Ka} & \multicolumn{2}{c}{Cl + Ga}
         & \multicolumn{2}{c}{Cl + Ka}\\
\cline{2-3} \cline{4-5} \cline{6-7}
                 & $\chi^2$/d.o.f.  & best
               & $\chi^2$/d.o.f.  & best
               & $\chi^2$/d.o.f.  & best \\
 \hline
  RSM &  39.6/2 & & 50.2/2  &   & 41.9/2  & \\
 \hline
 \hline
 $S_{17}$ & 35.0/1 & 0.50  & 30.5/1 & 0.12 & 22.8/1 & 0.21\\
 \hline
 $T_c$    &18.6/1  & 0.950 & 10.9/1   & 0.929  & 9.4/1  & 0.939  \\
 \hline
 $S_{33}^{res}$ & 13.3/1 & 6.8  & 8.18/1  & 22.2  & 8.87/1  & 13   \\
 \hline
 \hline
 $T_c$      & 12.9   & 0.987 & 7.3    & 0.975 & 7.5   & 0.972 \\
 $S_{33}^{res}$  &        & 4   &        & 6   &       &  3 \\
 \hline
 $S_{33}^{res}$   & 8.2$^{\mbox{\dag}}$  & 150  & 
6.4$^{\mbox{\dag}}$   & 200 &7.2  & 198  \\
 $S_{17}$    &          &  5.0  &          & 3.3 &         & 4.1  \\
 \hline
 $T_c$    & $7.8^{\mbox{\dag}}$   & 
0.890 & $5.3^{\mbox{\dag}}$    & 0.874  & $4.8^{\mbox{\dag}}$    & 0.880 \\
 $S_{17}$ &           & 5.0  &            & 5.0    &            & 5.0  \\
 \hline
 \hline
 $T_c$       &  4.9$^{\mbox{\dag}}$ & 0.936 & 
3.4$^{\mbox{\dag}}$  & 0.928 & 3.6$^{\mbox{\dag}}$  & 0.922 \\
 $S_{33}^{res}$   &          & 8   &          & 11   &          & 6   \\
 $S_{17}$    &          & 5.0   &          & 5.0   &          & 5.0  \\
% \hline
% \hline
 \end{tabular}
 \end{table}
\begin{table}
%\footnotesize
\caption[bb]{Comparison among solar models obtained with different versions
of the FRANEC code. The labels (a) to (e) corresponds to the models
defined in the Appendix A. Our best Standard Solar Model is (e).
The last column shows the helioseismological results.
}
\begin{tabular}{lr@{}lr@{}lr@{}lr@{}lr@{}lr@{}lr@l}
%\hline
%\hline
 &\multicolumn{2}{c}{(a)}
                   &\multicolumn{2}{c}{(b)}
                              &\multicolumn{2}{c}{(c)}
                                         &\multicolumn{2}{c}{(d)}
                                              &\multicolumn{2}{c}{(e)}
                                            &\multicolumn{2}{c}{Helioseism.}\\
 &\multicolumn{2}{c}{}
                   &\multicolumn{2}{c}{}
                              &\multicolumn{2}{c}{}
                                         &\multicolumn{2}{c}{}
						 &\multicolumn{2}{c}{our best}
                                                     &\multicolumn{2}{c}{}\\
\hline
t$_\odot$[Gyr]             &    
4.&57          &      4.&57     &  4.&57     &    4.&57    &  4.&57 & & \\
L$_\odot$[\unitaerg]       &    
3.&846        &      3.&843   &  3.&844      &    3.&843   &  3.&844  & &\\
R$_\odot$[10$^{10}$ cm]    &    
6.&961         &      6.&963    &  6.&959    &    6.&959   &  6.&960  &  & \\
(Z/X)$_{photo}$            &    
0.&0245        &      0.&0245   &  0.&0245   &    0.&0245  &  0.&0245 &   &\\
$\alpha$                   &    
2.&023         &      1.&774    &  1.&786    &    1.&904   &  1.&901 &   & \\ 
\hline
X$_{in}$                   &    
0.&699         &      0.&718    &  0.&722    &    0.&711   & 0.&711  &   &\\
Y$_{in}$                   &    
0.&284         &      0.&265    &  0.&261   &    0.&269    & 0.&269  &  &\\
Z$_{in}$                   &    
0.&0171       &      0.&0176  &  0.&0177  &    0.&0198     & 0.&0198 &   & \\
\hline
X$_{photo}$                &    
0.&699         &      0.&718    &  0.&722    &    0.&743   &  0.&744  &   & \\
Y$_{photo}$                &    0.&284         &      0.&265   &  
0.&261     &    0.&238   &  0.&238 & 0.233 -- 0.268\\
Z$_{photo}$                &    
0.&0171       &      0.&0176  &  0.&0177     &    0.&0182 &  0.&0182&   &\\
\hline
R$_b$/R$_\odot$              &    0.&738         &      0.&726    &  
0.&728    &    0.&716  &  0.&716 & 0.710 -- 0.716\\
$T_b$[10$^6$ K]            &    
1.&99          &      2.&10     &  2.&08     &    2.&17    &  2.&17 &     & \\
$c_b$[10$^7$ cm s$^{-1}$]  &    2.&11          &      2.&16     &  
2.&16     &    2.&22    &  2.&22 & 2.21 -- 2.25 \\
\hline
$T_c$[10$^7$ K]            &    
1.&555         &     1.&545     &    1.&542 &    1.&569  & 1.&569 & &\\
$\rho_c$[100 gr cm$^-3$]   &    
1.&524         &      1.&472    &    1.&470  &    1.&514   & 1.&518 & &\\
Y$_c$                      &    
0.&63          &       0.&61    &    0.&61   &    0.&63    & 0.&63 & & 
%\hline
% OPACITY                   &  OP&AL            &      
%OP&AL     &  OP&AL     & OP&AL   &OP&AL& &\\
%EOS                        &   S&88            &       
%S&88     &  OP&AL     & OP&AL   &OP&AL& &\\
%\hline
%\hline
\label{modellinoi}
\end{tabular}
\end{table}
\begin{table}
\centering
\caption[bb]{ Neutrino fluxes and signals  obtained with different versions
of the FRANEC code. Labels (a) to (e) correspond to the models
defined in  Appendix A. Our best prediction is (e).}
\begin{tabular}{lr@{}lr@{}lr@{}lr@{}lr@{}lr@{}lr@{}lr@{}l}
%\hline
%\hline
 &\multicolumn{2}{c}{(a)}
                   &\multicolumn{2}{l}{(b)}
                              &\multicolumn{2}{c}{(c)}
                                         &\multicolumn{2}{c}{(d)}
						 &\multicolumn{2}{c}{(e)}\\
 &\multicolumn{2}{c}{}
                   &\multicolumn{2}{l}{}
                              &\multicolumn{2}{c}{}
                                         &\multicolumn{2}{c}{}
 						 &\multicolumn{2}{c}{our best}\\
\hline
$\fipp$ [\unitanb]     &     60.&17      &    60.&37  &  60.&66    & 59.&76
&  59.&92 \\
$\fipep$[\unitanb]     &      0.&14     &    0.&14    &   0.&14  &  0.&14
&  0.&14 \\
$\fibe$ [\unitanb]     &     4.&58      &    4.&22    &   4.&09   &  4.&71
&  4.&49 \\
$\fin $  ~[\unitanb]   &     0.&39      &    0.&36    &   0.&35  &  0.&52
&  0.&53 \\
$\fio $  ~[\unitanb]   &     0.&33      &    0.&30    &   0.&29  &  0.&45
&  0.&45 \\
$\fib$  ~[\unitasb]    &     4.&73      &    4.&18    &   3.&95   &  5.&37
&  5.&16 \\
\hline
S$_{Ga}$[SNU]          &     126&     &    121&   & 120&     &130&      & 128&\\
S$_{Cl}$[SNU]          &     6.&9     &    6.&2   &   5.&9   &  7.&7    &  7.&4
%\hline
%\hline
\label{flussinoi}
\end{tabular}
\end{table}
\begin{table}
\caption[cccc]{
Astrophysical $S$-factors 
[MeV barn] and their derivatives with respect to energies $S'$ [barn]
for the RSM \cite{BP95} and for our models.}
\begin{tabular}{l l l l}
%\hline
%\hline
            &     RSM               &  models (a-d)   &  
model (e)  \\           
            &                        &                 &  
our best  \\             
\hline
$S(0)_{11}$ &  3.89$\times 10^{-25 }$&  3.89 $\times 10^{-25}$ &  
3.89 $\times 10^{-25}$\\ 
$S'(0)_{11}$&  4.52$\times 10^{-24}$ &  4.52 $\times 10^{-24}$ &  
4.52 $\times 10^{-24}$\\
$S(0)_{33}$ &  4.99                  &  5.00                   &  5.1 \\  
$S'(0)_{33}$& -0.9                   &  -0.9                   &  3.0 \\  
$S(0)_{34}$ &  5.24 $\times 10^{-4}$ &  5.33 $\times 10^{-4}$  &  
5.1   $\times 10^{-4}$\\ 
$S'(0)_{34}$& -3.1  $\times 10^{-4}$ & -3.10 $\times 10^{-4}$  & 
-4.23  $\times 10^{-4}$\\ 
$S(0)_{17}$ &  2.24 $\times 10^{-5}$ &  2.24 $\times 10^{-5}$  &  
2.24  $\times 10^{-5}$\\ 
$S'(0)_{17}$& -3.00 $\times 10^{-5}$ & -3.00 $\times 10^{-5}$  & 
-3.00 $\times 10^{-5}$\\ 
\hline
$S(0)_{{}^{12}\protect\text{C}+p} $ &1.45 $\times 10^{-3}$ &
1.40 $\times 10^{-3}$ & 1.40 $\times 10^{-3}$\\ 
$S'(0)_{{}^{12}\protect\text{C}+p}$ &2.45 $\times 10^{-4}$&  
4.24 $\times 10^{-3}$ & 4.24 $\times 10^{-3}$\\ 
$S(0)_{{}^{13}\protect\text{C}+p} $ &5.50 $\times 10^{-3}$& 
5.50 $\times 10^{-3}$ & 5.50 $\times 10^{-3}$\\ 
$S'(0)_{{}^{13}\protect\text{C}+p}$ &1.34 $\times 10^{-2}$&
1.34$\times 10^{-2}$ &1.34$\times 10^{-2}$\\ 
$S(0)_{{}^{14}\protect\text{N}+p} $ &3.29 $\times 10^{-3}$&  
3.32$\times 10^{-3}$ & 3.32$\times 10^{-3}$\\ 
$S'(0)_{{}^{14}\protect\text{N}+p}$ &-5.91$\times 10^{-3}$& 
-5.91$\times 10^{-3}$ &-5.91$\times 10^{-3}$ \\ 
$S(0)_{{}^{15}\protect\text{N}(p,\gamma){}^{16}\protect\text{O}}$& 
6.40$\times 10^{-2}$&6.40$\times 10^{-2}$& 6.40$\times 10^{-2}$ \\ 
$S'(0)_{{}^{15}\protect\text{N}(p,\gamma){}^{16}\protect\text{O}}$&
3.00$\times 10^{-2}$&3.00 $\times 10^{-2}$& 3.00 $\times 10^{-2}$ \\ 
$S(0)_{{}^{15}\protect\text{N}(p,\alpha){}^{12}\protect\text{C}}$&
7.80$\times 10$&  7.04  $\times 10$ &  7.04  $\times 10$\\ 
$S'(0)_{{}^{15}\protect\text{N}(p,\alpha){}^{12}\protect\text{C}}$&
3.51$\times 10^{2}$&  4.21$\times 10^{2}$ &4.21$\times 10^{2}$\\ 
$S(0)_{{}^{16}\protect\text{O}+p}$    &  9.40$\times 
10^{-3}$&9.40$\times 10^{-3}$ &9.40$\times 10^{-3}$\\ 
$S'(0)_{{}^{16}\protect\text{O}+p}$& -2.30$\times 
10^{-2}$ &-2.30$\times 10^{-2}$ & -2.30$\times 10^{-2}$  
%\hline
%\hline
\label{cross}
\end{tabular}
\end{table}

\newpage
%
%%%%%%%%%%%%%%%%%%FIGURES
\begin{figure}%[htb]
 \caption[cc]{The pp chain.
 The probability of the different branches are from
the Reference Solar Model \cite{BP95}.
 The neutrino energies E$_\nu$ are also 
 indicated.
 }
\label{catenapp}
\end{figure}
\begin{figure}%[htb]
\caption[cc]{The CNO cycle.}
\label{catenaCNO}
\end{figure}
\begin{figure}
\caption[cc]
{The photospheric helium mass fraction Y$_{photo}$ and the depth
of the convective zone (R$_b$/R$_\odot$):\\
a) as constrained by helioseismology (the dotted rectangle), see Sect.
\ref{structure};\\
b) as predicted by solar models without diffusion,  open circles from
top to bottom correspond to  \cite{DS96,CD,RCVD96,TCL,P94,BP95,Ciacio};\\
c) as predicted by solar model with 
helium diffusion,   full squares from top
to bottom  correspond to \cite{CD,BP92,P94});\\
d) as predicted by solar model with helium and heavy 
elements diffusion: the full circles, from top to bottom correspond to  
\cite{RCVD96,COX,P94,Ciacio,DS96}) and the full diamond indicates the 
RSM \cite{BP95}.
}
\label{figbarbara}
\end{figure}

\begin{figure}%[htb]
\caption[cc]{The solar neutrino spectrum, from \cite{Bahcall1989}.
For continuous sources,
the differential flux is in cm$^{-2}$ s$^{-1}$ MeV$^{-1}$.
 For the lines, the 
total flux is in cm$^{-2}$ s$^{-1}$.
}
\label{spettro}
\end{figure}
\begin{figure} %[htb]
\caption[cc]{For the indicated components, $df$ is the fraction of neutrinos 
 produced inside the sun within $dR$.
On the bottom (top) scale the radial (mass) coordinate is indicated.
}
\label{profilo}
\end{figure}
\begin{figure}
\caption[lmts1ex]{
The \oB and \sBe neutrino fluxes, consistent with
the luminosity constraint and experimental results, for
standard neutrinos.
The dashed (solid) lines correspond to central ($\pm1 \sigma$)
experimental values for Chlorine (Cl), Gallium (Ga) and
KAMIOKANDE (Ka), see Eqs. (\ref{Scl},\ref{Sga},\ref{ska1}).
The hatched area corresponds to the region within
$2\sigma$ from each experimental result. The diamond
represents the prediction of 
the Reference Solar Model \cite{BP95}, and the bars
the estimated uncertainties.
}
\label{intb}
\end{figure}

\begin{figure}%[htb]
\caption[cc]{
Same as in Fig. \ref{intb}, for the \oB and 
\sBe + CNO neutrino fluxes. The luminosity constraint is 
supplemented with the
 estimates
for $\fipep/ (\fipep+\fipp)$ and $\fin/(\fin+\fio)$, from  the RSM \cite{BP95}.
The prediction of the RSM model  \cite{BP95} (full diamond) is shown, together 
with those of other solar models
including diffusion of helium and heavy elements (full circles), corresponding 
from right to left to \cite{P94,RCVD96,Ciacio,DS96}.
Solar model calculations without diffusion are represented by open circles, 
corresponding
to \cite{SDF,CL,CESAM,TCL}, again from right to left.
}
\label{intb2}
\end{figure}
\begin{figure}%[htb]
 \caption[cc]{The fate of \sBe nuclei.}
 \label{padre}
\end{figure}
%
%figure capitolo 4
\begin{figure}%[htb]
\caption[cc]{For $x=S_{pp}/S_{pp}^*=2$ the behaviour of 
several structure parameters,
$\O_i(x,m)/\O_i(m)^*$,
 as a function of 
 the mass coordinate
% (M/M$_\odot$)
 in the whole internal radiative region.
 The considered structural parameters are: radius (R), density ($\rho$), 
 temperature (T), pressure (P) and the hydrogen mass fraction (X).}
\label{omospp}
\end{figure}
\begin{figure}%[htb]
 \caption[cc]{Same as in Fig.~\ref{omospp}, for $opa$ = 0.7.
  }
\label{omopa}
\end{figure}
\begin{figure}%[htb]
\caption[cc]{Same as in Fig. \ref{omospp}, for (Z/X)/(Z/X)$^{SSM}$=0.5.}
\label{omoz}
\end{figure}
\begin{figure} %[htb]
\caption[cc]{The temperature profiles $T(m)$ normalized to $T(m)^*$
for models with the indicated values of the solar age. 
%SSM indicates the CDF93 solar model.
}
\label{tempeta}
\end{figure}
\begin{figure}%[htb]
\caption[cc]{Same as in Fig. \ref{omospp}, for the model with 
t$_\odot$ = 4 Gyr.}
\label{omoeta}
\end{figure}
\begin{figure}%[htb]
\caption[Temp]{The temperature profiles $T(m)$ normalized to $T^{SSM}$(m) 
for a few representative non-standard solar models, from \cite{PRD}}
\label{fig5}
\end{figure}
\begin{figure}%[htb]
\caption[beha]{The behaviour of $\fipp$, $\fibe$, and $\fib$ as a function 
 of the central temperature $T_c$ when varying $S_{pp}$, opacity, Z/X and 
 age, from \cite{PRD}.}
\label{fig7}
\end{figure}
\begin{figure}%[htb]
\caption[peak]{Relations among the temperatures $T$ at the $^7$Be and $pep$
peak production zones (R/R$_\odot=0.06$ and R/R$_\odot=0.09$, respectively) and
the central temperature $T_c$ in non-standard solar models. Data from
numerical calculations are shown with the same symbols as
in Fig.~\ref{fig7}, while full lines show the homology relations
$T_i= T_c \,( T_i^*/T_c^*)$, from \cite{PRD}.}
\label{fig13}
\end{figure}
\begin{figure}[htb]
\caption[dati33]{The \tHe(\tHe,\qHe)2p reaction. Experimental data
for the S-factor as a function of CM energy (lower scale). The corresponding
temperature $T_9(=T\cdot10^{-9}) $, such that $E_o(T_9)=E_{cm}$, is also 
indicated (upper 
scale). The full curve corresponds to Eq.~(\ref{poli33}). The Gamow peak 
for the solar center is indicated by the arrows. The listed  symbols 
corresponds to data in~\cite{GO54,KR87,DW74,DW71,WA66,BA67}, from 
top to bottom.}
\label{datis33}
\end{figure}
\begin{figure}
\caption[dati34]{The \tHe(\qHe,$\gamma$)\sBe reaction. Experimental data
for the S-factor as a function of CM energy (lower scale). The corresponding
temperature $T_9$, such that $E_o(T_9)=E_{cm}$, is also indicated
(upper scale). The full (dashed) curve corresponds to Eq. ~(\ref{poli34}) 
(\ref{expo34}). 
The listed  symbols corresponds to data 
in~\cite{HO59,PA63,NA69,KR82,OS84,AL84,HI88,OS84,RO83}, from top to bottom.} 
\label{datis34}
\end{figure}
\begin{figure}
\caption[tc]{The central temperature $T_c$ as a function of $S_{33}$.}
\label{temps33}
\end{figure}
\begin{figure}
\caption[tc]{ The flux of (a) $^7$Be neutrinos
and (b)  $^8$B neutrinos as a function of 
$\chi=S_{34}/S_{33}^{1/2}$ for several values of $S_{33}$, 
as calculated by using 
the FRANEC code. The dashed line corresponds to Eq.~(\ref{scala3He}),
from \cite{AA}.}
\label{flussis33}
\end{figure}
\begin{figure}
\caption[tc]{Proton capture (a) and Coulomb dissociation (b) reactions.}
\label{diagrammi}
\end{figure}
\begin{figure}
\caption[tc]{The central temperature $T_c$ as a function of $S_{17}$.}
\label{TCS17}
\end{figure}
\begin{figure}
\caption[schermi]{Enhancement factors along the solar profile. The results 
of weak screening \cite{salp} (dashed curves), Graboske \etal~1973~\cite{gdgc}
(dot-dashed curves), Mitler 1977~\cite{mitler} (solid curves) and 
Carraro \etal~1988~\cite{koonin}(dotted curves) are shown, 
from~\cite{schermonoi}.}
\label{figffactors}
\end{figure}
%
% figure cap 6
\begin{figure}
 \caption[dpndnc]{
  Dependence of the main solar fluxes on  $T_c$, $S_{17}$ and 
$S_{33}$ (this latter parameterized
  by its zero-energy resonant contribution $S_{33}^{res}$).
  The coefficients $Q_i$ are defined in Eq.~(\ref{qf}).
  }
  \label{dependence}
\end{figure}
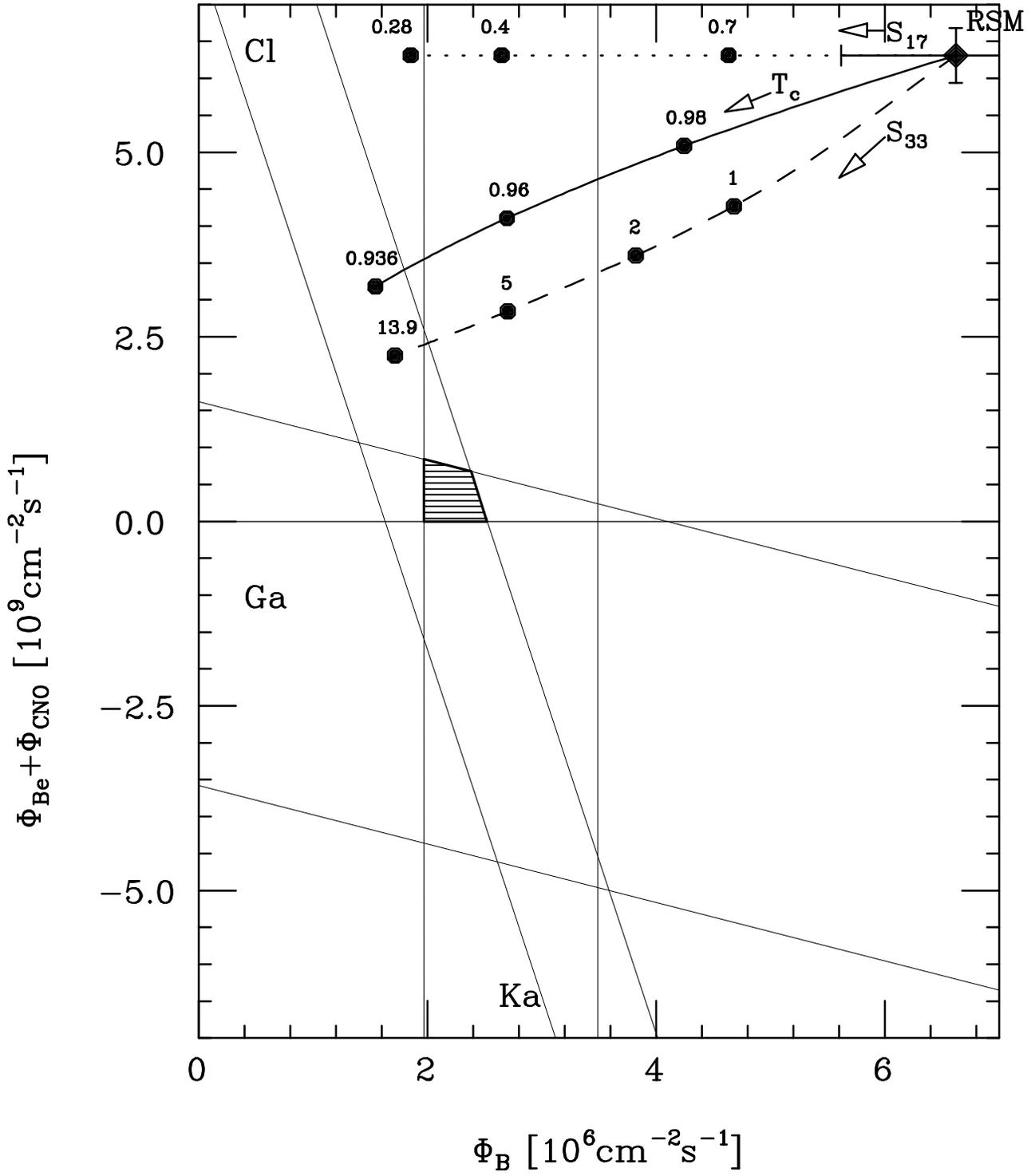
\begin{figure}[htb]
 \caption[ff2]{\sBe plus CNO neutrino fluxes vs. \oB neutrino flux.
 The three stripes, labeled Ga, Cl and Ka, confine the regions allowed at the 
 $2\sigma$ level by the three current experimental data (GALLEX + SAGE,
 Chlorine, and KAMIOKANDE) with the constraint due to the luminosity 
 sum rule. The hatched area emphasizes the interception of the three regions.
 The diamond shows the RSM prediction.
 The solid (dashed, dotted) line shows the effect of decreasing $T_c$
 (increasing the resonant part of $S_{33}$, decreasing $S_{17}$).
 Dots indicate the fluxes at specific values of the parameters
 shown by the label in units of their RSM values.}
 \label{fig2}
 \end{figure}
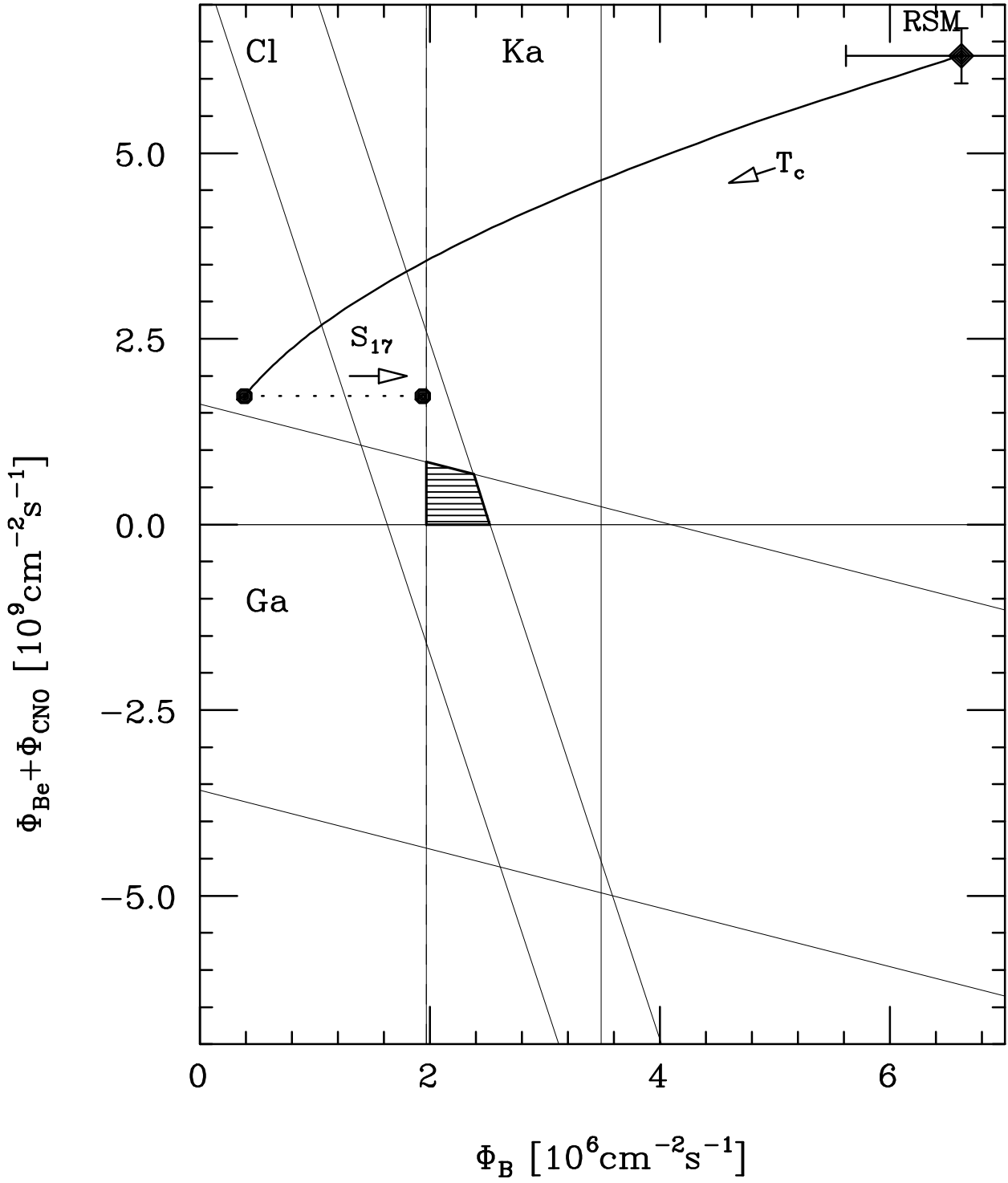
 \begin{figure}
 \caption[ff3]{Similar to Fig.~\ref{fig2}, but this time two parameters are 
 changed in the same model. The temperature is decreased down to $T_c=0.88$ 
 (solid curve) and, then, $S_{17}$ is increased up to 5 times its RSM value 
 (dotted line).}
 \label{fig3}
 \end{figure}
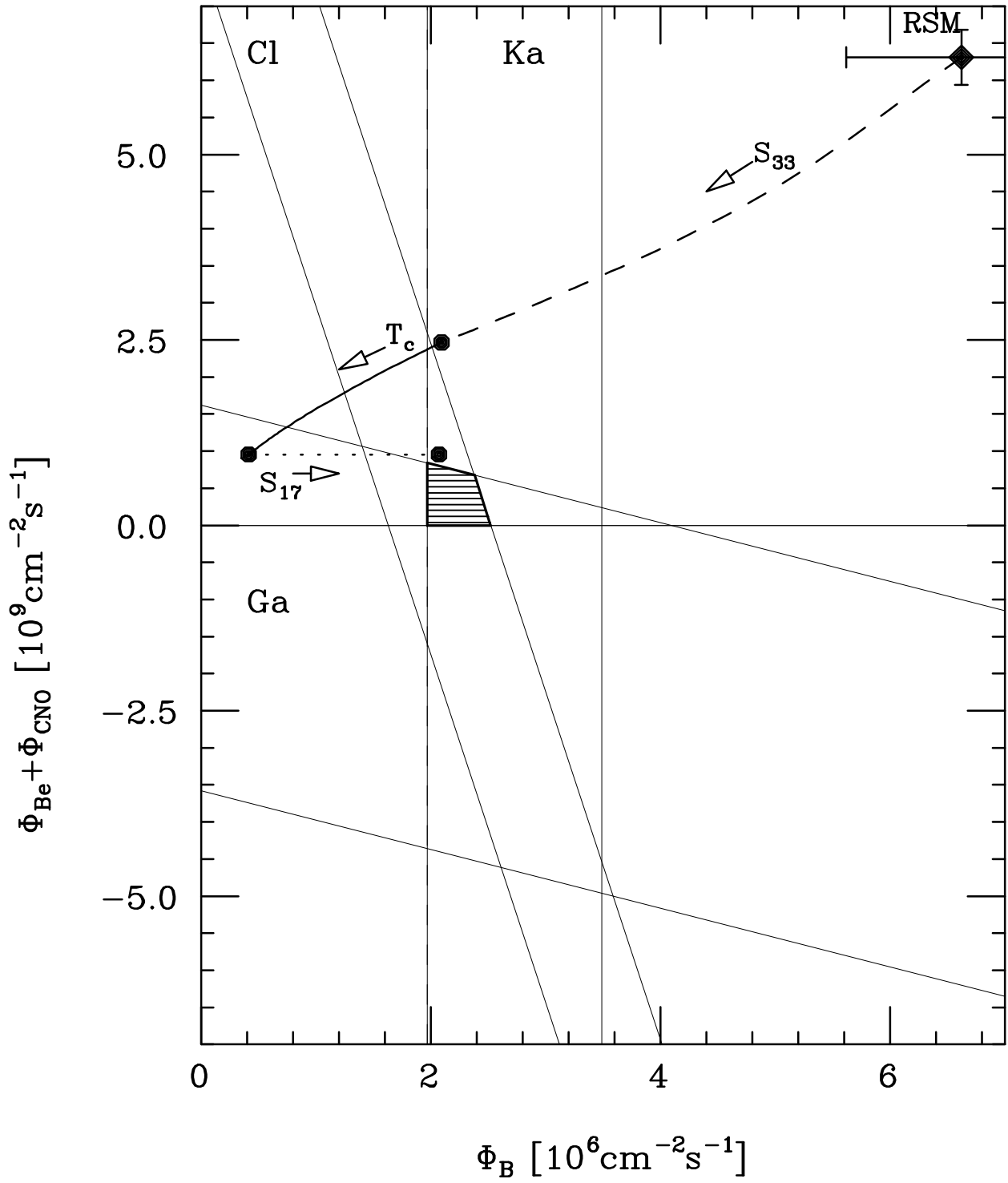
 \begin{figure}
 \caption[ff4]{
 Similar to Fig.~\ref{fig2}, but this time all three parameters are 
 changed in the same model. First $S_{33}^{res}$ is increased from
 zero up to 9 times the RSM value of $S_{33}$ (dashed curve), then $T_c$ 
 is decreased down to 0.929 (solid curve) and, finally, $S_{17}$ is
 increased up to 5 times its RSM value (dotted line). The order of
 the transformations is obviously inconsequential.
 }
 \label{fig4}
 \end{figure}
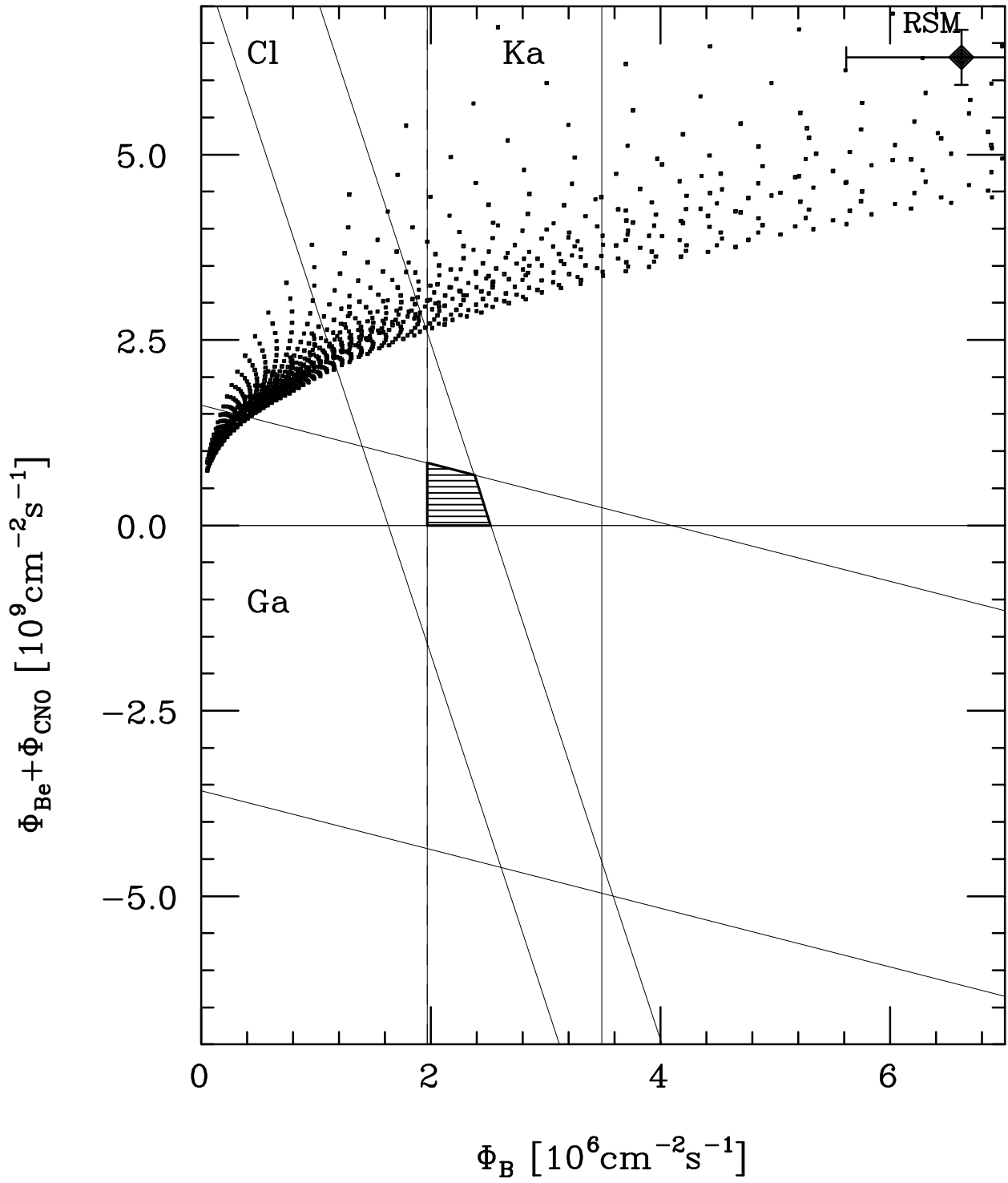
 \begin{figure}
 \caption[ff4]{The dotted area shows the result
 of arbitrarily varying the enhancement factors $\fpp$, $\fhh$ and
 $\fpn$ up to $f=6$. We recall that in standard solar model calculations
 $f\lapprox 1.2$, see sect. \ref{cap5}.}
 \label{figschermo}
 \end{figure}
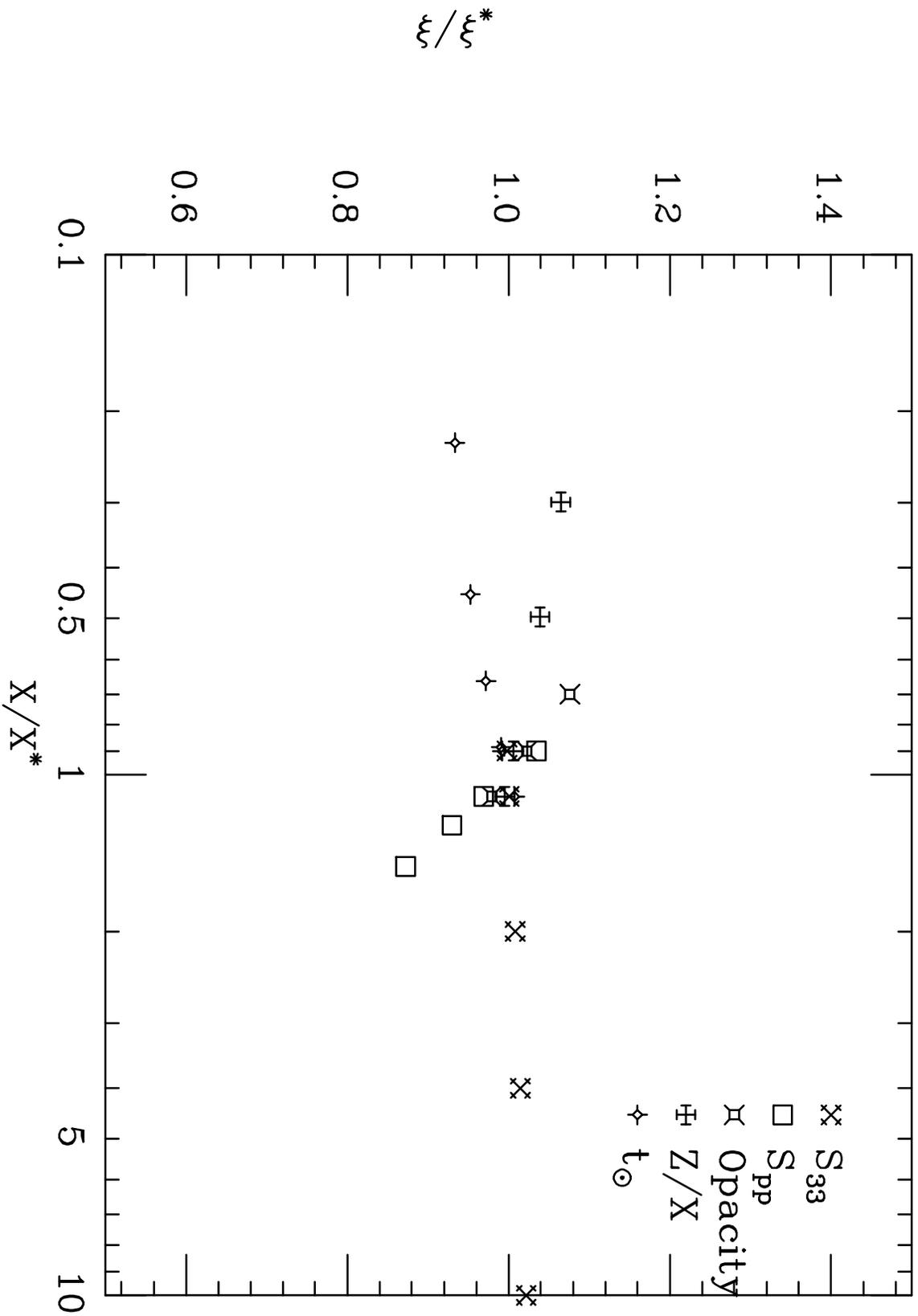
\begin{figure}
\caption[csicsi]{The behaviour of $\csi=\fipep/(\fipp+\fipep)$
 as one of the input 
parameters is varied by a scaling factor $ X/X^*$.}
\label{figapp1}
\end{figure}
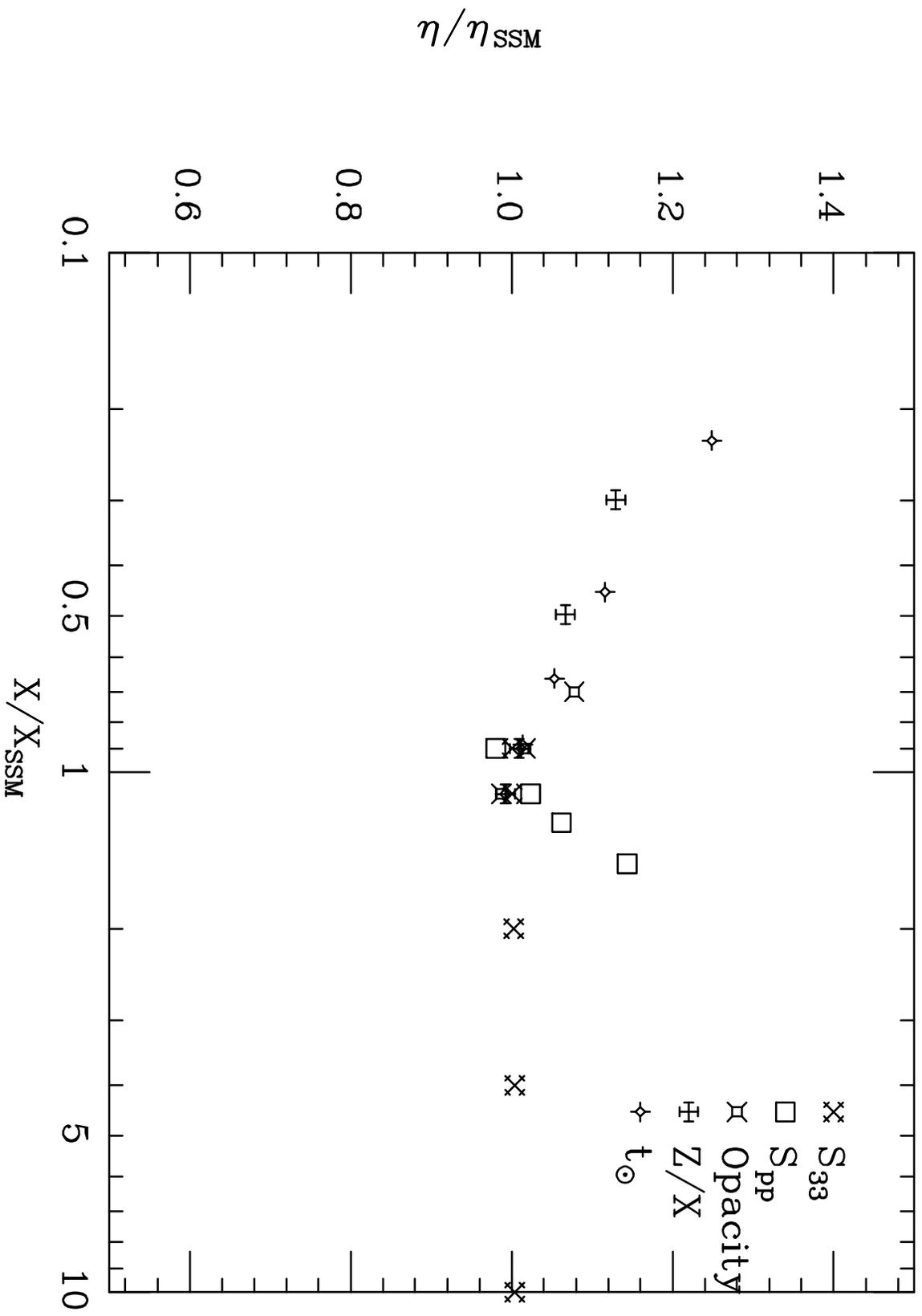
\begin{figure}
\caption[etaeta]{The behaviour of $\eta=\fin/(\fin+\fio)$ as one of the input 
parameters is varied by a scaling factor  $X/X^*$.}
\label{figapp2}
\end{figure}

%%%%%%%%%%%%%%%%%%
\end{document}